\documentclass[twocolumn]{aastex63}
\usepackage[caption=false]{subfig}

\received{} 
\revised{} 
\accepted{} 
\submitjournal{ApJ}

\shorttitle{Magnetic Fields and Filaments in NGC 1333}
\shortauthors{Doi {\em et al.}}
\graphicspath{{./}{figures/}}

\begin{document}

\title{The JCMT BISTRO Survey: Magnetic Fields Associated with a Network of Filaments in NGC 1333}

\correspondingauthor{Yasuo~Doi}
\email{doi@ea.c.u-tokyo.ac.jp}

\author[0000-0001-8746-6548]{Yasuo~Doi}
\affiliation{Department of Earth Science and Astronomy, Graduate School of Arts and Sciences, The University of Tokyo, 3-8-1 Komaba, Meguro, Tokyo 153-8902, Japan}

\author[0000-0003-1853-0184]{Tetsuo Hasegawa}
\affiliation{National Astronomical Observatory of Japan, National Institutes of Natural Sciences, Osawa, Mitaka, Tokyo 181-8588, Japan}

\author[0000-0003-0646-8782]{Ray S. Furuya}
\affiliation{Institute of Liberal Arts and Sciences, Tokushima University, Minami Jousanajima-machi 1-1, Tokushima 770-8502, Japan}

\author[0000-0002-0859-0805]{Simon Coud\'e}
\affiliation{SOFIA Science Center, Universities Space Research Association, NASA Ames Research Center, M.S. N232-12, Moffett Field, CA 94035, USA}
\affiliation{Centre de Recherche en Astrophysique du Qu\'ebec (CRAQ), Universit\'e de Montr\'eal, D\'epartement de Physique, C.P. 6128 Succ. Centre-ville, Montr\'eal, QC H3C 3J7, Canada}

\author[0000-0002-8975-7573]{{Charles L. H.} Hull}
\affiliation{National Astronomical Observatory of Japan, NAOJ Chile, Alonso de C\'ordova 3788, Office 61B, 7630422, Vitacura, Santiago, Chile}
\affiliation{Joint ALMA Observatory, Alonso de C\'ordova 3107, Vitacura, Santiago, Chile}
\affiliation{NAOJ Fellow}

\author{Doris Arzoumanian}
\affiliation{Instituto de Astrof\'isica e Ci{\^e}ncias do Espa\c{c}o, Universidade do Porto, CAUP, Rua das Estrelas, PT4150-762 Porto, Portugal}
\affiliation{Department of Physics, Graduate School of Science, Nagoya University, Furo-cho, Chikusa-ku, Nagoya 464-8602, Japan}

\author[0000-0002-0794-3859]{Pierre Bastien}
\affiliation{Institut de Recherche sur les Exoplan\`etes (iREx), Universit\'e de Montr\'eal, D\'epartement de Physique, C.P. 6128 Succ. Centre-ville, Montr\'eal, QC H3C 3J7, Canada}
\affiliation{Centre de Recherche en Astrophysique du Qu\'ebec (CRAQ), Universit\'e de Montr\'eal, D\'epartement de Physique, C.P. 6128 Succ. Centre-ville, Montr\'eal, QC H3C 3J7, Canada}

\author[0000-0003-4242-973X]{Michael Chun-Yuan Chen}
\affiliation{Department of Physics and Astronomy, University of Victoria, Victoria, BC V8P 1A1, Canada}

\author[0000-0002-9289-2450]{James Di Francesco}
\affiliation{Department of Physics and Astronomy, University of Victoria, Victoria, BC V8P 1A1, Canada}
\affiliation{Herzberg Astronomy and Astrophysics Research Centre, National Research Council of Canada, 5071 West Saanich Rd, Victoria, BC V9E 2E7, Canada}

\author[0000-0001-7594-8128]{Rachel Friesen}
\affiliation{National Radio Astronomy Observatory, 520 Edgemont Rd., Charlottesville, VA, 22903, USA}

% ...



\begin{abstract}
We present new observations of the active star-formation region NGC 1333 in the Perseus molecular cloud complex from the James Clerk Maxwell Telescope \emph{B}-Fields In Star-forming Region Observations (BISTRO) survey with the POL-2 instrument.
The BISTRO data cover the entire NGC 1333 complex $(\sim 1.5~\mathrm{pc} \times 2~\mathrm{pc})$ at 0.02 pc resolution and spatially resolve the polarized emission from individual filamentary structures for the first time.
The inferred magnetic field structure is complex as a whole, with each individual filament aligned at different position angles relative to the local field orientation.
We combine the BISTRO data with low- and high- resolution data derived from Planck and interferometers to study the multiscale magnetic field structure in this region.
The magnetic field morphology drastically changes below a scale of $\sim 1$ pc and remains continuous from the scales of filaments ($\sim 0.1$ pc) to that of protostellar envelopes ($\sim 0.005$ pc or $\sim 1000$ au).
Finally, we construct simple models in which we assume that the magnetic field is always perpendicular to the long axis of the filaments.
We demonstrate that the observed variation of the relative orientation between the filament axes and the magnetic field angles are well reproduced by this model, taking into account the projection effects of the magnetic field and filaments relative to the plane of the sky.
These projection effects may explain the apparent complexity of the magnetic field structure observed at the resolution of BISTRO data toward the filament network.
\end{abstract}



\section{Introduction}
\label{sec:introduction}

It has long been recognized that the interstellar medium (ISM) is full of structures that can be identified as spatially elongated ``filaments'' \citep[e.g.,][for a review]{1979ApJS...41...87S,1987ApJS...63..645U,1994ApJ...423L..59A,2008ApJ...680..428G,2009A&A...504..415S,2012A&ARv..20...55H}.
Recent observations of thermal dust emission using Herschel revealed the omnipresence of filaments in the ISM, and more importantly their direct connection to star-formation activity \citep[e.g.,][]{2010A&A...518L.102A,2010A&A...518L.106K,2010A&A...518L.103M,2010A&A...518L.100M,2010A&A...518L.104M,2011A&A...529L...6A,2012A&A...541A..63P}.
Most $(>70\%)$ of the prestellar cores and Class 0 young stellar objects (YSOs) in low-mass star-forming clouds are located on filaments, especially in regions of the filaments that are thermally supercritical and thus gravitationally unstable \citep[e.g.,][]{2010A&A...518L.102A,2014prpl.conf...27A,2015A&A...584A..91K,2016MNRAS.459..342M}.
Additionally, YSOs show an age dependence with their distance to the nearest filament, which is consistent with the scenario that YSOs are born within these filaments and then drift away after birth with random relative velocities of $\sim 0.1 ~ \mathrm{km ~ s^{-1}}$ (\citealp{2015PASJ...67...50D}; also see \citealp{2016A&A...590A...2S}).

All of these observations suggest that interstellar filaments play an essential role in the star-formation process.
Understanding the formation and evolution of these filaments is therefore crucial to understand the whole process of star formation.

Although the dominant mechanisms for filament formation and evolution are still under debate, many theoretical studies and numerical simulations suggest that interstellar magnetic fields (\emph{B}-fields, hereafter) may play a significant role \citep[e.g.,][]{1998ApJ...506..306N,2008ApJ...679L..97K,2011ApJ...728..123K,2008ApJ...687..354N,2011MNRAS.414.2511V,2013A&A...556A.153H,2013ApJ...774..128S,2017MNRAS.465.2254K,2018PASJ...70S..53I}.
Dense clouds are formed through collisional interactions in the ISM and the associated shock-compressed fluid dynamics.
In this process, the \emph{B}-field makes turbulent flows in the ISM more coherent along its field lines, naturally producing filamentary ISM structures that stretch in directions perpendicular to the mean \emph{B}-field lines (see  \citealp{2019FrASS...6....5H} for a recent review).

The plane-of-sky (POS) component of \emph{B}-fields in dense molecular clouds can be traced by polarimetric observations of thermal continuum emission from interstellar dust particles \citep{1993prpl.conf..279H,2007JQSRT.106..225L,2008MNRAS.388..117H,2009ApJS..182..143M,2012ARA&A..50...29C}.
Aspherical dust particles irradiated by starlight are charged up by the photoelectric effect, as well as spun up as a result of radiative torques (RATs; \citealp{1996ApJ...470..551D}; \citealp{1997ApJ...480..633D}; \citealp{2007MNRAS.378..910L}; \citealp{2008ApJ...676L..25L}; \citealp{2008MNRAS.388..117H}; \citealp{2014MNRAS.438..680H}; \citealp{2016ApJ...831..159H}; \citealp{2019ApJ...883..122L}).
These spinning particles are aligned with their rotation axes (i.e., their minor axes) parallel to the \emph{B}-field orientation, resulting in preferential thermal emission polarized in the direction perpendicular to the field lines \citep{1966ApJ...144..318S,1988QJRAS..29..327H}.

Polarized dust emission thus can provide a direct trace of the morphology of the interstellar \emph{B}-field.
For example, some studies based on Planck and BLASTPol observations (e.g.,  \citealp{2016A&A...586A.135P,2016A&A...586A.136P,2016A&A...586A.138P,2017A&A...603A..64S,2019ApJ...878..110F}) conclude that \emph{B}-fields are observed to be mostly perpendicular to the long axes of dense filaments.
Several ground-based polarimetry studies support these conclusions \citep{2017ApJ...846..122P,2017ApJ...842...66W,2018ApJ...859..151L,2019ApJ...883...95S}.

We note that the \emph{B}-field angle derived from polarization observations is the true 3D angle projected onto the POS.
To understand the actual structure of the \emph{B}-field and its relationship with surrounding ISM, we need to take into account their respective 3D structures (e.g., \citealp{2015ApJ...807...47T,2016A&A...586A.136P,2018A&A...614A.100T}).
Moreover, Planck and BLASTPol observations have limited spatial resolutions $(> 2'\negthinspace.5)$ that are not sufficient to trace the \emph{B}-field orientation within filaments.

Despite the importance of high-resolution polarimetry of filaments, observational studies of polarized dust emission that can spatially resolve individual star-forming regions had previously been limited to only the brightest parts owing to sensitivity limitations \citep[e.g.,][]{2009ApJS..182..143M}.
The recent addition of a polarimeter \citep[POL-2;][]{2011ASPC..449...68B,2016SPIE.9914E..03F} to the Submillimeter Common-User Bolometer Array 2 (SCUBA-2) camera \citep{2013MNRAS.430.2534D,2013MNRAS.430.2513H} on the James Clerk Maxwell Telescope (JCMT) now enables us to trace global \emph{B}-field structures in star-formation regions in greater detail (i.e., with spatial resolutions comparable to or below the spatial scale of prestellar cores).

To investigate the role of \emph{B}-fields in the star formation process in light of its association with filaments, we use submillimeter polarimetry observations from the B-fields in STar-forming Region Observations (BISTRO) survey, which utilizes SCUBA-2/POL-2 at the JCMT.
A full description of the survey is given by \citet{2017ApJ...842...66W}. This survey is designed to cover a variety of nearby star-forming regions, with an initial focus on the Gould Belt molecular clouds.
The regions published so far are
Orion A \citep{2017ApJ...846..122P},
M16 \citep{2018ApJ...860L...6P},
$\rho$ Ophiuchus A \citep{2018ApJ...859....4K},
$\rho$ Ophiuchus B \citep{2018ApJ...861...65S},
$\rho$ Ophiuchus C \citep{2019ApJ...877...43L},
Perseus B1 \citep{2019ApJ...877...88C},
and
IC 5146 \citep{2019ApJ...876...42W}.
In this paper, we present new observational results of the NGC 1333 star-forming region in the Perseus molecular cloud obtained by the BISTRO survey.

NGC 1333 is currently the most active site of ongoing star formation in the Perseus molecular cloud complex, and it is one of the most active star-formation regions within 300 pc of the Sun \citep{1987ApJS...63..645U,1996ApJ...473L..49B,2006A&A...451..539S,2008hsf1.book..308B,2008ApJ...683..822J,2008hsf1.book..346W}.
This region shows a strong concentration of interstellar material and is often regarded as a `hub' in the filamentary network of the Perseus region \citep{2009ApJ...700.1609M}.
Furthermore, NGC 1333 itself displays a complicated filamentary structure (\citealp{2017A&A...606A.123H}; also see Figures \ref{fig:I image} and \ref{fig:PI image} of this work).

\citet{2018ApJ...865...73O} estimated a distance to NGC 1333 of $293 \pm 22$ pc by using Gaia DR2 parallax measurements.
\citeauthor{2018ApJ...869...83Z} \citep[\citeyear{2018ApJ...869...83Z}; see also][]{2019ApJ...879..125Z} combined Gaia data with stellar photometry for individual CO velocity components and derived an average distance to NGC 1333 of
$299 \pm 3$ pc.
Following these results, we assume the distance to the region to be $299$ pc throughout this paper, which leads to a spatial resolution of 0.02 pc or 4200 au for the $14''\negthinspace\negthinspace.1$ JCMT beam at $850~\mu\mathrm{m}$ (FWHM).
The typical width of filaments suggested from Herschel observations is $\sim 0.1$ pc \citep{2011A&A...529L...6A,2012A&A...544A..14J,2014A&A...568A..98A,2015MNRAS.452.3435K,2016MNRAS.457..375F,2019A&A...621A..42A}, which can be well resolved by the JCMT beam.
NGC 1333 is, therefore, a suitable area to study the relationship between filaments and \emph{B}-fields over the whole star-forming area.
We aim to clarify this relationship in this paper.

This paper is organized as follows.
First, in Section \ref{sec:ObsReduce}, we describe our polarimetric observations of NGC 1333, the data reduction process, and verification of the data compared with other observations.
In Section \ref{sec:results}, we describe the spatial structure of the \emph{B}-field revealed by our BISTRO observations, with a focus on its relationship to the structure of the filaments.
In Section \ref{sec: multiscale}, we compare our data with low- and high- resolution data derived from Planck, interferometers, and optical and near-IR observations and discuss the multiscale distribution of the \emph{B}-field in and around NGC 1333.
In Section  \ref{sec: B association filament}, we introduce a simplifying model, in which the \emph{B}-field lines and the dense filaments are perpendicular to each other and differently oriented with respect to the POS.
We demonstrate that this simplifying model can reproduce observed relative orientation angles between filaments and the \emph{B}-field.
In Section \ref{close-relationship-between-filaments-and-b-field}, we discuss the close relationship between filaments and the \emph{B}-field revealed by our observation.
Finally, we summarize our results in Section \ref{sec:conclusions}.

\section{Observation and Data Reduction}\label{sec:ObsReduce}

\subsection{Observation}

We observed NGC 1333 in $850~\mu\mathrm{m}$ continuum using
SCUBA-2/POL-2 on the JCMT.
The JCMT has a primary diameter of 15 m and achieves an effective angular resolution of $14\arcsec\negthinspace\negthinspace.1$ at $850~\mu$m if we fit the beam with a single Gaussian profile \citep{2013MNRAS.430.2534D}.
Its new polarimeter POL-2 on the bolometer array SCUBA-2 \citep{2013MNRAS.430.2513H} achieves a superior sensitivity compared with the previous-generation instrument SCUBAPOL (SCUPOL; \citealp{2011ASPC..449...68B,2016SPIE.9914E..03F}).

We scanned the sky with POL-2 at a speed of $8 \arcsec\ \mathrm{s}^{-1}$ and a data sampling rate of 200 Hz with a POLCV\_DAISY spatial scanning pattern, which covers a field that is roughly $11\arcmin$ in diameter.
For this paper, the flux calibration factor (FCF) of POL-2 at 850 $\mu$m is assumed to be 725 Jy pW$^{-1}$ beam$^{-1}$ for each of the Stokes \emph{I}, \emph{Q}, and \emph{U} parameters.
This value was determined by multiplying the typical SCUBA-2 FCF of 537 Jy pW$^{-1}$ beam$^{-1}$ \citep{2013MNRAS.430.2534D} by a transmission correction factor of 1.35 measured in the laboratory and confirmed empirically by the POL-2 commissioning team using observations of the planet Uranus \citep{2016SPIE.9914E..03F}.

We observed NGC 1333 at two positions, namely, field 1 ($\alpha = 03^{\rm h}29^{\rm m}03.\hspace{-0.25em}^{\rm s}450,~\delta = 31^{\circ}14{\arcmin}34.\negthinspace\negthinspace{\arcsec}70$) and field 2 ($\alpha = 03^{\rm h}29^{\rm m}06.\hspace{-0.25em}^{\rm s}350,~\delta = 31^{\circ}20{\arcmin}22.\negthinspace\negthinspace{\arcsec}50$).
\hspace{-2.0em}
\footnote{The data are available on the Canadian Astronomy Data Centre (CADC) archive (\url{http://www.cadc-ccda.hia-iha.nrc-cnrc.gc.ca/}) under the names "PERSEUS NGC 1333 FIELD 1" and "PERSEUS NGC 1333 FIELD 2."}
The POLCV\_DAISY footprint has a central $3'$-diameter region with uniform coverage, and then the coverage decreases approximately linearly to the edge at $5.\negthinspace'5$ from the center.
Since our two observed positions are $5.\negthinspace'83$ apart, we cover the whole NGC 1333 region with approximately uniform coverage with these two spatial scans.
We observed the two fields between 2017 August 16 and 2018 January 16 as a part of the BISTRO large program (project ID: M16AL004).
We observed field 1 in the earlier half of the observational campaign before 2017 November 25 and observed field 2 in the subsequent observational period.
Twenty exposures ranging from 42.2 to 42.7 minutes were devoted to both field 1 and field 2, for a total observational time of 28.3 hr.
The atmospheric opacity at 225 GHz ($\tau_{225}$) ranged between 0.03 and 0.07, among which five observations of field 1 and four observations of field 2 were performed under very dry weather conditions (Grade 1; $\tau_{225} < 0.05$), and others were performed under dry weather conditions (Grade 2; $0.05 \leqslant \tau_{225} < 0.08$).
We conducted the observations when the target elevation angle was $> 30^\circ$.
The noise level of the observed data has a loose correlation with the expected air emission.
We find $\sim 30\%$ difference of the noise level corresponding to the estimated range of the air emission $\tau_{225}\cdot \mathrm{sec}(z) = 0.04$--0.09, where $z$ is a zenith angle of the target.
Since the difference in the noise level is insignificant, we use all the observational data to estimate the polarized and the total intensity maps.

\subsection{Pipeline Data Reduction and Regridding}

We derived the spatial distribution of the Stokes \emph{I}, \emph{Q}, and \emph{U} parameters from their time-series measurements by using the \sc{Starlink} \rm{procedure} $pol2map$ \citep[][software version on 2018 November 17]{Parsons2018}, which is adapted from the SCUBA-2 data reduction procedure $makemap$ \citep{2013MNRAS.430.2545C}.
We estimated this distribution on a grid with $4\arcsec \times 4\arcsec$ pixels, which is a standard for this analysis pipeline.

The grid of the pipeline output is considerably smaller than the spatial resolution of SCUBA-2/POL-2 ($14''\negthinspace\negthinspace.1$).
To match the pixel scale of our map to the spatial resolution of SCUBA-2/POL-2 and improve the signal-to-noise ratios (S/Ns), we resample the \emph{I}, \emph{Q}, \emph{U} data with a $7''\negthinspace\negthinspace.05$ grid by applying a 2D least-squares fit of a second-order polynomial using a Gaussian kernel with $\mathrm{FWHM} = 14''\negthinspace\negthinspace.1$.
To propagate error values estimated by \emph{pol2map} through the nonlinear polynomial fitting and estimate the error of the resampled $I$, $Q$, and $U$ data, we perform a Monte Carlo simulation by assuming a Gaussian distribution for the \emph{pol2map}-estimated error.
We repeat the polynomial fitting 1000 times with Gaussian random errors at each resampled position.
We take the mean of 1000 samples as the estimated values and the standard deviation as the estimation error of each position.
The estimation error tends to be smaller for the signals with better S/Ns, as those signals generally give better-fitting results.
In other words, the error for low-S/N signals is nonlinearly enhanced.
As a result, we exclude these low-S/N signals from the subsequent discussion (see below).

Using the resampled $Q$ and $U$ data, we estimate polarization intensity $PI$ and its standard error $\left( \delta PI\right)$ as follows:
\begin{eqnarray*}
PI &=& \sqrt{Q^2 + U^2}~,\\
\delta PI &=& \frac{\sqrt{(Q^2 \cdot \delta Q^2 + U^2 \cdot \delta U^2)}}{PI}~.
\end{eqnarray*}
The $PI$ shown above is a biased estimator because the errors in $Q$ and $U$ are squared and thus offsets the derived $PI$.
We debias the $PI$ value to estimate the true $PI$ as follows:
\begin{eqnarray*}
PI_\mathrm{debiased} &=& \sqrt{PI^2 - \delta{PI}^2}~.
\end{eqnarray*}
If the S/N is high (e.g., $PI / \delta PI > 3$), the difference between the $PI$ and $PI_\mathrm{debiased}$ is negligibly small \citep{1974ApJ...194..249W,1993A&A...274..968N,2006PASP..118.1340V}.
In the following, we refer to $PI_\mathrm{debiased}$ as the polarized intensity and describe this as $PI$ for simplicity.

Based on this debiased $PI$, we estimate polarization fraction $P$, polarization angle $\psi$, and their standard errors $\left(\delta P~\mathrm{and}~\delta\psi \right)$ as follows:
\begin{eqnarray*}
P &=& \frac{PI}{I}~,\\
\delta P &=& P \times\sqrt{\left(\frac{\delta PI}{PI}\right)^2 +\left(\frac{\delta I}{I}\right)^2}~,\\
\psi &=& 0.5 \times \arctan\frac{U}{Q}~,\\
\delta \psi &=& 0.5 \times \frac{\sqrt{ \left(Q \cdot \delta U \right)^2 + \left(U \cdot \delta Q \right)^2 }}{PI^2}~.
\end{eqnarray*}

\begin{figure*}[tp]
\centering
\includegraphics[width=\linewidth]{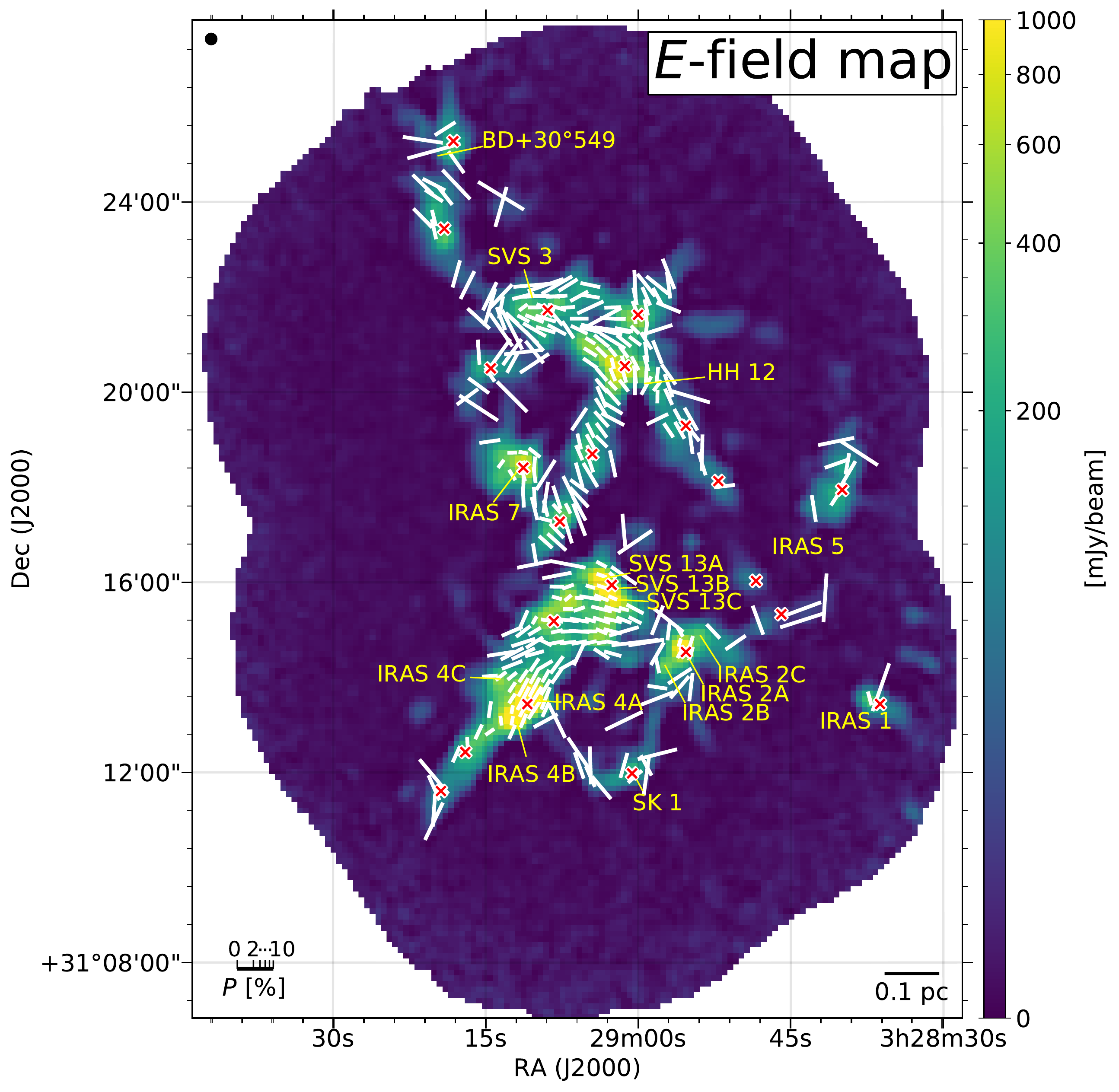}
\caption{
Total intensity and polarization position angles ($E$-field map) observed at $850~\mu\mathrm{m}$.
The color scale traces the Stokes $I$ total intensity.
Polarization position angles are shown for data points with $I \geqslant 25~\mathrm{mJy~beam^{-1}}~(I / \delta I > 10)$ and $PI / \delta PI \geqslant 3$.
The length of the line segments is proportional to $\sqrt{P}$.
A reference scale of $P$ is shown in the lower left corner of the figure, with a scale from 0 \% to 10 \% for every 2 \%.
The JCMT beam ($14''\negthinspace\negthinspace.1$) is shown in the upper left corner of the figure.
A reference scale for 0.1 pc is shown, in which we assume the distance to the source as 299 pc \citep{2018ApJ...865...73O, 2018ApJ...869...83Z}.
Names of main YSOs and infrared sources tabulated by \citet{2001ApJ...546L..49S} are indicated.
Positions of these sources are taken from SIMBAD (\citealp{2000A&AS..143....9W}; SK1, HH 12), \citeauthor{2012ApJS..201...12A} (\citeyear{2012ApJS..201...12A}; IRAS 5), and \citeauthor{2016ApJ...818...73T} (\citeyear{2016ApJ...818...73T}; other sources).
Red crosses are the positions of dense cores identified with 1.1 mm continuum emission \citep{2006ApJ...638..293E}.}
\label{fig:I image}
\end{figure*}

\begin{figure*}[tp]
\centering
\includegraphics[width=\linewidth]{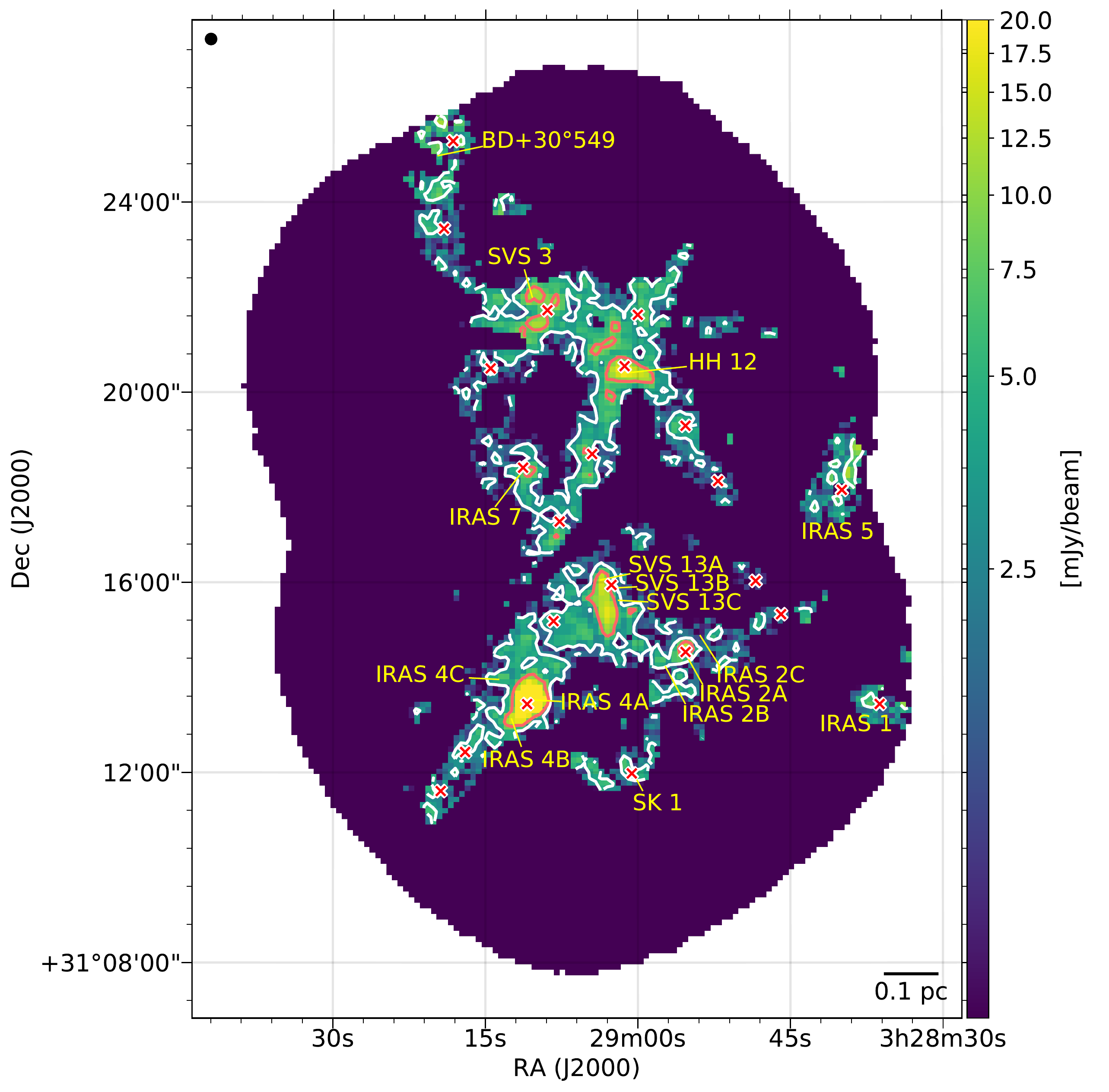}
\caption{Polarized intensity observed at $850~\mu\mathrm{m}$.
The JCMT beam ($14''\negthinspace\negthinspace.1$) is shown in the upper left corner of the figure.
Color scale is polarized intensity, $PI$. Contours are for $PI / \delta PI = 3$ (white contour) and $PI / \delta PI = 10$ (orange contour). Red crosses are the same as those in Figure \ref{fig:I image}.}
\label{fig:PI image}
\end{figure*}

We show the estimated spatial distributions of $I$ and $\psi$  in Figure \ref{fig:I image}, and that of $PI$ in Figure \ref{fig:PI image}.
Since the typical rms noise in a Stokes $I$ map for $I < 25$ mJy beam$^{-1}$ is 2.3 mJy beam$^{-1}$, we restrict our analysis in this paper to data where $I \geqslant 25$ mJy beam$^{-1}$ ($>10\sigma$ compared to the background noise level).
For the estimation of $\psi$, we require $PI / \delta PI \geqslant 3$ in addition to the $I \geqslant 25$ mJy beam$^{-1}$ threshold.
For the restricted data with $I \geqslant 25$ mJy beam$^{-1}$ and $PI / \delta PI \geqslant 3$, the typical rms noises are 1.1 mJy beam$^{-1}$ for $I$ and 0.9 mJy beam$^{-1}$ for $Q$ and $U$, respectively.
The minimum S/N value of $I$ of the restricted data is $(I/\delta I)_\mathrm{min} = 13.7$.

In the remainder of this paper, $I,~PI,~P$, and $\psi$ are our observed values in $850~\mu\mathrm{m}$ band unless otherwise stated.

\subsection{Comparison with Previous SCUBA Results}
\label{comparison-with-the-previous-scuba-results}

\begin{figure}[tp]
\centering
\includegraphics[width=\linewidth]{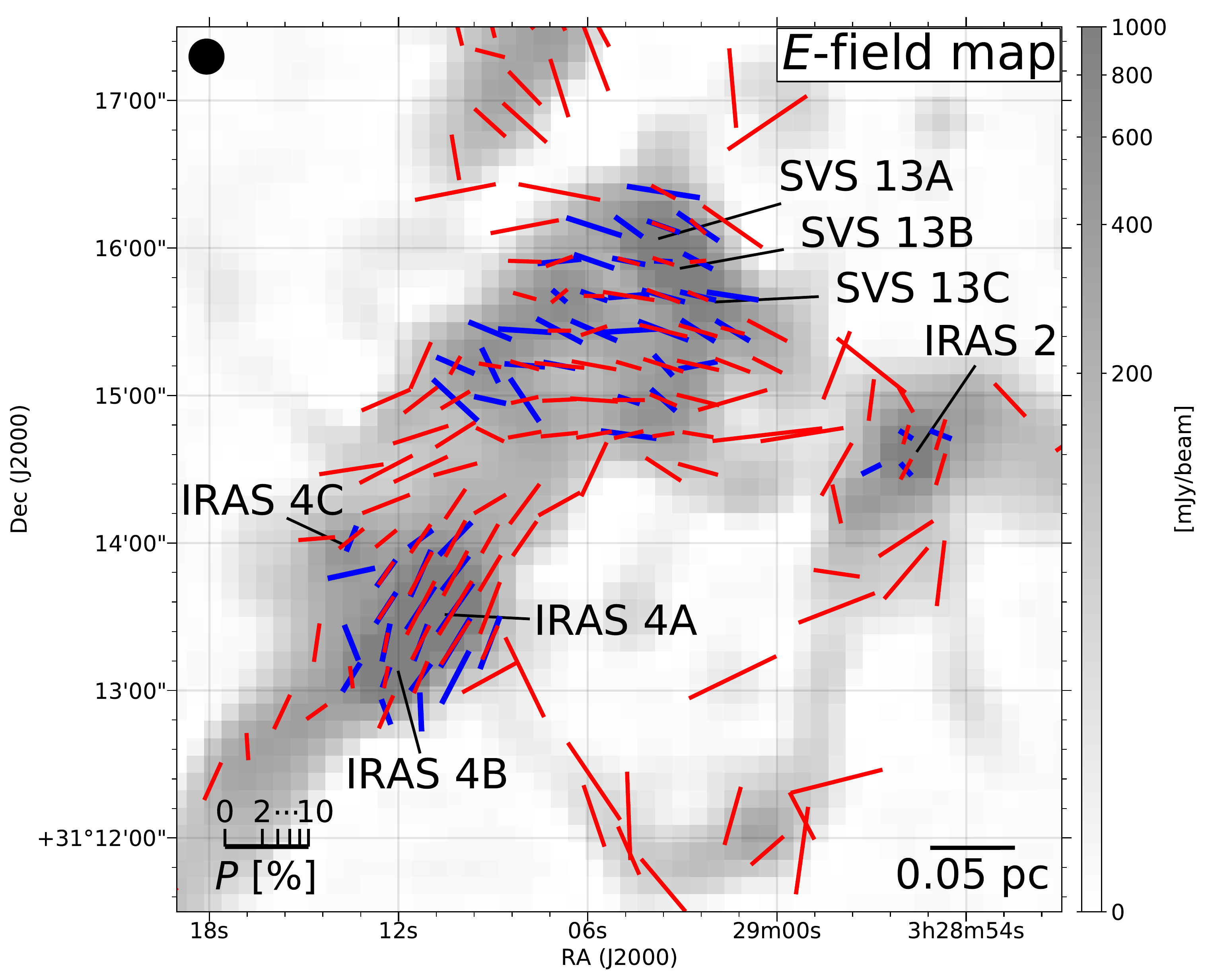}
\caption{Comparison of the two SCUBA polarization position angles observed by SCUPOL (blue line segments; \citealp{2004Ap&SS.292..509C,2009ApJS..182..143M}) and SCUBA-2/POL-2 (red line segments; this study).
Line segments are scaled by $\sqrt{P}$, as in Figure \ref{fig:I image}, for $I/\delta I \geqslant 10$ ($I \geqslant 2$ mV for SCUPOL and $I \geqslant 25$ mJy beam$^{-1}$ for SCUBA-2/POL-2) and $PI/\delta PI \geqslant 3$.
The background image is the SCUBA-2/POL-2 850 $\mu$m total intensity map.
Positions of main YSOs \citep{2016ApJ...818...73T} are indicated.
The JCMT beam ($14''\negthinspace\negthinspace.1$) is shown in the upper left corner,
and a reference scale of 0.05 pc is given in the lower right corner.
}
\label{fig:SCUPOL B}
\end{figure}

\citet{2004Ap&SS.292..509C} observed NGC 1333 using SCUPOL.
We compare our results with those obtained from archival data \citep{2009ApJS..182..143M} and check the consistency of the two data sets.
Figure \ref{fig:SCUPOL B} shows the spatial distribution of two polarimetric data sets.
The details of the comparison are described in Appendix \ref{sec:scupol}.

The observed \emph{Q} and \emph{U} values around IRAS 2 are nearly 0 for SCUPOL data, and we find a relative offset between SCUPOL and SCUBA-2/POL-2 of $\sim 9$ mJy beam$^{-1}$ (see Appendix \ref{sec:scupol} and Figure \ref{fig:SCUPOL_corr}).
This offset leads to the difference in polarization angles between the two datasets around IRAS 2 (Figure \ref{fig:SCUPOL B}).

By comparing our SCUBA-2/POL-2 data with interferometric observations with higher spatial resolutions \citep{2014ApJS..213...13H}, we find good consistency between the two datasets for all the regions including IRAS 2 (see Section \ref{sec: B continuity} and Figure \ref{fig: TADPOL}).
Thus, the discrepancy in position angle between SCUBA-2/POL-2 and SCUPOL found at IRAS 2 is attributed to the measured offsets in \emph{Q} and \emph{U} values in the SCUPOL observations.
We thus conclude that our BISTRO data from SCUBA-2/POL-2 reliably trace polarized radiation and thus the \emph{B}-field morphology in the observed region.

\subsection{Comparison with the Total \emph{I}, \emph{Q}, \emph{U} Values Observed by Planck}
\label{sec:Planck absolute}

The observed region in NGC 1333 has a size of $\sim 10' \times 20'$.
Consequently, our observations are not sensitive to the diffuse emission, whose spatial scales are larger than the size of the observed region.
Due to the need to remove the atmospheric signal, the \emph{pol2map} pipeline estimates and subtracts a background signal in the observed region.
This process further reduces the maximum spatial scale recovered to $\sim 5'$ and possibly is even lower \citep{2013MNRAS.430.2545C}.
In comparison, \citet{2018arXiv180104945P} measured the total value of \emph{Q} and \emph{U}, though with a low effective spatial resolution $\geqslant 10'$ \citep{2015A&A...576A.104P,2018arXiv180706212P}.
Thus, both BISTRO and Planck data have complementary spatial scales.

We compare our JCMT $850~\mu \mathrm{m}$ \emph{I}, \emph{Q}, and \emph{U} intensities with the Planck 353 GHz $(= 850~\mu \mathrm{m})$ observations to estimate the missing large-scale flux in our observational data.
The details are described in Appendix \ref{sec:Planck absolute App}.

We find that the missing flux in \emph{Q} and \emph{U} makes a difference in the estimated \emph{B}-field position angle at each position of $-0^\circ\negthinspace.2 \pm 8^\circ\negthinspace.2$ (the circular mean and the circular deviation).
Note that throughout this paper, the mean and the standard deviation values of position angles are the circular means and the circular standard deviations that are taking into account the $180^\circ$ degeneracy of the polarization pseudo-vectors.
The definitions of the circular mean and the circular deviation are given in Appendix \ref{sec:Directional}.
The estimated offset value is negligible compared to the differences of the \emph{B}-field orientation angles between individual filaments (see the discussion in Section \ref{sec: filaments and B}), which means that the offset does not change the observed \emph{B}-field morphology significantly.

\section{Results}
\label{sec:results}

\subsection{Spatial Distribution of \emph{I} and \emph{PI}}
\label{spatial-distribution-of-i-and-pi}

The estimated spatial distributions of $I$ and $PI$ are shown in Figures \ref{fig:I image} and \ref{fig:PI image}, respectively.
The distribution of $I$ shows an intricate filamentary structure throughout the observed region, which is in good agreement with results of previous JCMT/SCUBA-2 observations (e.g., the JCMT Gould Belt Survey; \citealp{2013MNRAS.429L..10H}).
The spatial structure is dominated by components extending both parallel and orthogonal to the northwest--southeast direction.
By eye, the physical scale of the filamentary structures we see is about 0.05--0.1 pc in width and about 0.3--0.5 pc in length.
We will give quantitative identification of filaments and estimation of their physical scales in Sections \ref{sec: feature ID} and \ref{sec: filaments and B}.

Emission from known young stellar objects (YSOs) is also detected in $I$.
The position and name of the main YSOs and infrared sources tabulated by \citet{2001ApJ...546L..49S} are shown in Figure \ref{fig:I image}.

The filamentary structure of $I$ is also well traced by $PI$ (Figure \ref{fig:PI image}).
Indeed, we successfully detected the whole network of filaments in NGC 1333 in $PI$ with an S/N of $\geqslant 3$. 
To our knowledge, these data are the first time that polarized emission from entire filaments in a star formation region has been detected with a high spatial resolution of 0.02 pc.
With these data, we can investigate in detail the relationship between filaments and the \emph{B}-field for the first time.

While the distributions of \emph{I} and \emph{PI} show overall consistency with each other, there are also marked differences.
For example, \emph{PI} emission is noticeable in the south of SVS 13, but there is no corresponding distribution in \emph{I}.
On the other hand, in the region southeast of SVS 13, the \emph{PI} emission is deficit compared to that of \emph{I}.
In the region northeast of SVS 13, we note a clear spatial break with the neighboring clamp.
This break is found in both \emph{I} and \emph{PI}, as well as in, e.g., $\mathrm{N_2H^+}$ line emission \citep{2017A&A...606A.123H}.
We reserve detailed investigations of these distributions for future studies.

\subsection{Spatial Distribution of \emph{B}-field}
\label{sec:spatial_distribution_B}

\begin{figure*}[tp]
\centering
\includegraphics[width=\linewidth]{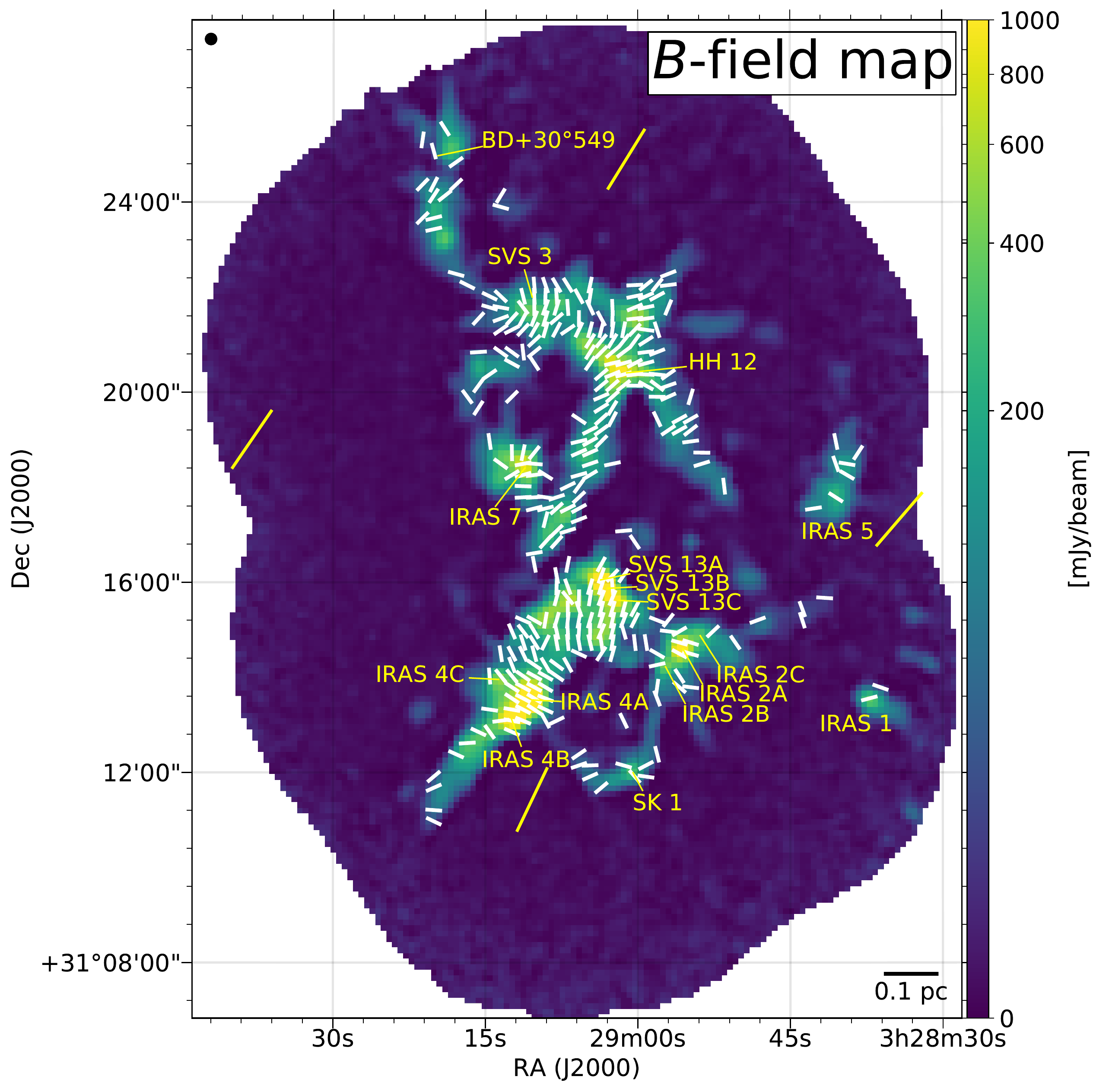}
\caption{Observed \emph{B}-field orientation (white line segments) overlaid on a color-scale map of Stokes $I$.
The \emph{B}-field orientation is assumed to be perpendicular to the observed polarization.
\emph{B}-field orientation is shown for the data points with $I \geqslant 25~\mathrm{mJy~beam^{-1}}~(I / \delta I > 10)$ and $PI / \delta PI \geqslant 3$.
The length of the line segments has been normalized to show only the orientation of the \emph{B}-field.
Yellow line segments are the \emph{B}-field orientation observed by the Planck satellite \citep{2018arXiv180706212P}, whose spatial resolution is set as $10'$ in this analysis.
The JCMT beam ($14''\negthinspace\negthinspace.1$) is shown in the upper left corner of the figure.
The spacing of the JCMT and Planck \emph{B}-field line segments is equal to the beam size of the observation.
Names of main YSOs and infrared sources tabulated by \citet{2001ApJ...546L..49S} are shown.
}
\label{fig:B image}
\end{figure*}

We show our observed \emph{B}-field orientation in Figure \ref{fig:B image} (white line segments).
In the following, we assume that the polarized dust emission traces the POS orientation of the \emph{B}-field due to RATs (Section \ref{sec:introduction}), rotated by $90 ^\circ$.
As shown in Figure \ref{fig:B image}, we reveal a complicated \emph{B}-field morphology associated with an intricate network of filaments in NGC 1333.

\begin{figure}[tp]
  \includegraphics[width=\linewidth]{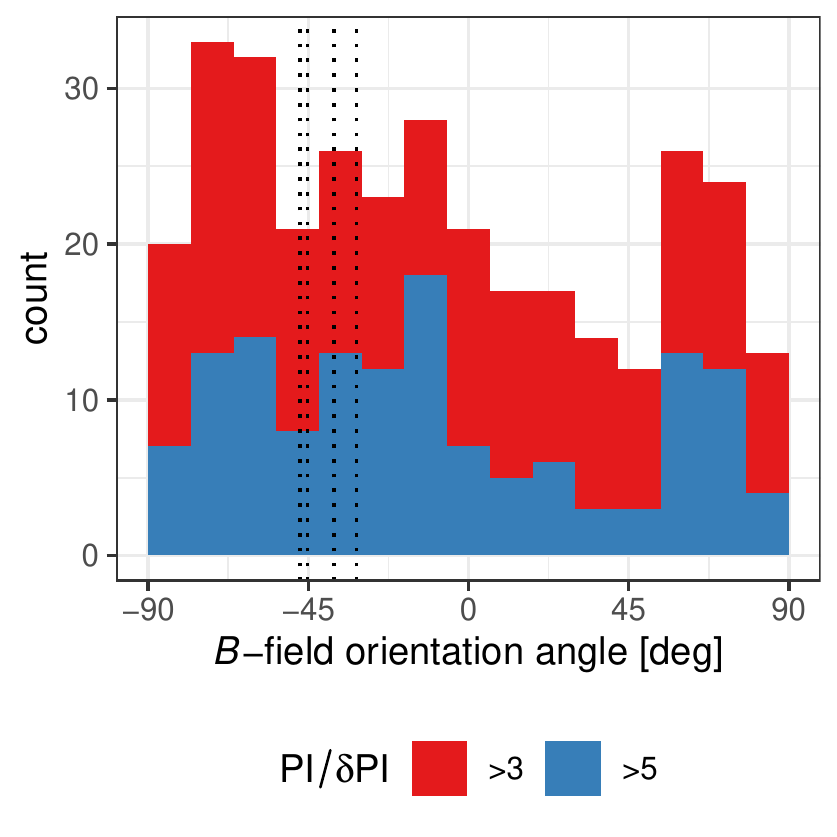}
  \caption{Histogram of the position angles for the \emph{B}-field orientations ($\psi + 90^\circ$) shown in Figure \ref{fig:B image}.
  The typical estimation error of $\psi~ (1\sigma)$ is $4^\circ\negthinspace.0$ for $PI/\delta PI > 5$ and $5^\circ\negthinspace.9$ for $PI/\delta PI > 3$.
  So we set the bin size of the histogram as $12^\circ$.
  Dotted lines indicate the \emph{B}-field orientation observed with Planck within the JCMT field of view (see the yellow line segments in Figure \ref{fig:B image}). The distribution of position angles is estimated as $-46^\circ \pm 58 ^\circ$ for the \emph{B}-field orientation observed by JCMT  and $-40 ^\circ \pm 7 ^\circ\negthinspace.3$ for that by Planck.}
\label{fig:ANG dist}
\end{figure}

The orientation of the \emph{B}-field shows a large diversity, which is shown as a broad distribution of the position angle histogram in Figure \ref{fig:ANG dist}.
The estimated circular mean and circular standard deviation of the orientation are $-46 ^\circ \pm 58 ^\circ$.
The typical estimation error of the \emph{B}-field orientation angle $(1\sigma)$ is $4^\circ\negthinspace.0$ for the data with $PI/\delta PI > 5$ and $5^\circ\negthinspace.9$ for the data with $PI/\delta PI > 3$.
The observed statistical scatter of the \emph{B}-field orientation is much larger than the estimation error.
The missing large-scale component that is not traced by the JCMT observations also cannot account for this large scatter, as the observation recovers almost all the total polarized emission (Section \ref{sec:Planck absolute}).
Thus, we conclude that the \emph{B}-field in NGC 1333 shows a large intrinsic diversity in its projected orientation.

A closer look at Figure \ref{fig:B image} gives us the impression that the \emph{B}-field orientation is not random on the smallest scale but instead appears to be correlated within individual filaments.
For example, the \emph{B}-fields around IRAS 4A, IRAS 4B, and IRAS 4C show perpendicular orientations with respect to the filament whose major axis is in a northwest--southeast direction.
The \emph{B}-fields around SVS 13A, SVS 13B, and SVS 13C are nearly orthogonal to the \emph{B}-field around IRAS 4A, IRAS 4B, and IRAS 4C, and also show nearly perpendicular orientations with respect to the filament whose major axis is in a northeast--southwest direction.
On the other hand, the \emph{B}-fields around HH 12 and its associated filament whose major axis is in a northwest--southeast direction show a nearly parallel orientation with respect to the filament.

Our earlier BISTRO observations of Orion indicate that \emph{B}-fields lie orthogonal to filaments where the filament is gravitationally supercritical, and aligned with the filament when the filament is gravitationally subcritical \citep{2017ApJ...846..122P,2017ApJ...842...66W}.
In the following, we investigate the relative offset angle between the major axes of filaments and the \emph{B}-field orientations, in light of the gravitational stability of individual filaments.

\subsection{Identification of Emission Features}
\label{sec: feature ID}

\begin{figure}[tp]
\centering
\includegraphics[width=\linewidth]{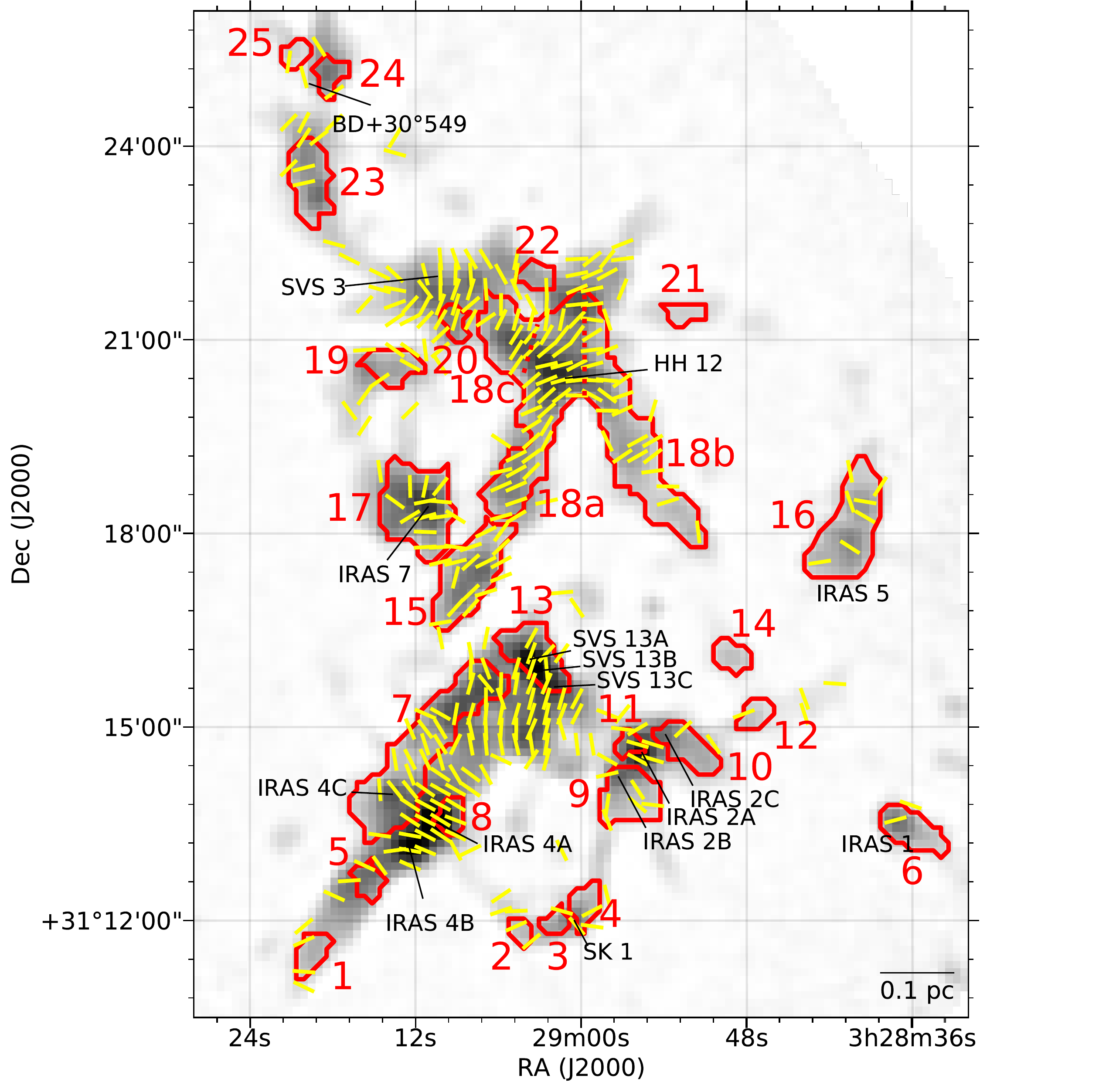}
\caption{Identified 24 ISM features in NGC 1333 as the result of the density-based clustering.
Details of the identification method are found in Appendix \ref{sec: dbscan}.
Feature \#18 is divided into three parts for the estimation of position angles in Section \ref{sec: filaments and B}.
Yellow line segments are the observed \emph{B}-field orientation.}
\label{fig:alpha3D_ID}
\end{figure}

To examine the alignment of the \emph{B}-field orientation with respect to the individual filaments described in Section \ref{sec:spatial_distribution_B} more quantitatively, we first need to identify ISM structures objectively.
We identify local emission features (features, hereafter) in the Stokes \emph{I} image.
In addition to the \emph{I} image, we refer to spectral line data and use radial velocity information to separate overlapping features projected on the POS.
For this purpose, we adopt the $\mathrm{N_2H^+}$ (1--0) line data taken at the IRAM 30 m Telescope with $30''$ spatial resolution and 0.08 km s$^{-1}$ spectral resolution \citep{2017A&A...606A.123H}.
We apply a density-based clustering method \citep{DBSCAN,dbcluster}, which is an algorithm that can identify coherent data points above noise in a multidimensional space, to identify continuous emission features in the R.A.--decl.--$V_\mathrm{LSR}$ datacube.
Figure \ref{fig:alpha3D_ID} shows the features identified with this method.
A detailed description of the method is given in Appendix \ref{sec: dbscan}, together with a justification for the choice of the spectral line.

As for feature \#18, we arbitrary separate it into nearly straight parts, as indicated in Figure \ref{fig:alpha3D_ID} (\#18a, \#18b, and \#18c) to estimate the linear elongation of these features in later analyses (Section \ref{sec: filaments and B}).
We exclude \#18c from the following analyses, as it is artificially defined, and hence it is difficult to estimate its actual orientation.

\subsection{Local \emph{B}-field Alignment to Filaments}
\label{sec: filaments and B}

Here we estimate the \emph{B}-field orientations measured within the identified individual features.
We analyze only the eight features with sizes $\geqslant 10$ independent beams so that we can estimate the mean position angles and the standard deviation of the \emph{B}-field with reasonable accuracy.

\begin{figure}[tp]
\centering
\includegraphics[width=\linewidth]{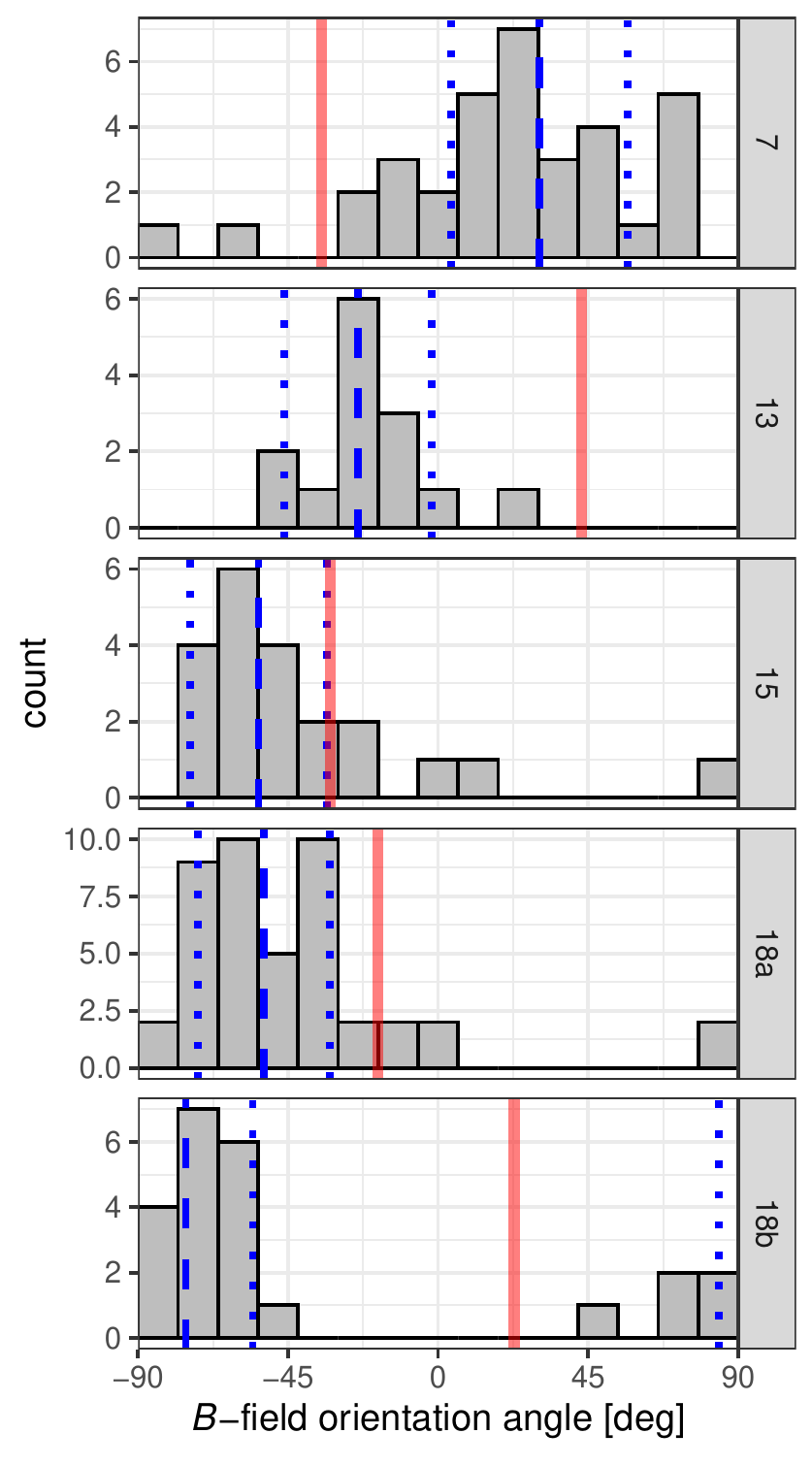}
\caption{Histograms of the \emph{B}-field orientations in individual filaments.
We set the bin size of the histogram as $12^\circ$ (see Figure \ref{fig:ANG dist}).
The number in each panel refers to a feature ID shown in Figure \ref{fig:alpha3D_ID}.
Blue dashed lines are the circular mean position angles, and blue dotted lines are the circular standard deviation of the position angles $\left( \pm 1\sigma \right)$.
Red solid lines are the position angle of the filaments.
See text for the evaluation of the filament position angle.
}
\label{fig:ANG_dist_filament}
\end{figure}

We find that \emph{B}-field orientations are self-consistent in particular directions for features \#7, \#13, \#15, \#18a, and \#18b.
In Figure \ref{fig:ANG_dist_filament}, we show histograms of the \emph{B}-field orientations measured within these features, together with their circular means (blue dashed lines) and circular standard deviations (blue dotted lines).
The circular standard deviations of these orientations are $20^\circ$--$26^\circ$ (see Table \ref{tab:filament param}).
As shown in Figure \ref{fig:alpha3D_ID}, those features with concentrated \emph{B}-field orientations have elongated spatial structures.
Hereafter, we call these features 'filaments.'
We estimate the position angles of the major axes of the filaments by least-squares linear fits to their spatial structures in $I$, weighted by $I$ intensity.
The estimated position angles are shown as red solid lines in Figure \ref{fig:ANG_dist_filament} (also see Table \ref{tab:filament param}).

\begin{figure}[tp]
\centering
\includegraphics[width=\linewidth]{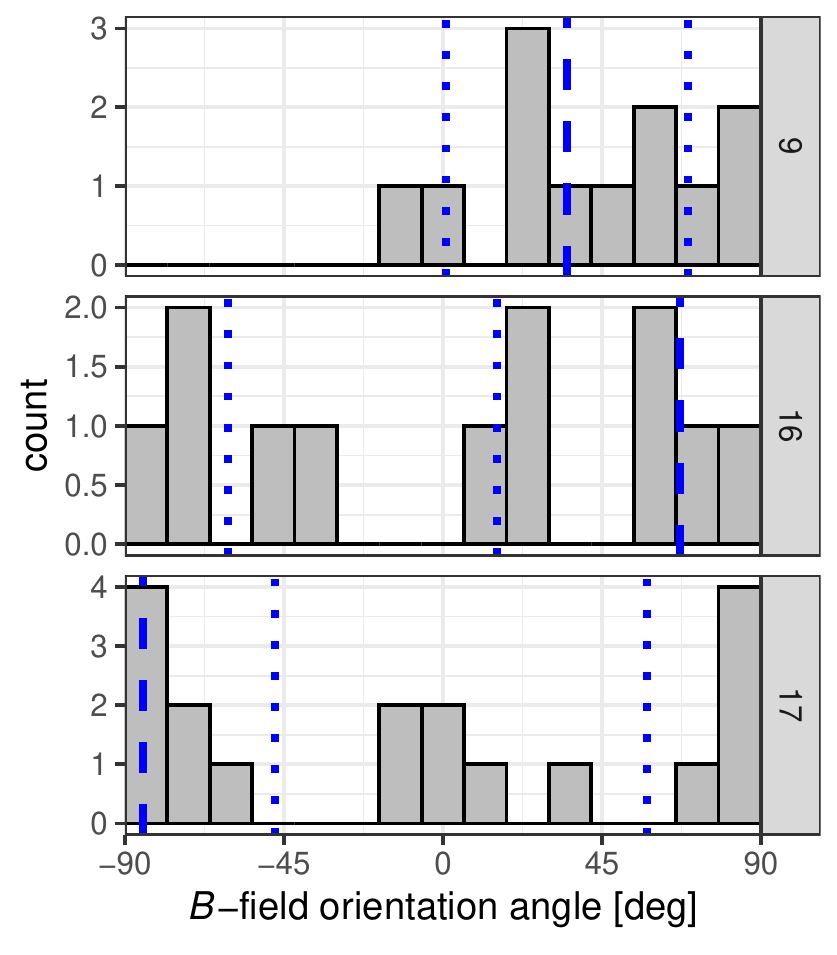}
\caption{Same as Figure \ref{fig:ANG_dist_filament} but for the features that show relatively scattered \emph{B}-field position angles (the circular standard deviation of the position angles $> 30 ^\circ$).}
\label{fig:ANG_dist_feature}
\end{figure}

\emph{B}-field orientations for other features, e.g., \#9, \#16, and \#17, show more scattered distributions.
In Figure \ref{fig:ANG_dist_feature}, we show histograms of the \emph{B}-field orientation measured within these features.
Unlike the features described earlier, these show less apparent concentration in their \emph{B}-field orientations.
The circular standard deviations of these orientations are $34^\circ$ for feature \#9, $52^\circ$ for feature \#16, and $37^\circ$ for feature \#17.
We note that these features tend to be spatially isolated and less elongated, as seen in Figure \ref{fig:alpha3D_ID}.
Thus, we cannot estimate well the position angles of these features.
The observed \emph{B}-field orientations in these features show radial (\#17, IRAS 7) or random (\#9, IRAS 2; \#16, IRAS 5) distributions.

\begin{table*}[t]
\caption{Estimated Parameters of Identified Filaments}
\label{tab:filament param}
\begin{center}
\begin{tabular}{lrrrrr}
\hline
\hline
ID \# \ldots & 7 & 13 & 15 & 18a & 18b\\
\hline
Filament position angle$^\mathrm{a}$ (deg) & $-34.9 \pm 0.3$ & $42.9 \pm 1.7$ & $-32.3 \pm 0.7$ & $-18.1 \pm 0.1$ & $22.8 \pm 0.1$\\
Length$^\mathrm{b}$ (pc) & 0.30 & 0.11 & 0.15 & 0.30 & 0.33\\
Width$^\mathrm{b}$ (pc) & 0.074 & 0.059 & 0.064 & 0.066 & 0.069\\
Column density$^\mathrm{c}$ ($10^{23}$ H-atom cm$^{-2}$) & $2.2 \pm 1.7$ & $5.4 \pm 5.3$& $1.2 \pm 0.8$& $2.1 \pm 1.7$& $1.0 \pm 0.8$\\ 
Mass$^\mathrm{d}$ ($M_\odot$) & 35 & 29 & 10 & 32 & 16\\
Line mass$^\mathrm{e}$ ($M_\odot~\mathrm{pc}^{-1}$) & 118 & 260 & 66 & 106 & 49\\
\emph{B}-field position angle$^\mathrm{f}$ (deg) & $30 \pm 26$ & $-24 \pm 22$ & $-54 \pm 20$ & $-52 \pm 20$ & $-76 \pm 20$\\
$|B$-$\mathrm{field} - \mathrm{filament}$ offset angle$|$$^\mathrm{g}$ (deg) & $65 \pm 26$ & $67 \pm 22$ & $22 \pm 20$ & $34 \pm 20$ & $82 \pm 20$\\
\hline
\multicolumn{6}{p{0.85\textwidth}}{\bf Notes.}\\
\multicolumn{6}{p{0.85\textwidth}}{$^\mathrm{a}$The position angle estimated by an $I$-weighted linear fit to their spatial structures.}\\ 
\multicolumn{6}{p{0.85\textwidth}}{$^\mathrm{b}$Length and width are the length of the filaments along their major and minor axes.}\\
\multicolumn{6}{p{0.85\textwidth}}{$^\mathrm{c}$The column density of each filament estimated by converting $\tau_{850\mu\mathrm{m}}$. The mean and the standard deviation values are shown. See text for the conversion factor between $\tau_{850\mu\mathrm{m}}$ and the column density.}\\
\multicolumn{6}{p{0.85\textwidth}}{$^\mathrm{d}$The mass of each filament estimated as the sum of the column density.}\\
\multicolumn{6}{p{0.85\textwidth}}{$^\mathrm{e}$The mass per unit length of each filament estimated as mass/length.}\\
\multicolumn{6}{p{0.85\textwidth}}{$^\mathrm{f}$The circular mean position angle of the \emph{B}-field pseudo-vectors inside each filament.}\\
\multicolumn{6}{p{0.85\textwidth}}{$^\mathrm{g}$The absolute angular difference between filament position angle and \emph{B}-field position angle.}\\
\end{tabular}
\end{center}
\end{table*}

Previous studies concluded that the \emph{B}-fields are observed to be mostly perpendicular to the main axis of dense filaments (e.g.,  \citealp{2016A&A...586A.135P,2016A&A...586A.136P,2016A&A...586A.138P,2017ApJ...846..122P,2017A&A...603A..64S,2017ApJ...842...66W,2018ApJ...859..151L,2019ApJ...878..110F,2019ApJ...883...95S}).
Three filaments in NGC 1333 (\#7, \#13, and \#18b) show perpendicular \emph{B}-field orientations, which is consistent with those studies.
In our filaments, however, the \emph{B}-fields are not always perpendicular to their major axes (e.g., \#15 and \#18a).
Here, we estimate the line mass of these filaments and their relative offset angle to the \emph{B}-field, to compare our observations with the previous studies.

The estimated parameters of the filaments are shown in Table \ref{tab:filament param}.
We estimate the length and width of a filament based on the average distance of the filament data points along its major and minor axes from the center (arithmetic mean position) of the filament.
These values correspond to half from the center of the filament to the endpoint.
We thus multiply these values by four to estimate the length and width of the filament.
The estimated widths are 0.06--0.07 pc, which is consistent with a typical observed width of filaments \citep{2011A&A...529L...6A,2012A&A...544A..14J,2013A&A...550A..38P,2014A&A...568A..98A,2015MNRAS.452.3435K,2019A&A...621A..42A}.

We make a crude estimate of the mass of each filament by referring to the dust optical depth at 850 $\mu$m based on our $I$ map.
We estimate the median value of $I$ in the background region ($I < 25$ mJy beam$^{-1}$; Section \ref{sec:ObsReduce}) and subtract that background value from the observed $I$.
We refer an estimate of the dust temperature, whose spatial resolution is $36''$, based on Herschel and Planck observations (\citealp{2016A&A...587A.106Z}).
The estimated dust temperatures range between 16.2 K and 19.2 K, and the derived optical depths at 850 $\mu$m ($\tau_{850\mu\mathrm{m}}$) range between $2\times10^{-4}$ and $2\times 10^{-2}$.
Thus, we assume an optically thin condition for the $I$ emission and estimate the column density by dividing $\tau_{850\mu\mathrm{m}}$ by the dust opacity at $850~\mu$m $(\sigma_{850\mu\mathrm{m}})$ as follows: $N_\mathrm{H} = \tau_{850\mu\mathrm{m}}/\sigma_{850\mu\mathrm{m}}$, where $\sigma_{850\mu\mathrm{m}} = 8.0 \times 10^{-27} ~\mathrm{cm^2~H^{-1}}$ \citep{2014A&A...571A..11P,2018ApJ...862...49N}.
The estimated column density of each filament is shown in Table \ref{tab:filament param}.

The total mass of each filament is estimated by summing up the derived column density encircled by the red contours in Figure \ref{fig:alpha3D_ID}.
The estimated total mass and the line mass of each filament are shown in Table \ref{tab:filament param}.
We estimate the line mass of the filaments as $\geqslant 49~M_\odot~\mathrm{pc^{-1}}$.
Although the mass estimation could be affected by local variation of dust temperature caused by embedded sources, especially for \#7 and \#13, we conclude that the filaments have line masses well above the critical value ($M_\mathrm{line,crit} = 2c_s^2/G \simeq 16~M_\odot~\mathrm{pc^{-1}}$ for $T_\mathrm{gas}=10$ K and $\simeq 32~M_\odot~\mathrm{pc^{-1}}$ for $T_\mathrm{gas}=20$ K; \citealp{1963AcA....13...30S}; \citealp{1964ApJ...140.1056O}; \citealp{1997ApJ...480..681I}).
This result is also compatible with that of \citet{2017A&A...606A.123H}.

We estimate the \emph{B}-field orientation of a filament as the circular mean and the circular standard deviation of the \emph{B}-field pseudo-vectors observed at the filament.
The projected offset angles between the \emph{B}-field and filaments tend to be large ($65^\circ$--$82^\circ$ for three filaments), which is consistent with the previous Planck and BLASTPol studies.
Some of them, however, are significantly small ($22^\circ$ for \#15 and $34^\circ$ for \#18a), though their line masses are estimated to be $> 60~M_\odot~\mathrm{pc}^{-1}$ (Table \ref{tab:filament param}).
They are typical gravitationally supercritical filaments, and thus their parallel orientation to the local \emph{B}-field is not in accordance with the Planck and BLASTPol observations.
We present a possible model to explain the interesting projected offset angle distributions that we find for the massive filaments in NGC 1333 in Section \ref{sec: B association filament}.

\section{Multiscale \emph{B}-field Structures}
\label{sec: multiscale}

Our polarization map covers a $1.5~\mathrm{pc}\times 2~\mathrm{pc}$ region with a 0.02 pc spatial resolution.
In this section, we compare the \emph{B}-field orientations from our polarization map with those from maps with broader spatial coverage or higher spatial resolution.

\subsection{Global and Local \emph{B}-field Observed with Planck, Optical, Near-infrared, and JCMT}\label{sec: comp with Planck}

\begin{figure*}[tp]
\centering
\includegraphics[width=\linewidth]{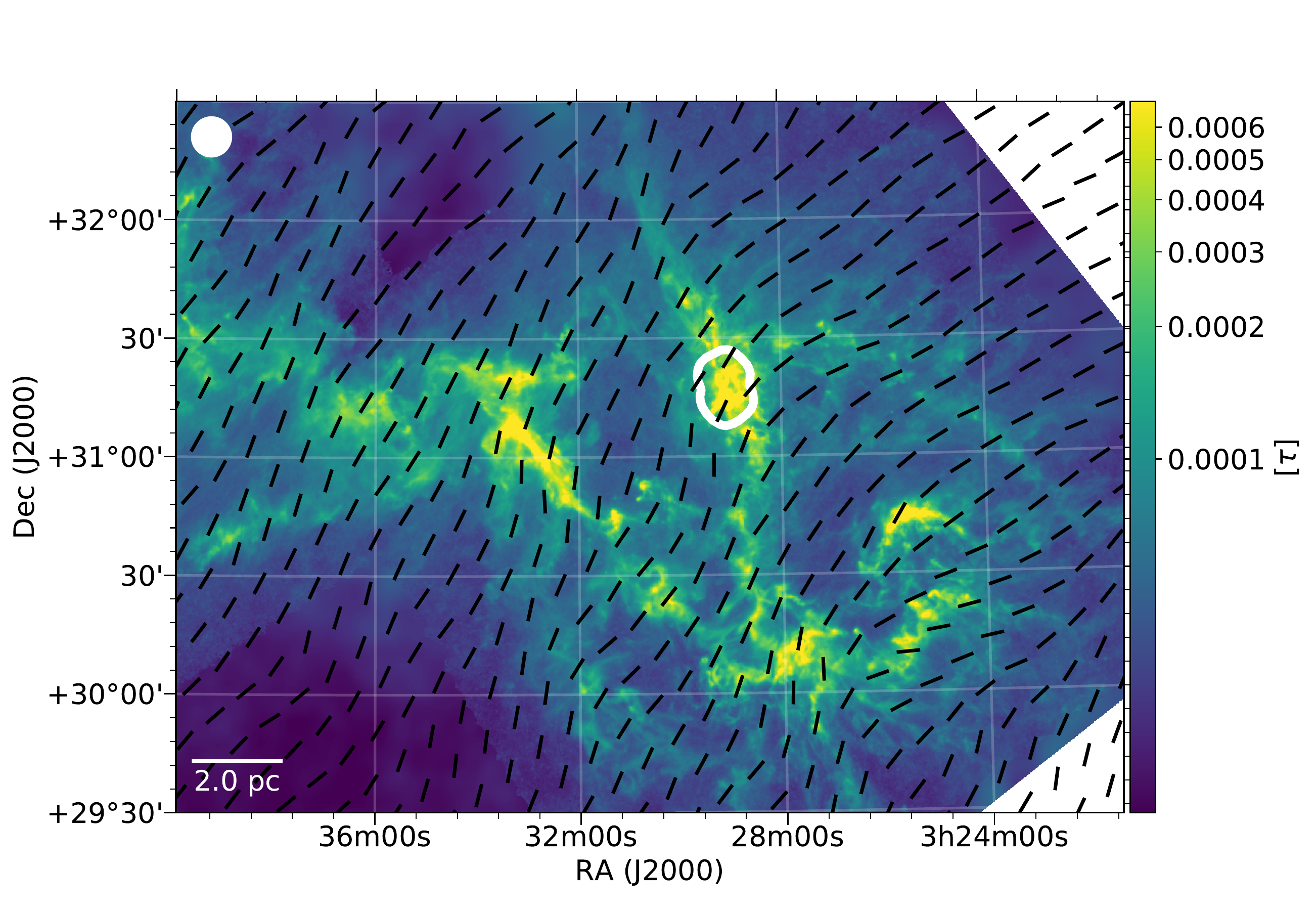}
\caption{
Planck observed \emph{B}-field orientations in the Perseus molecular cloud \citep{2018arXiv180706212P} are shown as black line segments.
Here, we set the spatial resolution of the Planck data as a $10'$ FWHM Gaussian to achieve good S/Ns.
The assumed beam ($10'$) is shown in the upper left corner of the figure.
Color scale is the dust optical depth at 353 GHz estimated by \citet{2016A&A...587A.106Z}.
The western half of the Perseus molecular cloud complex is shown in this figure.
Our observed area at NGC 1333 is marked in white.}
\label{fig:Herschel tau}
\end{figure*}

We show the large-scale \emph{B}-field observed across Perseus with Planck in Figure \ref{fig:Herschel tau}.
As shown in the figure, Planck observed a fairly uniform global \emph{B}-field for the whole Perseus molecular cloud, whose size is $> 10$ pc (\citealp{2018arXiv180706212P}).

The effective spatial resolution of the Planck data $(10')$ corresponds to $\sim 0.87$ pc at the distance of NGC 1333.
On the other hand, the spatial resolution of the BISTRO data with JCMT is $14''\negthinspace\negthinspace.1$, which corresponds to $\simeq 0.020$ pc, or $\simeq 43$ times finer than that of Planck.
Here, we compare the BISTRO \emph{B}-field across NGC 1333 with the global \emph{B}-field observed with Planck.

We show the Planck \emph{B}-field in Figure \ref{fig:B image} (yellow line segments) and its position angles in Figure \ref{fig:ANG dist} (black dotted lines), in addition to the BISTRO \emph{B}-field.
The Planck \emph{B}-field orientation shows a smoothly and slowly varying field distribution with a position angle of $-41 ^\circ \pm 10 ^\circ$ in our observed NGC 1333 area.
This distribution is in contrast with the much more diverse one observed with SCUBA-2/POL-2, which shows a significantly larger scatter ($-46 ^\circ \pm 58 ^\circ$).
As described in Section \ref{sec:spatial_distribution_B}, the large scatter in the \emph{B}-field observed with JCMT is due to neither the observational uncertainties nor the missing large-scale flux that is not detected by JCMT.
Thus, we conclude that the \emph{B}-field orientation in NGC 1333 becomes significantly more complex at spatial scales below 1 pc.

\begin{figure}[tp]
  \includegraphics[width=\linewidth]{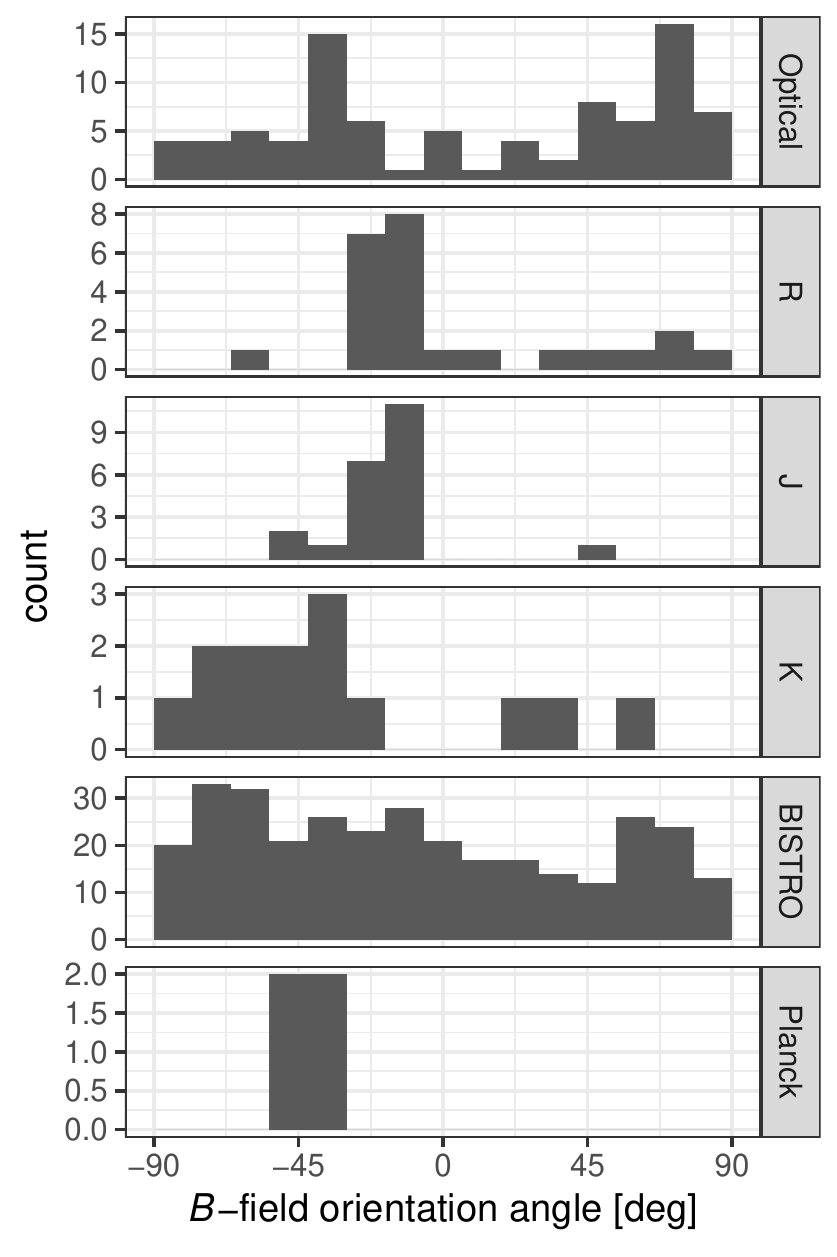}
  \caption{Histograms of the position angles for the \emph{B}-field orientations observed by optical and near-infrared polarimetry.
  Optical data (762.5 nm) are from \citet{1990ApJ...359..363G}.
  \emph{R}-band and \emph{J}-band data are from \citet{2011AJ....142...33A}.
  \emph{K}-band data are from \citet{1988MNRAS.231..445T}.
  Submillimeter data taken by BISTRO and Planck, rotated by $90^\circ$, are shown for comparison.
  Note that the spatial coverages are different for individual observations, especially \emph{R}-band and \emph{J}-band data by \citet{2011AJ....142...33A}, which cover only a small area south of the IRAS 4A and IRAS 4B (see text).}
\label{fig:ANG multi dist}
\end{figure}

Fluctuations in the \emph{B}-field structure in a dense molecular cloud on $< 1$ pc scales are also found by optical and near-infrared polarimetry as shown in Figure \ref{fig:ANG multi dist}.
Optical polarimetry of the Perseus molecular cloud around NGC 1333 by \cite{1990ApJ...359..363G} and other references therein shows two different populations of polarization position angles.
One population has a peak at $-39 ^\circ \pm 15 ^\circ$ \hspace{-0.25em}\footnote{The standard deviations given here are the arithmetic standard deviations estimated by \cite{1990ApJ...359..363G}. The arithmetic standard deviation and the circular standard deviation are not significantly different from each other if their values are $15^{\circ}$ and $25^{\circ}$ (see Appendix \ref{sec:Directional} and Figure \ref{fig:SD}).\label{footnote SD}} and agrees with the \emph{B}-field orientation obtained by Planck in the diffuse medium in the whole Perseus molecular cloud.
It is also approximately parallel to many filaments such as \#15 and \#18a (Figure \ref{fig:ANG_dist_filament}) and is characterized by a larger polarization percentages of $P > 1.5\%$ in the optical.
The other population has a broader peak in position angles at $60 ^\circ \pm 25 ^\circ$, \hspace{-0.25em}\textsuperscript{\ref{footnote SD}} has smaller polarization percentage values, and corresponds to field orientation with a perpendicular orientation to its major axis (see, e.g., filament \#7).
The two populations of vectors show no spatial distinction across the Perseus cloud complex.
They coexist in projection throughout the entire Perseus cloud.

The near-infrared polarimetry by \cite{2011AJ....142...33A} mainly covers a small area with a $5'$ diameter, south of the IRAS 4A and IRAS 4B protostars, while that by \cite{1988MNRAS.231..445T} covers a large area around NGC 1333.
Both observations show \emph{B}-field orientations peaking at $\simeq -15^\circ$ to $-30^\circ$, in agreement with the Planck data near that position but almost perpendicular to the \emph{B}-field at the protostars IRAS 4A and IRAS 4B.

Note that the Planck data measure the polarization from all dust grains visible in the effective $10'$-wide beam covered by Planck, but optical and near-infrared polarimetry measures dichroic extinction of light by dust in a pencil-beam along the line of sight (LOS) to individual stars background to the cloud.
This comparison with our JCMT map thus demonstrates that the field orientation inside dense molecular clouds shows considerable variation, which is in accordance with the finding mentioned above that the \emph{B}-field is significantly distorted on scales $< 1$ pc inside NGC 1333.

This distortion of the \emph{B}-field is likely due to the interaction of the field with gas flows associated with filament formation and evolution.
We will discuss this possibility in Section \ref{close-relationship-between-filaments-and-b-field}.

\subsection{Continuity of the \emph{B}-field at Scales between 0.1 pc and 1000 au}
\label{sec: B continuity}

\begin{figure*}[tp]
  \subfloat[IRAS 4A]{\includegraphics[clip,width=0.5\textwidth]{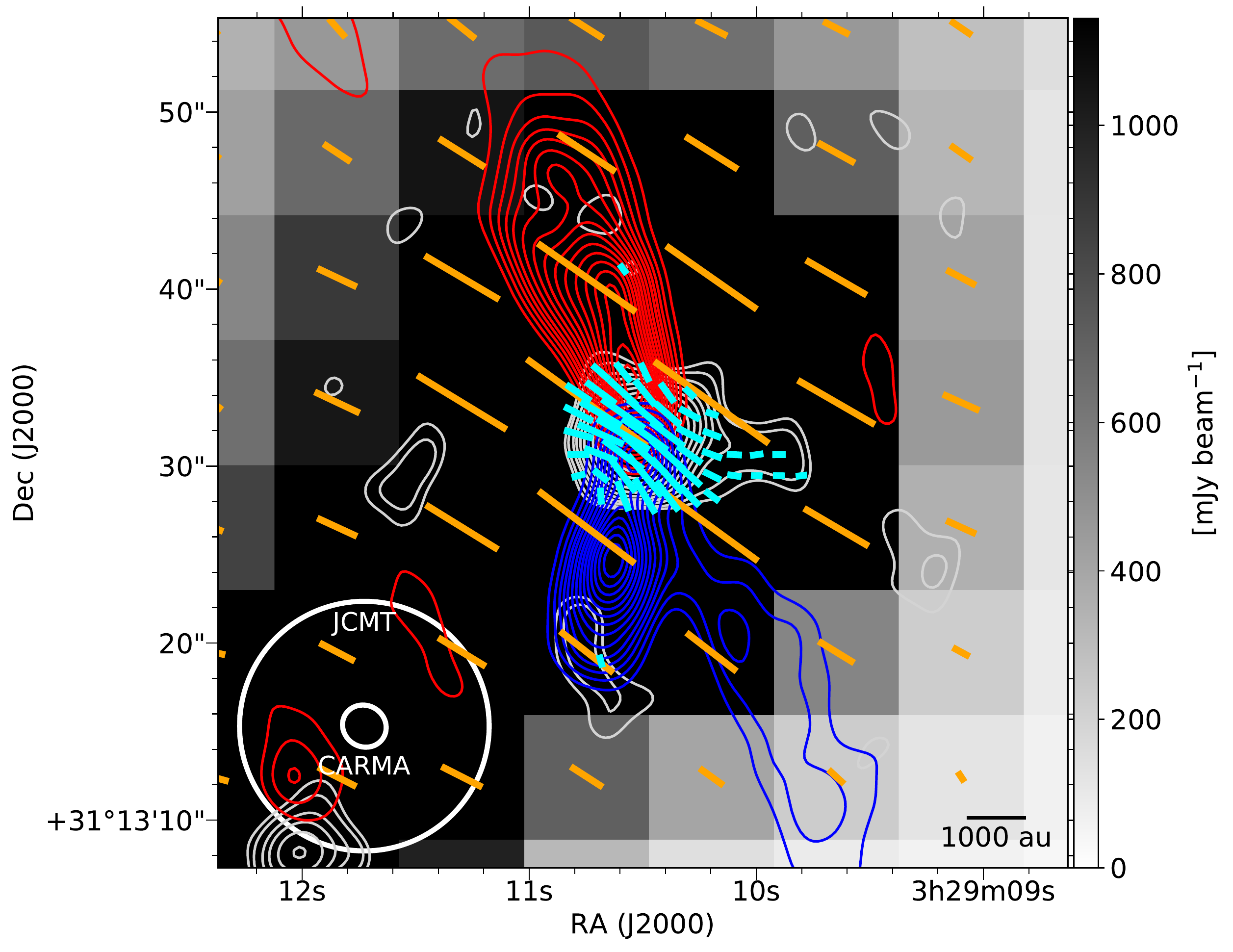}\label{fig:IRAS 4a image}}
  \subfloat[IRAS 4B]{\includegraphics[clip, width=0.5\textwidth]{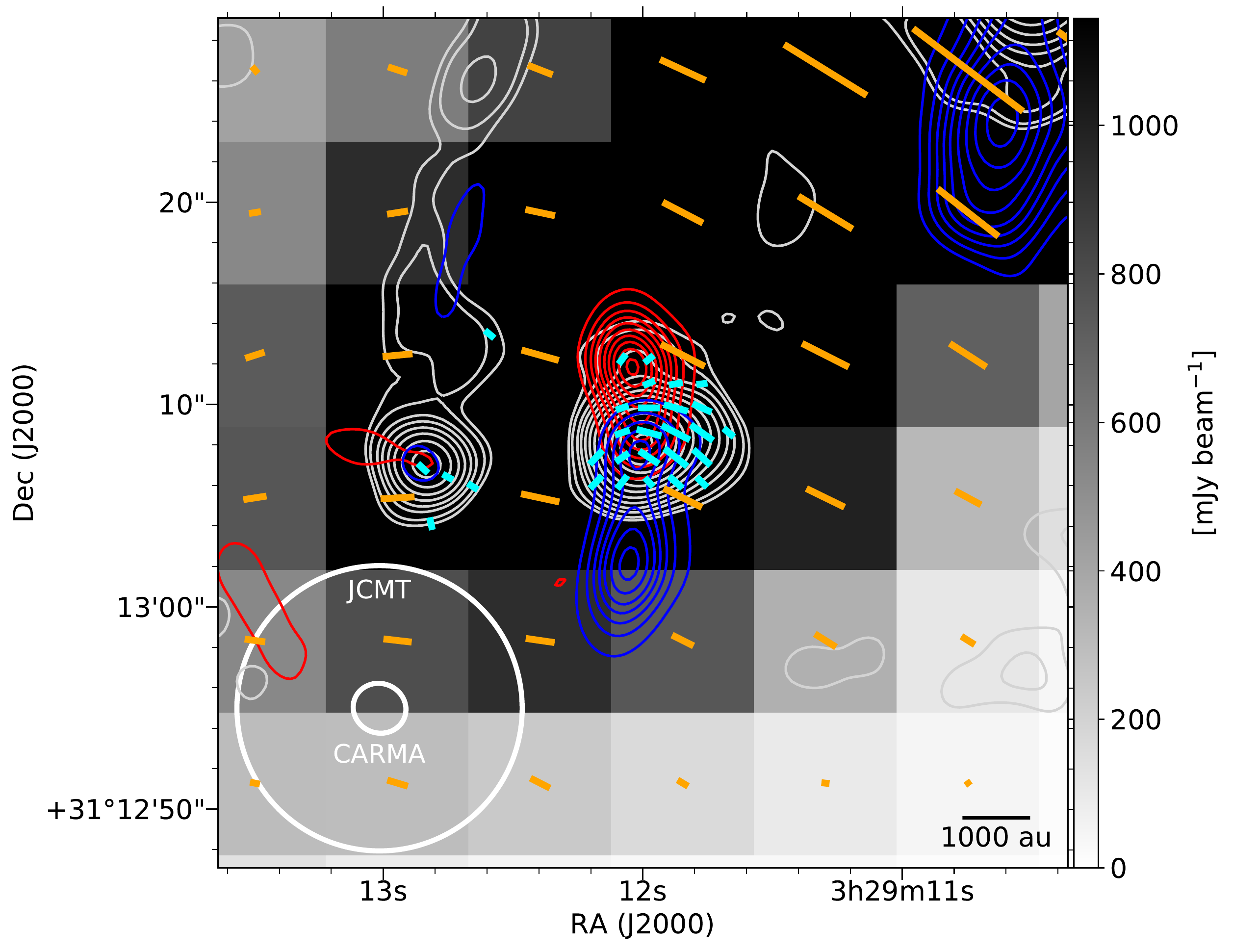}\label{fig:IRAS 4b image}}
\hspace{0.1\columnwidth}
  \subfloat[SVS 13A (left) \& 13B (right)]{\includegraphics[clip, width=0.5\textwidth]{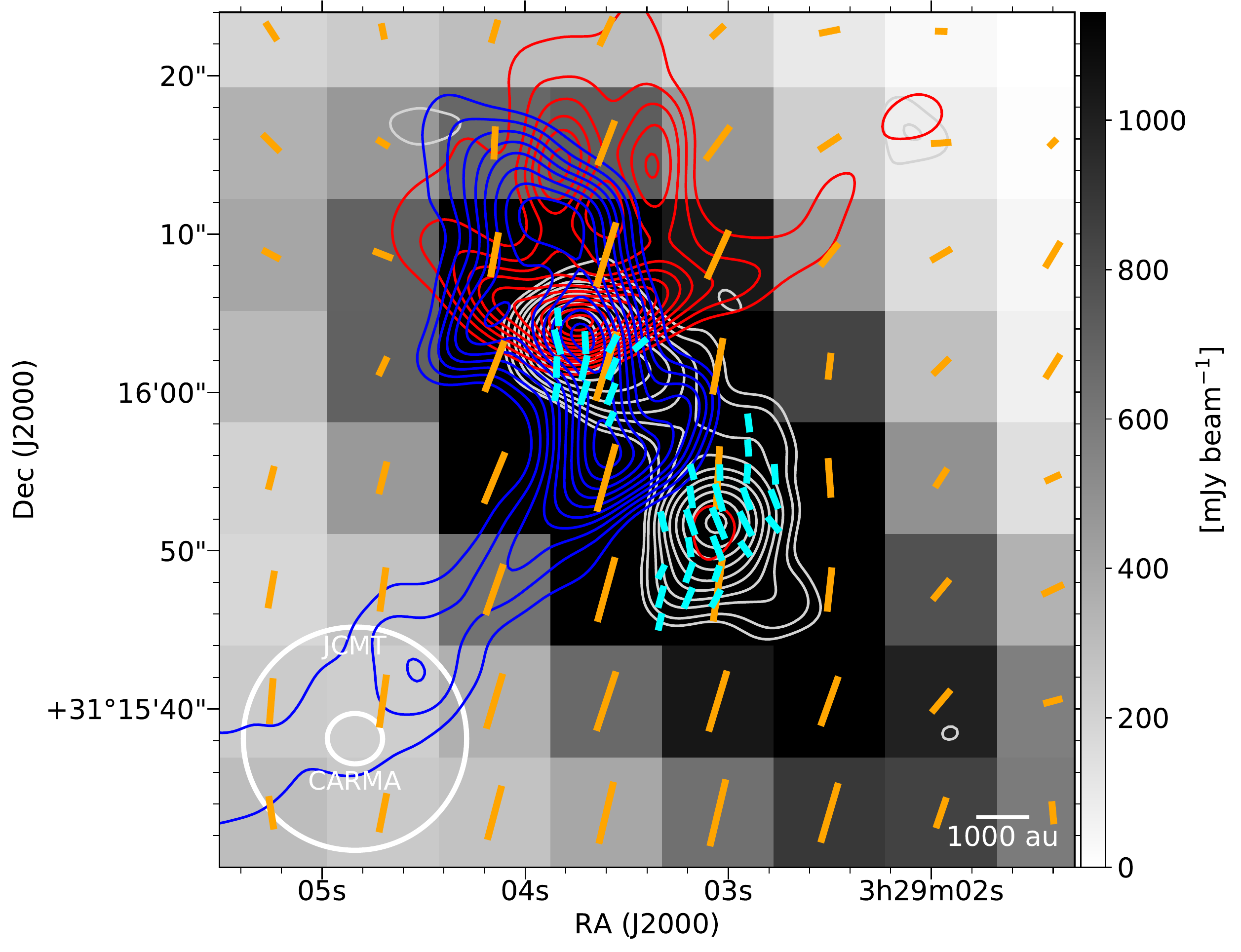}\label{fig:SVS 13 image}}
  \subfloat[IRAS 2A]{\includegraphics[clip, width=0.5\textwidth]{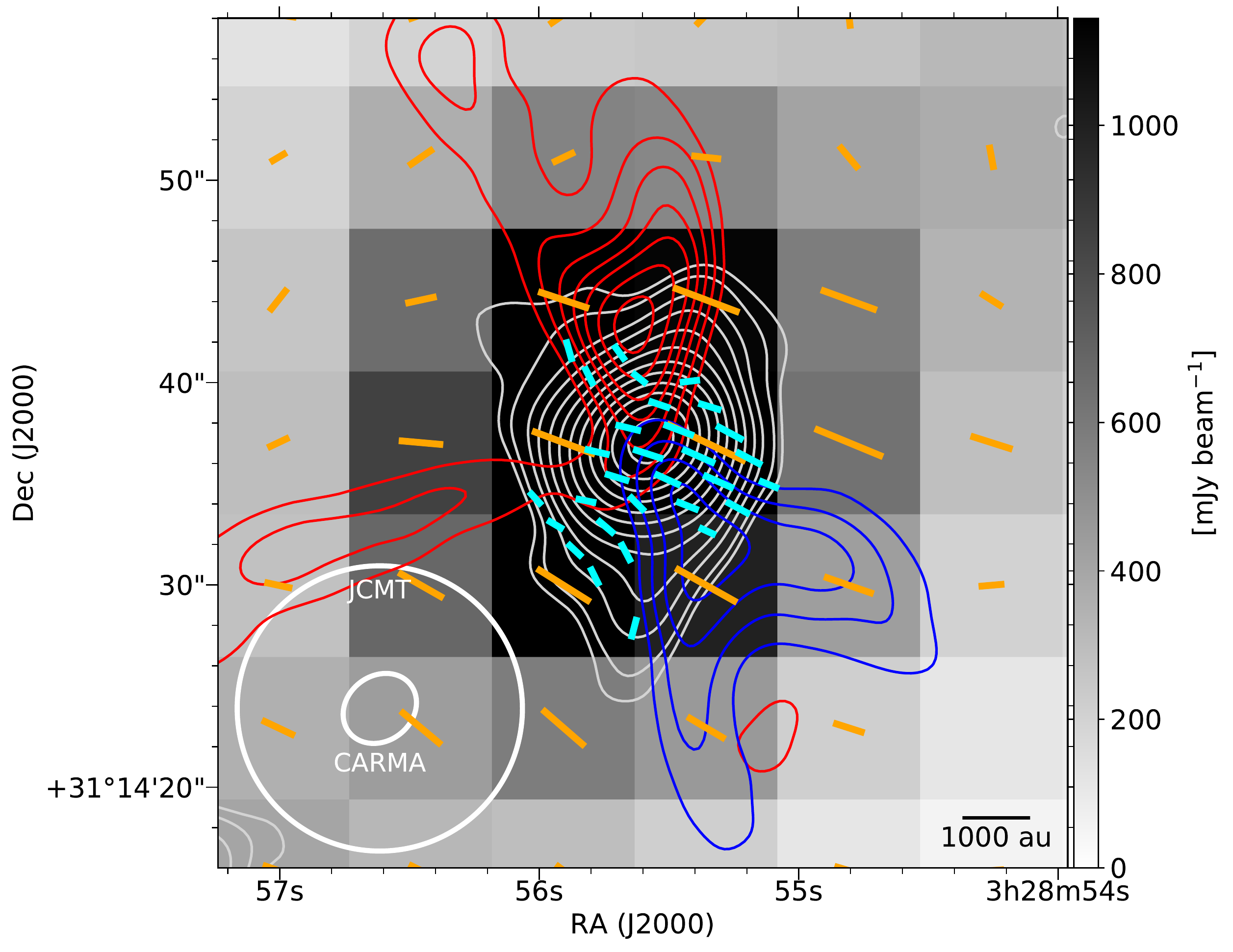}\label{fig:IRAS 2a image}}
  \caption{
  Inferred \emph{B}-field orientations from the JCMT (orange line segments; this study) and CARMA (blue line segments; \citealp{2014ApJS..213...13H}) toward five embedded YSOs in NGC 1333.
  Line segments are scaled by $\sqrt{PI}$ for $I \geqslant 25$ mJy beam$^{-1}~ (I/\delta I > 10)$ for the JCMT data.
  Background images are the SCUBA-2/POL-2 850 $\mu \mathrm{m}$ total intensity maps.
  White contours show the CARMA 1.3 mm continuum intensities with levels of 2, 3, 5, 7, 10, 14, 20, 28, 40, 56, 79, 111, 155, 217 $\times \sigma_I$, where $\sigma_I$ is the noise level of each image.
  Red and blue contours are the red- and blue-shifted outflows identified by CO (2 -- 1) integrated intensity maps from \citet{2014ApJS..213...13H} using contour levels of 4, 8, 12, 16, 20, 25, 30, \dots, 190, 195, 200 $\times \sigma_\mathrm{SL}$, where $\sigma_\mathrm{SL}$ is the rms noise level.
  White circles in the lower left corners show the beam sizes for the JCMT and CARMA.
  A reference scale of 1000 au is given in the lower right corners.
  See Table \ref{tab: TADPOL} for the beam sizes, $\sigma_I$, $\sigma_\mathrm{SL}$, and the velocity ranges for the outflows for the CARMA data.
  }
  \label{fig: TADPOL}
\end{figure*}

\begin{table*}[t]
\caption{Parameters of CO (2--1) emission maps \citep{2014ApJS..213...13H} Shown in Figure \ref{fig: TADPOL}}
\label{tab: TADPOL}
\begin{center}
\begin{tabular}{lccccc}
\hline
\hline
Region & Beam Size & \multicolumn{2}{c}{Noise Level} & \multicolumn{2}{c}{Velocity Range} \\
\cline{3-4}\cline{5-6}
& (arcsec) & Intensity & Velocity & Redshifted & Blueshifted \\
& & (mJy beam$^{-1}$) & (K km s$^{-1}$) & (km s$^{-1}$) & (km s$^{-1}$) \\
\hline
IRAS 4A & $2.52 \times 2.33$ & 10.9 & 2.41 & 11.2--~3.8 & -4.6-- -14.2 \\
IRAS 4B & $2.65 \times 2.45$ & ~7.3 & 3.12 & 22.5--~9.8 & ~3.4-- -12.5 \\
SVS 13A \& 13B & $3.49 \times 3.18$ & ~3.6 & 0.59 & 27.0--19.6 & -6.9-- -12.2 \\
IRAS 2A & $3.84\times 3.22$  & ~2.4 & 2.07 & 27.0--10.1 & ~2.6-- ~-5.8 \\  
\hline
\end{tabular}
\end{center}
\end{table*}

Here, we compare our observations with the \emph{B}-field orientations observed around YSOs by radio interferometers at higher spatial resolution.
In Figure \ref{fig: TADPOL}, we display the \emph{B}-field structure as measured at 1.3 mm around IRAS 4A, IRAS 4B, SVS 13A, SVS 13B, and IRAS 2A:  five young, embedded Class 0 YSOs observed as part of the TADPOL survey \citep{2014ApJS..213...13H} using the Combined Array for Research in Millimeter-wave Astronomy (CARMA; \citealp{2006SPIE.6267E..13B}).
See Table \ref{tab: TADPOL} for the parameters of TADPOL data shown in Figure \ref{fig: TADPOL}.
The spatial resolution of the BISTRO $(14''\negthinspace\negthinspace.1)$ data corresponds to $0.020$ pc or $4200$ au in these regions, while that of the TADPOL observation $(2''\negthinspace\negthinspace.4 - 3''\negthinspace\negthinspace.5)$ corresponds to 0.0035--0.0051 pc or 720--1000 au at the distance of NGC 1333.

\begin{figure}[t]
    \centering
        \includegraphics{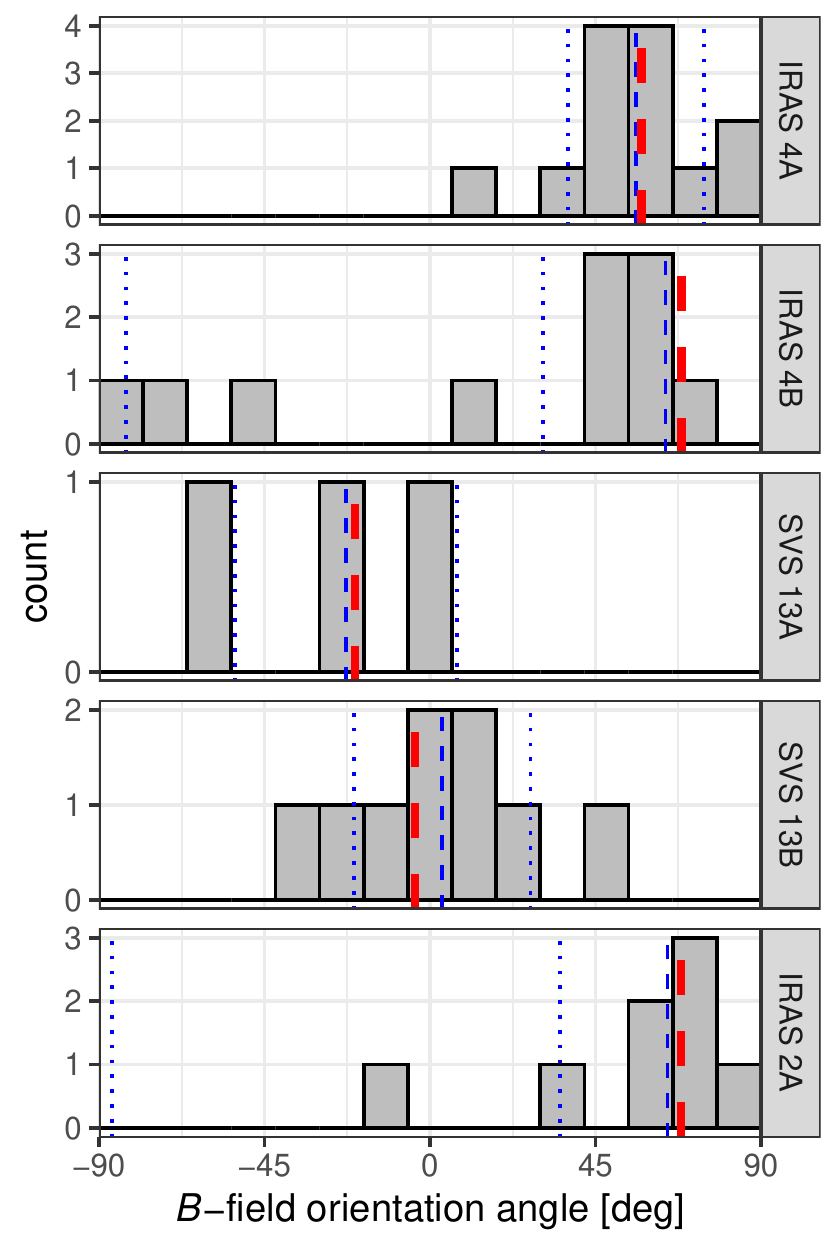}
    \caption{Comparison of \emph{B}-field orientations observed by TADPOL and BISTRO for individual YSOs.
    Histograms are the orientations observed by TADPOL.
    Spatially independent line segments among those given in Figure \ref{fig: TADPOL} are shown.
    Thin blue dashed lines are the circular mean orientation of the TADPOL \emph{B}-field, and thin blue dotted lines are the circular standard deviation around the mean value $(\pm 1~\sigma)$.
    Thick red dashed lines are the \emph{B}-field orientations at the YSO sources observed by BISTRO with $14''\negthinspace\negthinspace.1$ spatial resolution.}
    \label{fig:tadpol_hist}
\end{figure}

As found in Figure \ref{fig: TADPOL}, our observed \emph{B}-fields at $\geqslant 4200$ au scales are in good accordance with those observed by TADPOL, although our data exhibit a smoother distribution owing to the difference in spatial resolution.
To check this consistency in further detail, we show histograms of \emph{B}-field orientations observed with TADPOL in Figure \ref{fig:tadpol_hist}, together with \emph{B}-field orientations observed with BISTRO at the positions of various YSOs.
The comparison between the TADPOL and the BISTRO \emph{B}-fields is summarized in Table \ref{tab: TADPOLvsBISTRO}.

\begin{table*}[t]
\caption{Comparison between TADPOL and BISTRO \emph{B}-fields Shown in Figure \ref{fig: TADPOL}}
\label{tab: TADPOLvsBISTRO}
\begin{center}
\begin{tabular}{lcccc}
\hline
\hline
Region & \multicolumn{2}{c}{TADPOL} & \multicolumn{2}{c}{$\left|\mathrm{TADPOL} - \mathrm{BISTRO}\right|$}\\
\cline{2-3}\cline{4-5}
 & No. of Beams$^\mathrm{a}$ & \emph{B}-field Position Angle$^\mathrm{b}$ & \emph{B}-field Angle Difference$^\mathrm{c}$ & Significance$^\mathrm{d}$ \\
\hline
IRAS 4A & 13 & $~56^\circ\negthinspace.0 \pm 18^\circ\negthinspace.4$ & $1^\circ\negthinspace.5$ & $0.1\sigma$ \\
IRAS 4B & 11 & $~64^\circ\negthinspace.1 \pm 33^\circ\negthinspace.3$ & $4^\circ\negthinspace.3$ & $0.1\sigma$ \\
SVS 13A & ~3 & $-22^\circ\negthinspace.8 \pm 30^\circ\negthinspace.1~$ & $2^\circ\negthinspace.4$ & $0.1\sigma$ \\
SVS 13B & ~9 & $~~3^\circ\negthinspace.3 \pm 24^\circ\negthinspace.0$ & $7^\circ\negthinspace.4$ & $0.3\sigma$ \\
IRAS 2A & ~8  & $64^\circ\negthinspace.5 \pm 29^\circ\negthinspace.1$ & $3^\circ\negthinspace.7$ & $0.1\sigma$ \\  
\hline
\multicolumn{5}{p{0.81\textwidth}}{\bf Notes.}\\
\multicolumn{5}{p{0.81\textwidth}}{$^\mathrm{a}$The number of spatially independent TADPOL observations shown as cyan line segments in Figure \ref{fig: TADPOL}.}\\
\multicolumn{5}{p{0.81\textwidth}}{$^\mathrm{b}$The circular mean and the circular standard deviation of TADPOL \emph{B}-field position angles.}\\
\multicolumn{5}{p{0.81\textwidth}}{$^\mathrm{c}$The differences between the TADPOL \emph{B}-field position angles and the on-source \emph{B}-field position angles observed with BISTRO.}\\
\multicolumn{5}{p{0.81\textwidth}}{$^\mathrm{d}$Statistical significance of the angle difference (c), estimated as the angle difference divided by the circular  standard deviation of the TADPOL \emph{B}-field position angles.}\\
\end{tabular}
\end{center}
\end{table*}

The \emph{B}-field orientations of SVS 13A and SVS 13B show clear differences from those of other locations in NGC 1333.
The differences between the orientations of IRAS 4A, IRAS 4B, and IRAS 2A are not statistically significant, but they do show significantly different orientations from those of the global \emph{B}-field observed by Planck ($-40^\circ \pm 7^\circ\negthinspace.3$; Section \ref{sec: comp with Planck}).
The distribution of the \emph{B}-field orientations that differ for each region is similar to that observed by BISTRO (Section \ref{sec: filaments and B}).
The observed circular standard deviation of the \emph{B}-field orientation angle ($18^\circ$--$33^\circ$) is comparable to those in individual filaments observed with BISTRO ($20^\circ$--$26^\circ$; Table \ref{tab:filament param}).

The BISTRO-observed \emph{B}-field orientation angle at the position of each YSO with $14''\negthinspace\negthinspace.1$ spatial resolution agrees well with that observed by TADPOL (Figure \ref{fig:tadpol_hist} and Table \ref{tab: TADPOLvsBISTRO}).
This consistency is also found in the previous observations of IRAS 4A \citep{1995A&A...293L..61M,1995ApJ...448..346T,1999ApJ...525L.109G,2004Ap&SS.292..509C,2009ApJ...702.1584A} up to a maximum spatial resolution of $1''\negthinspace \negthinspace.56 \times 0''\negthinspace \negthinspace.99~(\simeq 500~\mathrm{au} \times 300~\mathrm{au})$ achieved by SMA \citep{2006Sci...313..812G}.
Considering the significant diversity of \emph{B}-field orientation at spatial scales below 1 pc, we thus note that these field variations are continuous and follow the larger \emph{B}-field structure as we go to smaller scales down to 1000 au (see also \citealp{2018A&A...616A.139G}).

\subsection{Misalignment of the \emph{B}-field and Outflows from YSOs}\label{sec: YSO misalignment}

\citet{2013ApJ...768..159H}, \citet[][and references therein]{2019FrASS...6....3H} examined the correlation between the direction of molecular outflows and \emph{B}-fields in YSOs for a compilation of 30 low- and intermediate-mass sources in various regions in the whole sky.
They found that the correlation is best explained by random orientations of outflows and \emph{B}-fields.
On the other hand, \citet{2018A&A...616A.139G} observed a sample of twelve low-mass Class 0 envelopes in nearby clouds using the SMA and pointed out that the envelope-scale \emph{B}-field is preferentially either aligned with or perpendicular to the outflow direction (e.g., \citealp{2016ARA&A..54..491B,2017ApJ...834..201L,2019FrASS...6...54P}).
Bipolar outflows are launched by the rotating accretion disk of the protostar and thus could be used to infer the orientation of the rotation axis (e.g., \citealp{2016ARA&A..54..491B,2017ApJ...834..201L,2020ApJ...890...44B,2019FrASS...6...54P}).
Accordingly, we can investigate further this correlation by studying how the rotation axes at the very centers of the NGC 1333 star-formation cores are aligned and possibly influenced by the \emph{B}-field orientation in the protostellar envelope (scale $\sim 1000$ au).

Here, we compare outflow orientations with the \emph{B}-fields observed with the JCMT toward all of the protostars with active outflows in NGC 1333.
An advantage of our analysis with respect to that of \citet{2019FrASS...6....3H} is that we can compare outflow versus \emph{B}-field orientation toward a large number (19) of protostars in a single star-forming region.

\begin{figure}[tp]
\centering
\includegraphics[width=\linewidth]{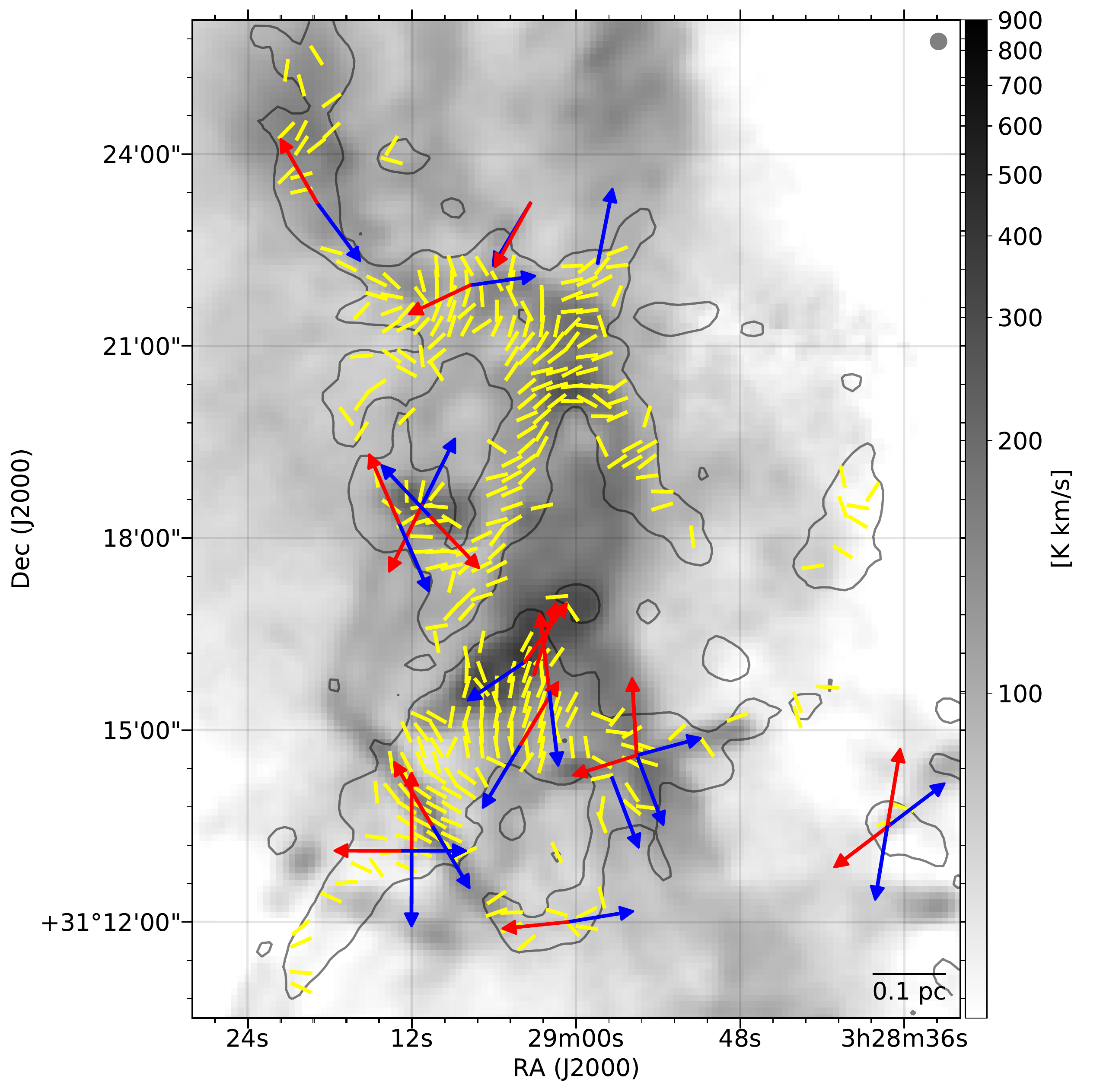}
\caption{CO molecular outflows in NGC 1333 \citep{2017ApJ...846...16S}.
Red and blue arrows are orientations of the red and blue components of the CO outflows.
The length of the arrows has been normalized to show only the orientation of the outflows.
An outflow at $\alpha = 03^{\rm h}29^{\rm m}03.\hspace{-0.25em}^{\rm s}331,~\delta = 31^{\circ}23 \arcmin 14.\negthinspace\negthinspace \arcsec 573$ is not included in our analysis because of the insignificant S/N of our polarimetry at the position ($I < 25$ mJy beam$^{-1}$).
Yellow line segments are the observed \emph{B}-field orientation.
Contours are Stokes $I$ with the level of 25 mJy beam$^{-1}$.
Gray scale is the CO (3--2) integrated intensity map observed by HARP \citep{2009A&A...502..139H}.
The beam of the HARP observation ($15''$) is shown in the upper right corner of the figure.}
\label{fig:B_vs_CO}
\end{figure}

\citet{2017ApJ...846...16S} examined the correlation of 57 position angles between CO (2--1) molecular outflows, observed by SMA at 1.3 mm wavelength, and local filaments for the entire Perseus cloud, including NGC 1333. They concluded that their correlation indicates random orientations of outflows with respect to filaments.
Among these, we find 19 outflows in our observational field (Figure \ref{fig:B_vs_CO}), including five outflows shown in Figure \ref{fig: TADPOL}.
We estimate the \emph{B}-field orientation at each position of YSOs from our JCMT data with $14''\negthinspace\negthinspace.1$ spatial resolution.
We assume that the BISTRO observations, though their spatial resolutions are $\sim 4200$ au, can trace the \emph{B}-field orientation down to $\sim 1000$ au scale based on the discussion in Section \ref{sec: B continuity}.

\begin{figure}[tp]
  \includegraphics[width=\linewidth]{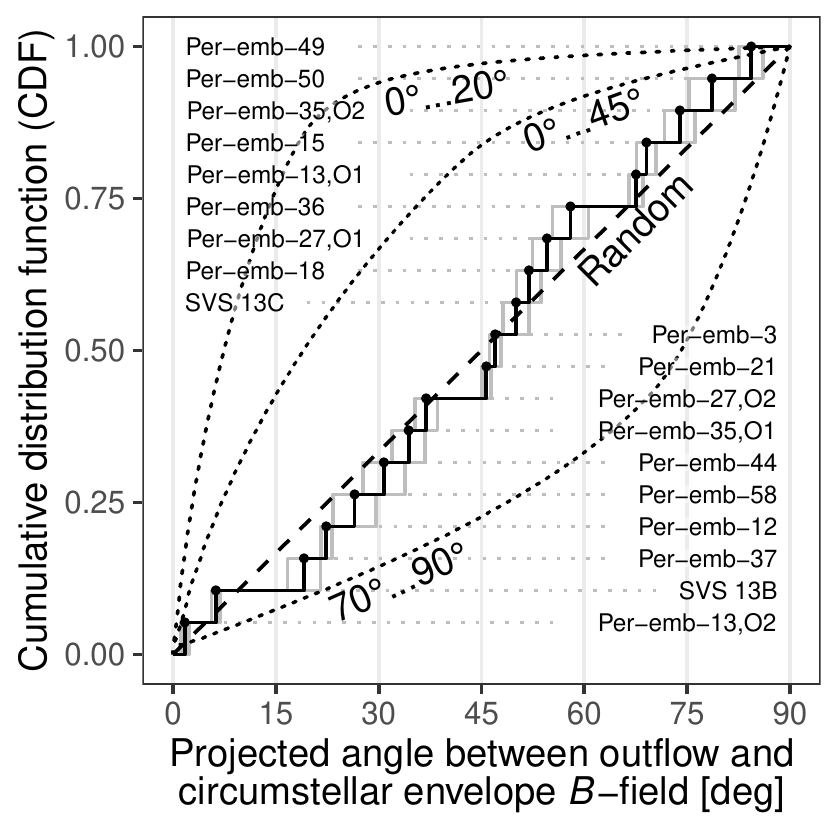}
  \caption{CDF of the projected offset angles between the orientation angles of bipolar outflows \citep{2017ApJ...846...16S} and the \emph{B}-field as observed by the JCMT (this work).
The stepped curve shows the CDF estimated from the observed offset angles (solid line) and its $\pm 1 \sigma$ errors due to the errors of the offset angles (thin lines).
Names of the 19 sources tabulated in \citet{2017ApJ...846...16S} are also shown.
The dotted curves are expected CDFs when the \emph{B}-fields and outflows are oriented within $20 ^\circ$, $45 ^\circ$, and $70 ^\circ$ -- $90 ^\circ$ of one another, respectively. 
The dashed straight line is the CDF for random orientation.}
\label{fig:BISTRO_vs_CO_outflow}
\end{figure}

Figure \ref{fig:BISTRO_vs_CO_outflow} shows the cumulative distribution function (CDF) of the projected angles between the outflows from \citet{2017ApJ...846...16S} and the \emph{B}-field orientations we measure with the JCMT.
As shown in the figure, the CDF is consistent with random orientations.
This random distribution is in accordance with previous studies \citep{2004A&A...425..973M,2007MNRAS.382..699C,2010ApJ...716..893P,2011ApJ...743...54T,2013ApJ...768..159H,2014ApJS..213...13H,2019FrASS...6....3H} and thus reconfirms the detachment of rotation axes at the centers of the star-forming cores from the larger-scale \emph{B}-fields at $\sim 1000$ au protostellar envelope scales.

\section{\emph{B}-field Associated with Filaments in a Three-dimensional Space}
\label{sec: B association filament}

Theoretical studies suggest that dense filaments form perpendicularly to the local \emph{B}-field (\citealp{2019FrASS...6....5H} for a review).
Since the observed position angles of \emph{B}-fields and filaments are the results of projection onto the POS, it is important to understand the true 3D morphologies of such structures as a first step \citep{2015ApJ...807...47T}.
Here, we make the simplifying assumption that the \emph{B}-field lines are indeed perpendicular to the dense filaments in NGC 1333 (Section \ref{sec: filaments and B}) and show that our observed offset angles can be explained by considering different inclination angles for these filaments with respect to the POS.
Note that our model presented here is purely geometrical and does not take into account physical properties of filaments.

\subsection{Effect of the 3D Orientation of the \emph{B}-field and the Filament on the Observed Projected \emph{B}-field Angle}
\label{sec: simple filament and B}

\begin{figure}[tp]
\centering
\includegraphics[width=\linewidth]{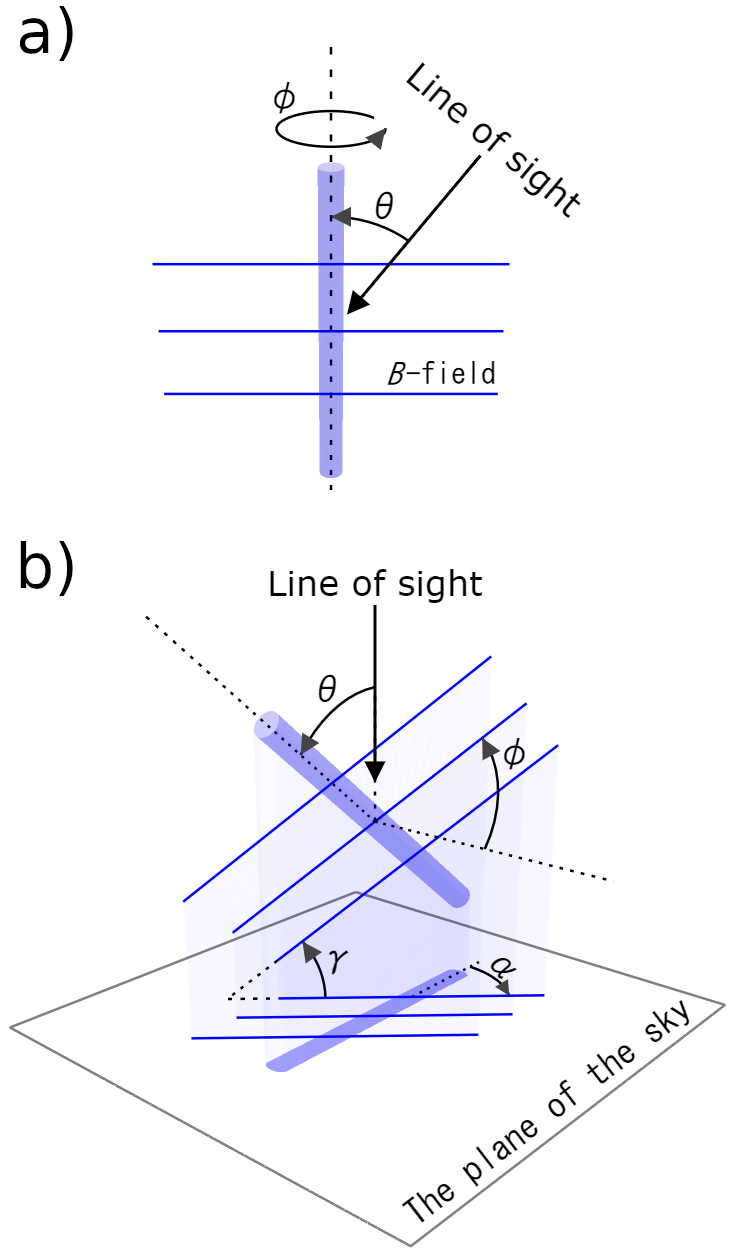}
\caption{Schematic drawing of the 3D structure of the filament and the \emph{B}-field, and their projection to the POS, as introduced by \citet{2015ApJ...807...47T}.
(a) Definitions of angles $\theta$ and $\phi$.
A filament and the \emph{B}-field are orthogonal in the model.
The angle $\theta$ is the inclination angle of the filament with respect to the LOS.
The angle $\phi$ is a rotation angle of the filament about its long axis.
The origin of the angle $\phi$ is defined as the angle when the \emph{B}-field is on the POS (see panel (b)).
(b) Definitions of angle $\alpha$ and $\gamma$, and their relation to $\theta$ and $\phi$.
The angle $\alpha$ is a relative orientation angle between the filament and the \emph{B}-field projected to the POS.
The angle $\gamma$ is an inclination angle of the \emph{B}-field with respect to the POS.
}
\label{fig:filament 3D}
\end{figure}

Figure \ref{fig:filament 3D} shows our assumed configuration of a filament and a \emph{B}-field in a 3D space.
\citet{2015ApJ...807...47T} showed that a projected offset angle between a \emph{B}-field and a filament ($\alpha$ in Figure \ref{fig:filament 3D}) can be different from $90^\circ$ even when they are perpendicular with each other in a 3D space, if we observe a filament that is inclined relative to the POS.

In Figure \ref{fig:filament 3D}, we indicate the definition of relative angles that were introduced by \citet{2015ApJ...807...47T}.
The angle $\theta$ is the relative inclination angle of a filament with respect to the LOS.
The angle $\phi$ is the rotation angle about the long axis of the filament.
 \citet{2015ApJ...807...47T} defined $\phi=0^\circ$ when a \emph{B}-field, which is orthogonal to the filament, is parallel with respect to the POS and perpendicular with respect to the LOS (see Figure \ref{fig:filament 3D}b).
These two angles ($\theta$ and $\phi$) are the two independent parameters that determine the orientation of a set of the filament and the \emph{B}-field in a 3D space.

The relative orientation angle between the filament and the \emph{B}-field we observe is the projected angle onto the POS, as indicated in Figure \ref{fig:filament 3D}b.
\citet{2015ApJ...807...47T} defined this projected offset angle as $\alpha$.
The angle $\alpha$ is  estimated as a function of
$\theta$ and $\phi$ as follows:
\begin{equation}
   \alpha = \frac{\pi}{2} - \arctan\left[\tan(\phi) \cdot \cos(\theta)\right]~. 
\label{eq:alpha}
\end{equation}

The angle $\gamma$ is the relative inclination angle of the \emph{B}-field with respect to the POS.
$\gamma$ is a dependent parameter of $\theta$ and $\phi$, as is the case of $\alpha$.
The following equation expresses the relationship between these parameters:
\begin{equation}
\sin \left( \gamma \right) = \sin \left( \phi \right) \cdot \sin \left( \theta \right)~.
\label{eq:gamma}
\end{equation}

As indicated in Equation (\ref{eq:alpha}), $\alpha$ can be significantly less than $90^\circ$ when $\left[\tan(\phi) \cdot \cos(\theta)\right] \gg 0$, in other words, $\theta < 90 ^\circ$ and $\phi > 0 ^\circ$.
In that case, Equation (\ref{eq:gamma}) indicates that $\gamma > 0^\circ$ unless $\theta = 0^\circ$.
That means that both the filament and the \emph{B}-field are inclined with respect to the POS if the observed $\alpha$ is smaller than $90^\circ$.
As a result, our model requires that a filament and the \emph{B}-field both be inclined with respect to the POS if nearly parallel orientation between the filament and the \emph{B}-field is observed.
If the relative orientation between the filament and the \emph{B}-field is orthogonal, our model requires that either or both the filament and the \emph{B}-field be in the POS.

\subsection{Probability Distribution of Offset Angle Projected to the POS}

Here we estimate the probability distribution of observed offset angle $\alpha$ if the combination of a filament and a \emph{B}-field shown in Figure \ref{fig:filament 3D} is randomly oriented in a 3D space.
We then discuss the consistency of this probability distribution with the values observed by BISTRO and by Planck.

Suppose a filament is on the POS and perpendicular with respect to the LOS $(\theta = 90^\circ)$.
In this case, $\alpha$ is always $90^\circ$ regardless of the value of $\phi$  (the rotation angle about the long axis of the filament).
Thus, the associated probability distribution shows a strong concentration at $90^\circ$.
On the other hand, if a filament is nearly parallel with respect to the LOS and perpendicular with respect to the POS $(\theta \simeq 0^\circ)$, $\alpha$ becomes any value depending on the value of $\phi$.
Thus, the associated probability distribution is uniform between $0^\circ$ and $180^\circ$ in this case.

\begin{figure}[tp]
\centering
\includegraphics{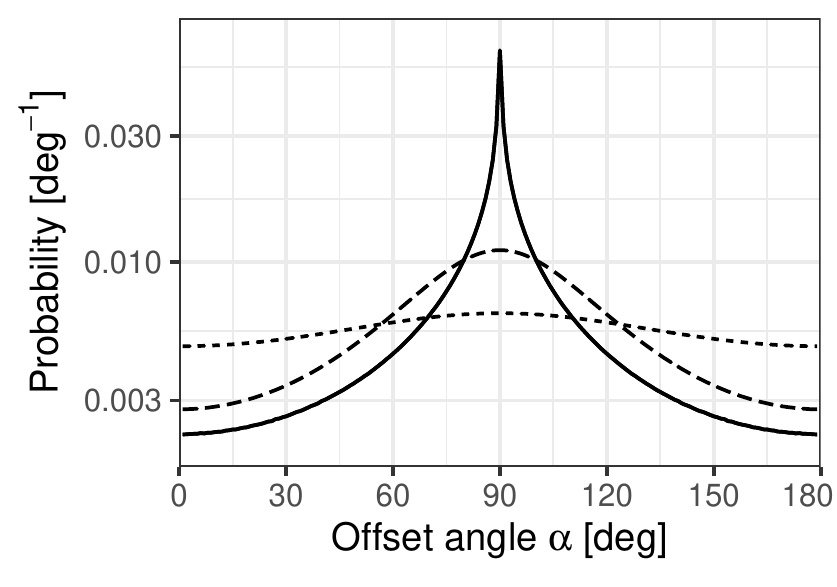}
\caption{Probability distributions of the projected offset angle $\alpha$ between the filament and the \emph{B}-field.
The solid line indicates the total probability assuming random orientation of both the filament and the \emph{B}-field while keeping their relative angle as $90 ^\circ$.
The dashed line shows the probability when $\theta = 60 ^\circ$.
Keeping the inclination angle of the \emph{B}-field with respect to the POS $\gamma = 30 ^\circ$ gives the same probability distribution.
The dotted line shows the probability for $\theta = 30 ^\circ$ or $\gamma = 60 ^\circ$.}
\label{fig:offset_func}
\end{figure}

For other values of $\theta$, the associated probability distributions are between these two extreme cases.
The distribution shows a loose concentration to $\alpha=90^\circ$ (the perpendicular projected orientation of the filament and the \emph{B}-field), and the degree of concentration increases at larger $\theta$.
In Figure \ref{fig:offset_func}, we show the probability distribution of $\alpha$ for $\theta=30^\circ$  (the dotted line in the figure) and $\theta=60^\circ$ (the dashed line in the figure).

When filaments are oriented randomly in a 3D space, the distribution of their inclination angle $\theta$ is more likely to be $90^\circ$ and less likely to be $0^\circ$ with a dependence $P(\theta) \propto \sin(\theta)$.
Taking into account this dependence, we estimate the probability distribution of $\alpha$ when the inclination and rotation angles ($\theta$ and $\phi$) of filaments are completely random, as shown in Figure \ref{fig:offset_func} as a solid line.
As shown in the figure, the $\alpha$ value shows a higher probability to be perpendicular than to be parallel.
The probability that $\alpha$ is observed to be perpendicular $(\alpha = 90 ^\circ \pm 30 ^\circ)$ is 66\%, while the probability that $\alpha$ is observed to be parallel $(\alpha = 0 ^\circ \pm 30 ^\circ)$ is 14\%.

This higher probability in perpendicular orientation is consistent with the preferentially perpendicular orientation of the \emph{B}-field with respect to the dense filaments found in Planck, BLASTPol, and ground-based observations \citep{2016A&A...586A.135P,2016A&A...586A.136P,2016A&A...586A.138P,2017ApJ...846..122P,2017A&A...603A..64S,2017ApJ...842...66W,2018ApJ...859..151L,2019ApJ...878..110F,2019ApJ...883...95S}.
Thus, our model, where we assume that a filament and the \emph{B}-field are orthogonal in a 3D space, can successfully reproduce these observational results.

We can also consider a random orientation of filaments in a globally uniform \emph{B}-field with a fixed inclination angle $\gamma$ to the POS.
In our model, the filament and the \emph{B}-field are orthogonal straight lines and thus interchangeable.
As a result, probability distributions for fixed $\theta$ shown in Figure \ref{fig:offset_func} (the dotted line for $\theta = 30^\circ$ and the dashed line for $\theta = 60^\circ$) also demonstrate probability distributions for fixed $\gamma$ with random orientation of filaments.
The dotted line corresponds to the case of $\gamma = 60^\circ$, and the dashed line corresponds to the case of $\gamma = 30^\circ$.

Referring to the Planck-observed $\alpha$ distribution of star-formation regions, we note that the level of concentration around $90 ^\circ$ is different from region to region (see probability distributions shown in Figures 3 and 4 of \citealp{2016A&A...586A.138P}).
This difference in concentration is consistent with our estimated probability distributions for different \emph{B}-field inclination angles (fixed-$\gamma$ cases in Figure \ref{fig:offset_func}).
Thus, the difference could be attributable to the difference in the inclination angle of the large-scale local \emph{B}-field.

In NGC 1333, we find that three out of five cases are perpendicular and one or two cases are parallel (Table \ref{tab:filament param}).
This result is roughly consistent with the probability expected from random orientation of both filaments and the \emph{B}-field.
Also, the case of $\gamma = 30^\circ$ (the dashed line in Figure \ref{fig:offset_func}) gives nearly the same probability as $P[\alpha = 90 ^\circ \pm 30 ^\circ] = 55\%$ and $P[\alpha = 0 ^\circ \pm 30 ^\circ] = 18\%$.

We thus claim that filament formation in a global \emph{B}-field that has $\sim 30^\circ$ inclination with respect to the POS is a plausible scenario to explain the observed distribution of polarization vectors associated with filaments in NGC 1333.
The considerable variation in the \emph{B}-field orientations of individual filaments within NGC 1333 suggests that the \emph{B}-field may locally be modified by filaments, most probably due to their formation and evolutionary processes.
Our model thus indicates that filaments, which may form in random orientations, are likely to modify the \emph{B}-field locally but keep the relative orientation between the local \emph{B}-field and filaments perpendicular to each other.

\section{Discussion}
\label{close-relationship-between-filaments-and-b-field}

Our data reveal for the first time a complex \emph{B}-field structure across the entire star formation region of NGC 1333.
This network is found on scales smaller than the smooth distribution of the global \emph{B}-field probed by Planck (Section \ref{sec: comp with Planck}).
For scales of $\sim 1$--0.01 pc, the complex \emph{B}-field structure observed by BISTRO shows overall consistency with archival interferometric observations with higher spatial resolutions (Section \ref{sec: B continuity}).
Here, we discuss possible causes of this increasing complexity of the \emph{B}-field at smaller scales.

One possible cause is a dynamical interaction of the molecular outflows from YSOs, which may perturb the surrounding ISM and thus affect the \emph{B}-field morphology.
For example, it has been proposed that molecular outflows can disrupt the parent clouds around YSOs and terminate star formation ('stellar feedback'; e.g., \citealp{2013MNRAS.435.1701C}).

In Figure \ref{fig: TADPOL}, we plot CO molecular outflows associated with some NGC 1333 YSOs, together with the local \emph{B}-fields observed by both BISTRO and TADPOL.
We find, however, no sign of gas interaction in the observed \emph{B}-field morphologies shown in Figure \ref{fig: TADPOL}.
This lack of interaction is also the case for the \emph{B}-field morphology of the entire NGC 1333 region (Figure \ref{fig:B_vs_CO}; see also \citealp{2010ApJ...716..893P}), although many outflows are found in NGC 1333 \citep{2000A&A...361..671K,2009A&A...502..139H,2010ApJ...715.1170A,2010MNRAS.408.1516C,2013ApJ...774...22P}.
Two exceptions, however, include the warped \emph{B}-field morphologies seen around HH 12 and SVS 3, which could be due to interactions with local ISM, i.e., an outflow from SVS 13B (HH 12; e.g., \citealp{2008hsf1.book..346W}) and a reflection nebula around SVS 3, respectively.

Another possible cause is a deformation of the \emph{B}-field associated with individual filaments, as we demonstrated that the observed local \emph{B}-field intersects with individual filaments with uniform offset angles in a 3D space (Section \ref{sec: filaments and B}).
Furthermore, we recognize that the filaments and local \emph{B}-fields are presumably perpendicular to each other (Section \ref{sec: B association filament}).

\begin{figure}[tp]
  \includegraphics[width=\linewidth]{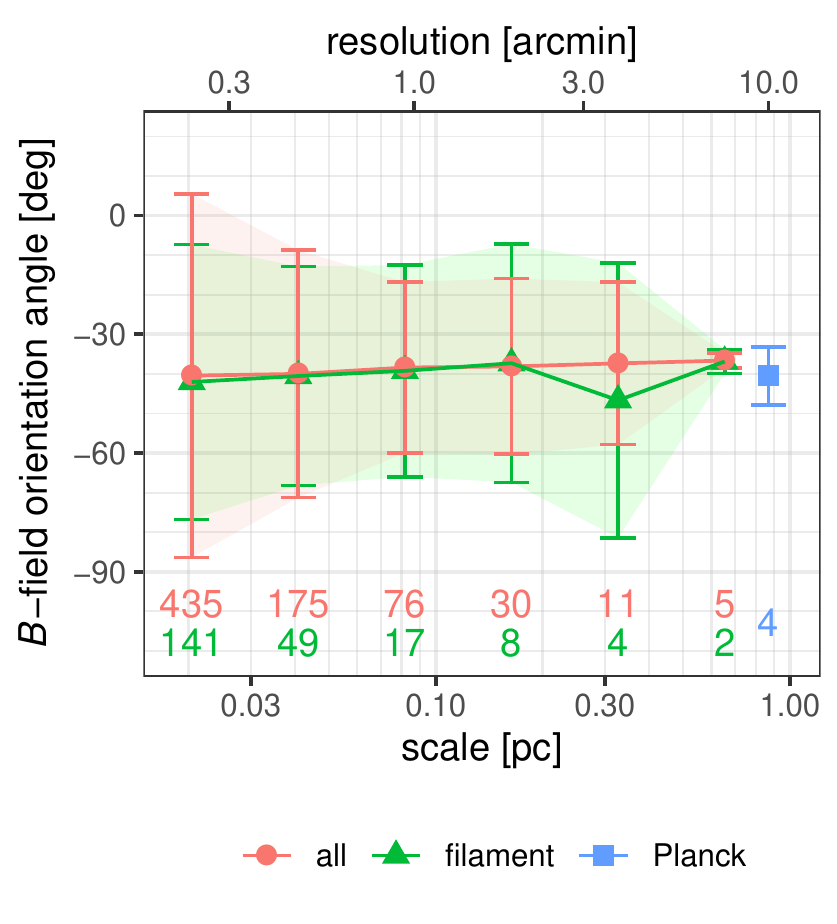}
  \caption{Scale dependence of the circular mean and the circular standard deviation ($\pm 1 \sigma$) of the \emph{B}-field orientation in NGC 1333.
Here we add the estimated missing large-scale flux of $Q$ and $U$ (Section \ref{sec:Planck absolute}) to our observed $Q$ and $U$ values to evaluate the \emph{B}-field orientation of larger spatial scales ($=$ lower spatial resolutions).
We apply Gaussian-smoothing to $Q$ and $U$ values and estimate the \emph{B}-field orientations at reduced spatial resolutions.
The points 'all' correspond to all the independent JCMT observations (white line segments in Figure \ref{fig:B image} and smaller number of beams for reduced spatial resolutions), while the points 'filament' correspond to the beams selected from 'all' whose central positions are on the filaments (\#7, \#13, \#15, and \#18 in Figure \ref{fig:alpha3D_ID}).
The \emph{B}-field orientation observed by Planck ($10'$ spatial resolution; yellow line segments in Figure \ref{fig:B image}) is also shown.
Note that the Planck data cover a wider field than the BISTRO data (see Figure \ref{fig:B image}).
The colored labels at the bottom of the figure indicate the number of independent beams.}
\label{fig:csd res}
\end{figure}

We show a spatial scale dependence of the angular dispersion of the \emph{B}-field orientations in NGC 1333 in Figure \ref{fig:csd res}.
The angular deviation is significant at spatial scales below $\sim 0.3$ pc, which corresponds to the typical length of the filaments (Table \ref{tab:filament param}).
This trend becomes more prominent if we select the data points whose central positions are on the filaments in NGC 1333 ('filament' in Figure \ref{fig:csd res}).
The angular dispersions of 'filament' show no further considerable increase at smaller scales, suggesting that the \emph{B}-field associated with filaments mainly deforms at the scale of the filament length and keeps its structure below that scale.

Indeed, the observed spatial scale dependence of the \emph{B}-field structure in NGC 1333 is consistent with a proposed mechanism for creating a filamentary cloud.
In an ideal MHD regime, gas flows along the \emph{B}-field onto the cloud much faster than perpendicular to the field.
As a result, long structures can be formed even in a highly supersonic environment, with the \emph{B}-fields acting as both the guiding rails of gas flow and the reinforcement.
The outcome is a long filamentary cloud with a \emph{B}-field oriented roughly perpendicular to the long axis of the cloud (e.g., \citealp{2015A&A...580A..49I,2018PASJ...70S..53I,2019MNRAS.485.4509L}).
These studies show that the projection effect can lead to incorrect interpretations of the physical shape of the clouds.

In fact, large-scale shock compression of the Perseus molecular cloud, including NGC 1333, is suggested by other observations.
The LOS velocity distributions of H\small{I} and CO lines suggest compression of the molecular cloud by an expanding ISM shell associated with the Per OB2 association \citep{1974IAUS...60..115S,2006A&A...451..539S,2019A&A...623A..16S}.
Furthermore, the LOS \emph{B}-field morphology around the molecular cloud (\citealp{2018A&A...614A.100T}; see also \citealp{2019A&A...632A..68T}) hints at deformation of the large-scale \emph{B}-field by the compressed molecular cloud.

Deformation of the \emph{B}-field at the filament length scale and the uniform \emph{B}-field orientations associated with individual filaments suggest that the formation and evolutionary process of filaments cause significant changes in the \emph{B}-field morphology with respect to the global \emph{B}-field observed by Planck.
Once a filament is formed, the filament and its \emph{B}-field maintain a constant angular orientation down to protostellar cores.
The existence of filaments in NGC 1333 that are misaligned with each other suggests that the compression mechanism may have acted multiple times with different overall orientations \citep{2015A&A...580A..49I}, which caused the complicated configuration of the \emph{B}-field in NGC 1333.

Our model, in which we assume the perpendicular orientation of filaments and the associated \emph{B}-field, is in good accordance with our observations.
The model, however, does not exclude the possibility that filaments and the \emph{B}-field may have relative orientations that differ from perpendicular.
By checking the applicability of our model to other regions where observational data of sufficient spatial resolution are available, we can better examine the plausibility of this model.

A diffuse ISM cloud, which is subcritical against magnetic pressure, becomes supercritical if its column density exceeds a threshold value of $N_\mathrm{H} \sim 10^{21} - 10^{22}~\mathrm{cm}^{-2}$ \citep{2012ARA&A..50...29C}.
The ISM in this column density range can be the site of filament formation.
We should keep in mind that our observations with the JCMT are limited to regions of $N_\mathrm{H} > 10^{23}~\mathrm{cm}^{-2}$, which are much higher than the threshold value.
To achieve a thorough understanding of the filament formation process, we should aim at making direct observations of the \emph{B}-field in the ISM with column densities down to $10^{21}~\mathrm{cm}^{-2}$ at high spatial resolution.
HAWC+ on board SOFIA pioneers polarimetry of low column density gas around bright star-forming regions (e.g., \citealp{2019ApJ...872..187C,2019ApJ...882..113S}).
Extensive observations can be conducted by future space-borne facilities (e.g., \citealp{2019PASA...36...29A,2019SPIE11115E..0QL}).

\section{Conclusions}
\label{sec:conclusions}

We performed submillimeter polarimetric observations using SCUBA-2/POL-2 at the JCMT and revealed the POS projections of the \emph{B}-field of the active star formation region NGC 1333 as a part of the BISTRO survey.
Our observations cover spatial scales of about 0.02--1 pc, which are crucially important for the formation of the filaments in the star-forming ISM.
These data mark the first time that the \emph{B}-field structure across an entire star formation region has been revealed on these spatial scales.

We draw the following conclusions:

\begin{enumerate}
\item
  We detect polarized emission \emph{PI} from an intricate network of filaments in the observed region $(\sim 1.5~\mathrm{pc} \times 2~\mathrm{pc})$ with a column density above $\sim 10^{23} ~\mathrm{H~cm}^{-2}$.
\item
  While the observations by Planck revealed a rather uniform and slowly varying \emph{B}-field structure over the entire Perseus molecular cloud, our observations show a highly complex \emph{B}-field structure in NGC 1333.
  This difference cannot be attributed to the missing flux in the JCMT observations.
  We instead propose that the \emph{B}-field changes its intrinsic structure on scales $< 1$ pc.
\item
  The observed \emph{B}-fields around active YSOs with the JCMT (4200 au resolution) show overall consistency with higher spatial resolution interferometric data (1000 au resolution), indicating that the \emph{B}-field structure remains broadly continuous between $\sim 1$ pc and 1000 au spatial scales. 
\item
  We find no correlation between the \emph{B}-field position angles near YSOs as traced by BISTRO and the rotation axes of the YSOs as inferred from molecular outflows.
\item
  The \emph{B}-fields associated with individual filaments show a uniform orientation angle that is not equal to the orientation angles of the global \emph{B}-field or those of other filaments.
  Projected offset angles of local \emph{B}-fields to filaments are also different from filament to filament, ranging from orthogonal to nearly parallel. 
\item
  We successfully reproduce the observed variety of offset angles by using a simple model, in which the \emph{B}-field and the long axis of a filament are perpendicular to each other in a 3D space.
  Since the observed offset angle is a projection of the true state of affairs onto the POS, the observed angle can be significantly narrower than $90^{\circ}$.
  It can even be nearly parallel ($0^{\circ}$) if both the \emph{B}-field and the filament are significantly inclined with respect to the POS.
  Random orientations of the \emph{B}-field and filaments in a 3D space can reproduce the observed distribution of the offset angle.
  A \emph{B}-field that has a constant inclination angle of $\sim 30^\circ$ with respect to the POS and a random orientation of filaments can also reproduce consistently the observed distribution of the offset angle.
\item
  We demonstrate that observed offset angle is more likely to be perpendicular than to be parallel, even if the filament and the \emph{B}-field are perpendicular with each other in a 3D space but randomly oriented with respect to the LOS.
  This result is consistent with previous observations using Planck and BLASTPol that showed that offset angles tend to be perpendicular.
  Random orientations of filaments in a \emph{B}-field that have a constant inclination angle with respect to the POS show different probability distributions of the offset angle, ones that are relatively less likely to be perpendicular if the \emph{B}-field has a larger inclination angle with respect to the POS.
  This different inclination angle of the \emph{B}-field can be an explanation of the differences of the offset angle distribution between regions found by Planck observations.
\end{enumerate}

\acknowledgments

\newpage

The authors thank the anonymous referee for their advice and suggestions, which improved the manuscript.
The James Clerk Maxwell Telescope is operated by the East Asian Observatory on behalf of the National Astronomical Observatory of Japan; Academia Sinica Institute of Astronomy and Astrophysics; the Korea Astronomy and Space Science Institute (KASI); the Operation, Maintenance and Upgrading Fund for Astronomical Telescopes and Facility Instruments, budgeted from the Ministry of Finance (MOF) of China and administrated by the Chinese Academy of Sciences (CAS); and the National Key R\&D Program of China (No. 2017YFA0402700).
Additional funding support is provided by the Science and Technology Facilities Council of the United Kingdom and participating universities in the United Kingdom and Canada.
SCUBA-2 and POL-2 were built through grants from the Canada Foundation for Innovation.
This research used the facilities of the Canadian Astronomy Data Centre operated by the National Research Council of Canada with the support of the Canadian Space Agency.
This research has also made use of the SIMBAD database and of NASA's Astrophysics Data System Bibliographic Services.
Part of the data analysis was carried out on the open-use data analysis computer system at the Astronomy Data Center, ADC, of the National Astronomical Observatory of Japan.
This research has been supported by Grants-in-Aid for Scientific Research (25247016, 18H01250, 19H01938) from the Japan Society for the Promotion of Science and a crowdfunding ran by academist project No. 25 (\href{https://academist-cf.com/}{https://academist-cf.com/projects/25}).
Simon Coud\'e is supported by the Universities Space Research Association, which operates the SOFIA Science Center under contract NNA17BF53C with the National Aeronautics and Space administration.
C.L.H.H. acknowledges the support of both the NAOJ Fellowship and JSPS KAKENHI grant 18K13586.
Doris Arzoumanian acknowledges support by FCT/MCTES through national funds (PIDDAC) by the grant UID/FIS/04434/2019.
F.K. and L.F. acknowledge support from the Ministry of Science and Technology of Taiwan, under grant MoST107-2119-M-001-031-MY3 and from Academia Sinica under grant AS-IA-106-M03.
C.W.L. is supported by the Basic Science Research Program through the National Research Foundation of Korea (NRF) funded by the Ministry of Education, Science and Technology (NRF-2019R1A2C1010851).
M.K. was supported by Basic Science Research Program through the National Research Foundation of Korea (NRF) funded by the Ministry of Science, ICT \& Future Planning (No. NRF-2015R1C1A1A01052160).
Di Li acknowledges support by NSFC grant No. 11725313.
W.K. was supported by the New Faculty Startup Fund from Seoul National University and by the Basic Science Research Program through the National Research Foundation of Korea (NRF-2016R1C1B2013642).

\facilities{JCMT (SCUBA-2/POL-2), Planck, CARMA, SMA, IRAM}
\software{Starlink \citep{2014ASPC..485..391C}, astropy \citep{2013A&A...558A..33A}, dbscan \citep{JSSv091i01}}





\appendix

\section{Comparison with Previous SCUBA Results}\label{sec:scupol}

The detection of the polarized emission with SCUPOL \citep{2004Ap&SS.292..509C} was limited to the regions around SVS 13, IRAS 2, and IRAS 4, which are particularly bright objects (see Figure \ref{fig:SCUPOL B}).
In contrast, our SCUBA-2/POL-2 data trace the global distribution of $I$ and $PI$ across the entirety of NGC 1333, meaning that we have achieved a significant improvement in sensitivity from SCUPOL to SCUBA-2/POL-2.
To make a direct comparison of the two datasets, we regrid the $I$, $Q$, and $U$ data of SCUPOL with the same procedure that was applied to our SCUBA-2/POL-2 data and obtain $I$, $Q$, and $U$ maps with the same spatial grids.
Note that archival data of SCUPOL are raw data in units of volts and not corrected for the variation of the FCF values, which may vary by $\sim 20$\% between individual exposures \citep[see Section 3 of][]{2009ApJS..182..143M}.
The uncertainty of each sample is not archived.
Therefore, we assume constant uncertainty for $I$, $Q$, and $U$ values, respectively, and apply intensity-weighted fits for regridding.
We estimate the observational error as the statistical scatters of fitting residuals for the regridding.

\begin{figure*}[tp]
\centering
\includegraphics[width=\linewidth]{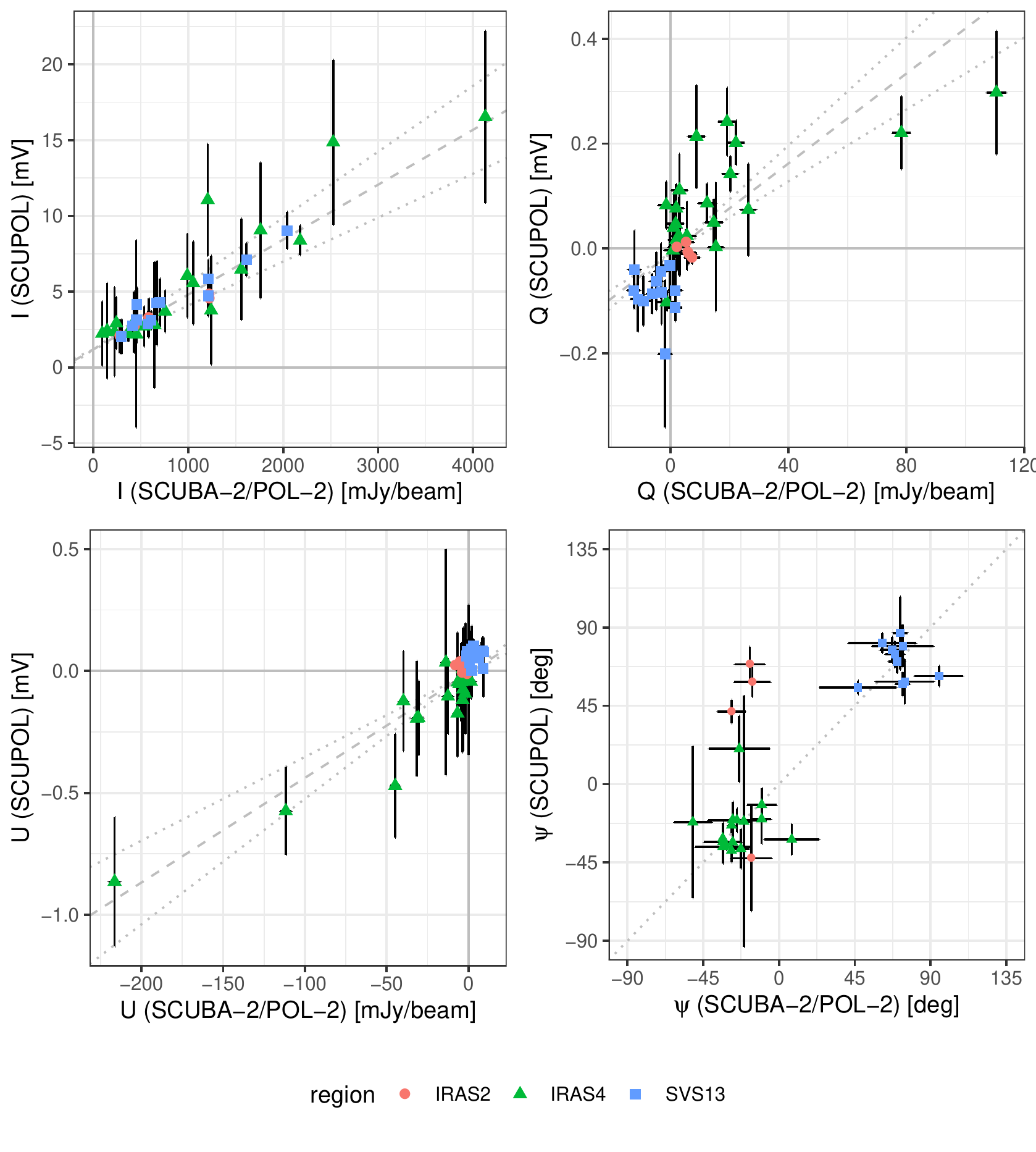}
\caption{Correlations of $I$, $Q$, $U$, and polarization angle $\psi$ between SCUBA-2/POL-2 and SCUPOL data. The error bars represent $3~\sigma$ errors.
Note that error bars of SCUBA-2/POL-2 measurements may be hidden behind the symbols.
The SCUPOL data are taken from \citet{2009ApJS..182..143M}.
See text for the details of the intensity and error estimation of SCUPOL data.
The dashed lines in the $I$, $Q$, and $U$ panels indicate the best-fit linear correlations, and dotted lines represent $\pm 20$\% variation of the conversion factor between the two datasets.}
\label{fig:SCUPOL_corr}
\end{figure*}

We show the correlations between SCUPOL and SCUBA-2/POL-2 in Figure \ref{fig:SCUPOL_corr} to check the consistency of the two datasets.
The estimated correlations are
\begin{eqnarray}
I_\mathrm{SCUBA-2/POL-2}~[\mathrm{mJy~beam^{-1}}] &=& (227\pm18)\cdot I_\mathrm{SCUPOL} + (-156\pm68)~\mathrm{[mV]}, \nonumber\\
Q,U_\mathrm{SCUBA-2/POL-2}~[\mathrm{mJy~beam^{-1}}] &=& (165\pm12)\cdot Q,U_\mathrm{SCUPOL} + (1\pm2)~\mathrm{[mV]}. \nonumber
\end{eqnarray}
The estimated FCF values for SCUPOL data ($227 \pm 18$ for $I$ and $165 \pm 12$ for $Q$ and $U$) are within the 20\% variation of the expected FCF value for SCUPOL ($207 \pm 13~\mathrm{Jy~beam^{-1}~V^{-1}}$; \citealp{2002MNRAS.336...14J}).
See Section \ref{comparison-with-the-previous-scuba-results} for the discussion on the offsets in \emph{Q} and \emph{U} between SCUBA-2/POL-2 and SCUPOL and the resultant \emph{B}-field position angle difference around IRAS 2.

\section{Comparison with the Total $I$, $Q$, $U$ Values Observed by Planck}
\label{sec:Planck absolute App}

To estimate the missing large-scale flux that is not recovered by JCMT, we compare our JCMT $850~\mu \mathrm{m}$ \emph{I}, \emph{Q}, and \emph{U} intensities with the Planck 353 GHz $(= 850~\mu \mathrm{m})$ observations \citep{2018arXiv180104945P}.
We assume a circular beam with a diameter of $14''\negthinspace\negthinspace.1$ for the JCMT data and convert them to units of surface brightness $[\mathrm{MJy~sr}^{-1}]$.
For the Planck data, we transform the original Planck $Q$ and $U$ data that are given in Galactic coordinates to the values in equatorial coordinates to make the comparison.
We refer to Table 2 of \citet{2018arXiv180104945P} for the intensity unit conversion and color correction.
The correction factor assumes a modified blackbody spectrum with a spectrum index of $\beta_d = 1.5$ and a dust temperature of $T_d = 19.6$ K.

The spatial extent of our observed area is not enough to correlate the JCMT and Planck data, as the effective spatial resolution of Planck polarimetry data $(10')$ is comparable to the size of our observation region.
Therefore, we estimate the Stokes \emph{I}, \emph{Q}, and \emph{U} intensities at the center of our observed region $(\mathrm{R.A.} = 3^\mathrm{h}29^\mathrm{m}00^\mathrm{s},~\mathrm{decl.} = 31^\circ17'00'')$ with $10'$ spatial resolution for both BISTRO and Planck data and estimate the offsets.

The estimated values are\vspace{-1ex}
\begin{eqnarray}
I_{Planck} = 23~\mathrm{MJy~sr}^{-1},~
&&I_\mathrm{JCMT} = 20~\mathrm{MJy~sr}^{-1},~\nonumber\\
Q_{Planck} = -0.13~\mathrm{MJy~sr}^{-1},~
&&Q_\mathrm{JCMT} = 0.008~\mathrm{MJy~sr}^{-1},~\nonumber\\
U_{Planck} = 0.42~\mathrm{MJy~sr}^{-1},~
&&U_\mathrm{JCMT} = 0.03~\mathrm{MJy~sr}^{-1}.\nonumber
\end{eqnarray}\vspace{-3ex}

The spatially smoothed \emph{I}, \emph{Q}, and \emph{U} of the JCMT are always smaller in absolute values compared to those of Planck, supporting that the observation by JCMT does not recover the large-scale component of the emission.
The estimated offsets then become\vspace{-1ex}
\begin{eqnarray}
I_{Planck\mathrm{-JCMT}} &=& 2.69~\mathrm{MJy~sr}^{-1}\nonumber\\
&=& 9.9~\mathrm{mJy~(14''\negthinspace\negthinspace.1\text{-}beam)}^{-1},\nonumber\\
Q_{Planck\mathrm{-JCMT}} &=& -0.14~\mathrm{MJy~sr}^{-1}\nonumber\\
&=& -0.51~\mathrm{mJy~(14''\negthinspace\negthinspace.1\text{-}beam)}^{-1},\nonumber\\
U_{Planck\mathrm{-JCMT}} &=& 0.39~\mathrm{MJy~sr}^{-1}\nonumber\\
&=& 1.42~\mathrm{mJy~(14''\negthinspace\negthinspace.1\text{-}beam)}^{-1} .\nonumber
\end{eqnarray}\vspace{-3ex}

The estimated offsets of \emph{I} are significant ($\sim 4~\sigma$), while those of \emph{Q} and \emph{U} are comparable to the uncertainties in our data. For example, the median values of the uncertainties are $\delta Q = 0.88~\mathrm{mJy~beam}^{-1}$ and $\delta U = 0.91~\mathrm{mJy~beam}^{-1}$, respectively.
Thus, the offset values estimated above correspond to $0.58~\sigma$ and $1.56~\sigma$, respectively, or less than two times the observational errors.
The difference between the \emph{B}-field position angles with and without considering the offset values is $-0^\circ\negthinspace.2 \pm 8^\circ\negthinspace.2$ (the circular mean and the circular deviation) if we restrict our analysis to the data with $PI / \delta PI \geqslant 3$, which is applied to the discussion of $\psi$ throughout this paper.

\section{Definitions of the Circular Mean and the Circular Standard Deviation of Polarization Pseudo-vectors in Directional Statistics}
\label{sec:Directional}

To summarize the statistical distributions of polarization position angles, we need to take into account the $180^\circ$ degeneracy of the polarization pseudo-vectors.
This accounting is especially necessary where the pseudo-vectors have large-angle variations, which is the case for NGC 1333.
For example, if we estimate the standard deviation of randomly distributed angles, the estimated deviation saturates at $\pi/\sqrt{12}~\mathrm{[rad]} =51^\circ\negthinspace.96$ and does not represent the actual angular variation \citep{SERKOWSKI1962289,2010ApJ...716..893P}.

We can utilize directional statistics to avoid this difficulty (see also \citealp{2019ApJ...878...10T}).
In the directional statistics, each angle $\theta_i$ is represented by a unit vector whose phase angle is $\theta_i$.
We can sum the individual vectors and estimate the mean angle as the direction of the resultant vector.
Thus, the circular mean position angle $\bar{\theta}$ is defined as follows:
\begin{eqnarray*}
    \tan \left(2\bar{\theta}\right) &=& \Sigma_{i=1}^n \sin(2\theta_i) / \Sigma_{i=1}^n \cos(2\theta_i)~.
\end{eqnarray*}
Note that we need to take into account the $180^\circ$ degeneracy of the pseudo-vectors instead of the $360^\circ$ degeneracy of normal vectors, and thus we multiply the individual angles by 2.

The mean resultant length of the composite vector $\bar{R}$, which is the length of the resultant vector divided by the number of vectors $n$, is defined as follows:
\begin{eqnarray*}
    \bar{R} &=& n^{-1}\sqrt{\left(\Sigma_{i=1}^n \cos(2\theta_i)\right)^2 + \left(\Sigma_{i=1}^n \sin(2\theta_i)\right)^2}~.
\end{eqnarray*}
If the distribution of angles $\theta_i$ has no deviation ($\theta_i \equiv \mathrm{const.}$ and unit vectors are totally aligned), the mean resultant length $\bar{R} = 1$.
On the other hand, $\bar{R}=0$ if $\theta_i$ are totally random.
Thus, $\bar{R}$ can be used as an indicator of the angle deviation.

When the distribution of $\theta_i$ follows a wrapped normal distribution, which is a normal distribution around the unit circle, $\bar{R}$ is expressed as follows \citep[e.g.,][]{1999Directional}:
\begin{eqnarray*}
    \bar{R} &=& \exp\left(-\frac{(2\nu)^2}{2}\right)~,
\end{eqnarray*}
where $\nu$ is the standard deviation of the normal distribution.
Note that we multiply $\nu$ by 2 in this equation because we estimate the angle variation of $2\times \theta_i$.

The circular standard deviation $\nu$, which is the standard deviation of the position angles, is then defined as follows:
\begin{eqnarray*}
\nu &=& \frac{\sqrt{-2\ln{\bar{R}}}}{2}~.
\end{eqnarray*}
This definition is useful because $\nu$ equals the (normal) standard deviation when the deviation is much smaller than the $180^\circ$ ambiguity of the pseudo-vectors.

\begin{figure}[tp]
\centering
\includegraphics{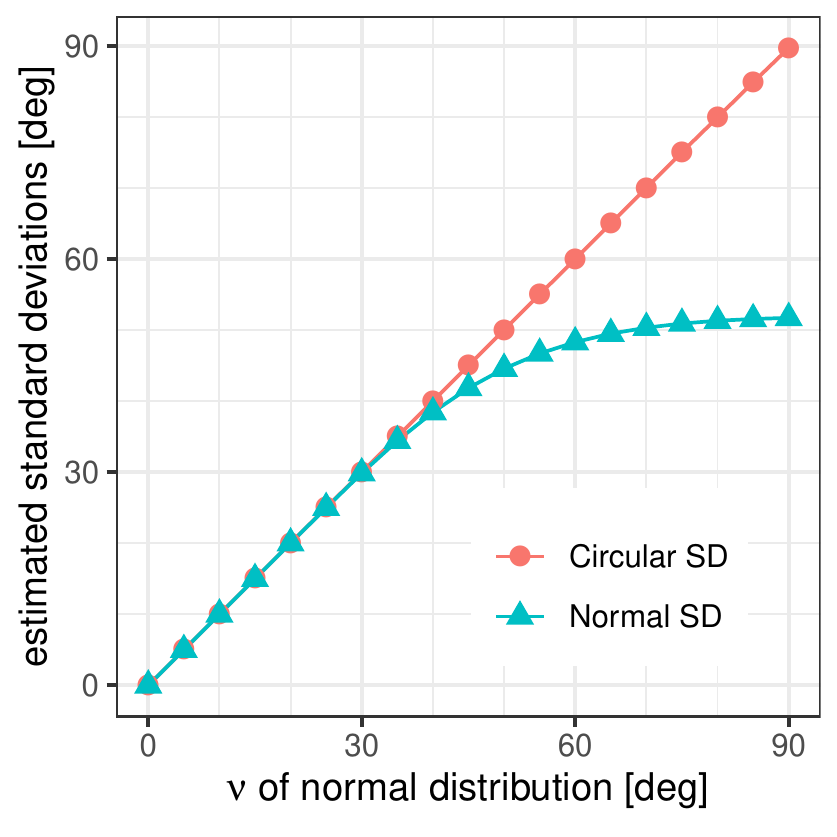}
\caption{Comparison between the standard deviation and the circular standard deviation.
We create the population distribution by randomly sampling angles that have $180^\circ$ ambiguity and estimate both the standard deviation ('Normal SD' in the figure) and the circular standard deviation ('Circular SD' in the figure) of the samples.
The population distribution follows a normal distribution with the standard deviation of $\nu$ and has $180^\circ$ ambiguities (a wrapped normal distribution).}
\label{fig:SD}
\end{figure}

We display a comparison between the normal standard deviation and the circular standard deviation in Figure \ref{fig:SD}.
While the normal standard deviation saturates at $52^\circ$, the circular standard deviation correctly estimates the standard deviation of the population distribution even if the deviation exceeds $50^\circ$.

\section{Identification of Emission Features Using Density-based Clustering}
\label{sec: dbscan}

We identify emission features in the observed region by applying a clustering analysis to a 3D distribution of ISM emission.
The spatial structure of the ISM is well traced by the observed $850 ~ \mu \mathrm{m}$ continuum intensity ($I_{850\mu \mathrm{m}}$; see Figure \ref{fig:B image}).
To separate overlapping ISM features projected on the POS, we utilize LOS velocity information from molecular line emission.

\begin{figure}[tp]
\centering
\includegraphics{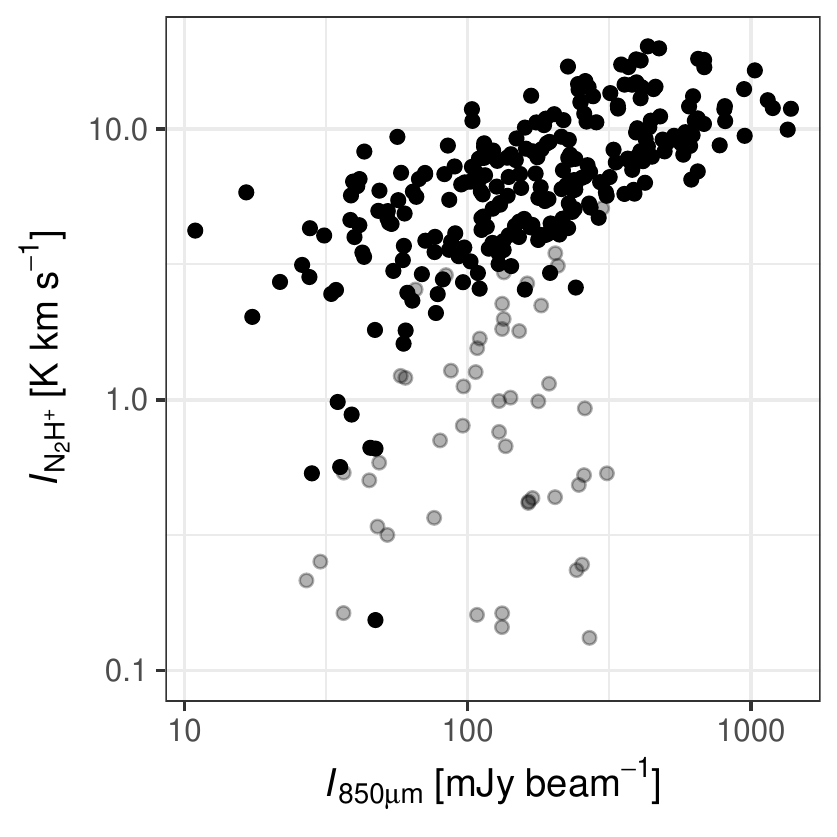}
\caption{Correlation between the $850~\mu\mathrm{m}$ continuum intensity and the integrated intensity of the $\mathrm{N_2H^+}$ line \citep{2017A&A...606A.123H}.
Gray filled circles are the data points around the two B-type stars BD $+30^\circ 459$ and SVS 3 ($< 2'$ from one of the two).
Black filled circles are the data points from other regions.}
\label{fig:N2Hp corr}
\end{figure}

\citet{2017A&A...606A.123H} estimated the 3D structure of the ISM in the NGC 1333 region by using $\mathrm{N_2H^+}$ molecular line data.
The data have a spatial resolution of $30''$ and a spectral resolution of 0.08 km s$^{-1}$.
We show the correlation of $I_{850\mu \mathrm{m}}$ and an integrated intensity of $\mathrm{N_2H^+}~(I_\mathrm{N_2H^+})$ in Figure \ref{fig:N2Hp corr}.
Note that we smooth $I_{850\mu \mathrm{m}}$ to the spatial resolution of $I_\mathrm{N_2H^+}$ $(30\arcsec)$, and show the data where we estimate the \emph{B}-field position angles $(PI / \delta PI \geqslant 3)$ in this figure.

As seen in Figure \ref{fig:N2Hp corr}, most of the data points show tight correlation between $I_{850\mu \mathrm{m}}$ and $I_\mathrm{N_2H^+}$ (black filled circles in the figure).
On the other hand, there are some data points that are out of the correlation (gray filled circles in the figure).
The deviating points are the data in the reflection nebula around two B-type stars BD $+30^\circ459$ and SVS 3 (\citealp{1990Ap&SS.166..315C}; \citealp{2008AJ....135.2496C}; see Figure \ref{fig:I image}), where $\mathrm{N_2H^+}$ ions are dissociated and thus no significant emission is detected.

Thus, we conclude that the $\mathrm{N_2H^+}$ emission is a good tracer of the ISM also traced by $I_{850\mu \mathrm{m}}$ except for the region in the reflection nebula, and use $I_\mathrm{N_2H^+}$ as a tracer of the LOS distribution of the ISM as was done by \citet{2017A&A...606A.123H}.
Note that we evaluate the spatial structure of ISM traced by $I_{850\mu \mathrm{m}}$ with a $14''\negthinspace\negthinspace.1$ resolution, while the spatial resolution of $\mathrm{N_2H^+}$ data is $30''$.
To construct a 3D position-position-velocity (PPV) datacube, we regrid the $\mathrm{N_2H^+}$ data and estimate the line profile at each $I_{850\mu \mathrm{m}}$ data point.
The line profiles are scaled so that the integrated intensities are equal to $I_{850\mu \mathrm{m}}$ at each position.

\begin{figure}[tp]
\centering
\includegraphics{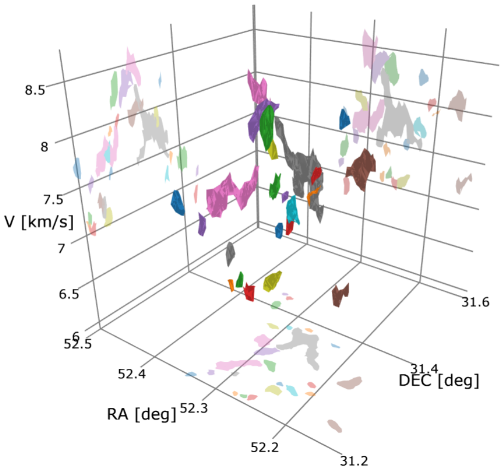}
\caption{ISM structure in NGC 1333 identified with the density-based clustering.
The identified structures are shown in RA-DEC-V$_\mathrm{LSR}$ 3D space with a different color for each structure.
The projection images of these structures onto position-position or position-velocity planes are shown with pale colors.}
\label{fig:alpha3D}
\end{figure}

To identify the 3D structure of the ISM, we apply a statistical cluster analysis on the obtained PPV datacube.
We utilize density-based clustering \citep{DBSCAN,dbcluster} to the data set.
It is a general statistical method that recently has begun to be applied to classify astronomical sources in a multidimensional data space (e.g., \citealp{2018MNRAS.481L..11B,2020MNRAS.491.2205B,2019MNRAS.489.4418J}).
This method can identify clusters in a multidimensional space not only for their crests but also for their spatial extent, without assuming that their underlying spatial structures are filamentary or clumpy.
Thus, this method is suitable to extract ISM structures from our PPV datacube.
We convert the PPV intensity data into discrete values in units of 5 $\mathrm{mJy~beam^{-1}}$, and apply the density-based clustering to this data set.
We show the result of this density-based clustering in Figure
\ref{fig:alpha3D}.
The main ISM structures in the region are successfully extracted.
A drawback of this method based on the LOS velocity is that it cannot identify ISM structures with large velocity dispersion. For example, we cannot identify the structure at and around the IRAS 4 complex, and the cloud that contains IRAS 2 is fragmented.
These misidentifications are attributable to perturbation by active YSOs.
Thus, the identified structures shown in Figure \ref{fig:alpha3D} represent quiescent ISM structures in the region.


\bibliography{paper1}{}

\begin{thebibliography}{}
\expandafter\ifx\csname natexlab\endcsname\relax\def\natexlab#1{#1}\fi
\providecommand{\url}[1]{\href{#1}{#1}}
\providecommand{\dodoi}[1]{doi:~\href{http://doi.org/#1}{\nolinkurl{#1}}}
\providecommand{\doeprint}[1]{\href{http://ascl.net/#1}{\nolinkurl{http://ascl.net/#1}}}
\providecommand{\doarXiv}[1]{\href{https://arxiv.org/abs/#1}{\nolinkurl{https://arxiv.org/abs/#1}}}

\bibitem[{{Abergel} {et~al.}(1994){Abergel}, {Boulanger}, {Mizuno}, \&
  {Fukui}}]{1994ApJ...423L..59A}
{Abergel}, A., {Boulanger}, F., {Mizuno}, A., \& {Fukui}, Y. 1994, \apjl, 423,
  L59, \dodoi{10.1086/187235}

\bibitem[{{Alves} {et~al.}(2011){Alves}, {Acosta-Pulido}, {Girart}, {Franco},
  \& {L{\'o}pez}}]{2011AJ....142...33A}
{Alves}, F.~O., {Acosta-Pulido}, J.~A., {Girart}, J.~M., {Franco}, G.~A.~P., \&
  {L{\'o}pez}, R. 2011, \aj, 142, 33, \dodoi{10.1088/0004-6256/142/1/33}

\bibitem[{{Alves de Oliveira} {et~al.}(2014){Alves de Oliveira}, {Schneider},
  {Mer{\'{\i}}n}, {Prusti}, {Ribas}, {Cox}, {Vavrek}, {K{\"o}nyves},
  {Arzoumanian}, {Puga}, {Pilbratt}, {K{\'o}sp{\'a}l}, {Andr{\'e}}, {Didelon},
  {Men'shchikov}, {Royer}, {Waelkens}, {Bontemps}, {Winston}, \&
  {Spezzi}}]{2014A&A...568A..98A}
{Alves de Oliveira}, C., {Schneider}, N., {Mer{\'{\i}}n}, B., {et~al.} 2014,
  \aap, 568, A98, \dodoi{10.1051/0004-6361/201423504}

\bibitem[{{Andr{\'e}} {et~al.}(2014){Andr{\'e}}, {Di Francesco},
  {Ward-Thompson}, {Inutsuka}, {Pudritz}, \& {Pineda}}]{2014prpl.conf...27A}
{Andr{\'e}}, P., {Di Francesco}, J., {Ward-Thompson}, D., {et~al.} 2014,
  Protostars and Planets VI, 27,
  \dodoi{10.2458/azu_uapress_9780816531240-ch002}

\bibitem[{{Andr{\'e}} {et~al.}(2010){Andr{\'e}}, {Men'shchikov}, {Bontemps},
  {K{\"o}nyves}, {Motte}, {Schneider}, {Didelon}, {Minier}, {Saraceno},
  {Ward-Thompson}, {di Francesco}, {White}, {Molinari}, {Testi}, {Abergel},
  {Griffin}, {Henning}, {Royer}, {Mer{\'{\i}}n}, {Vavrek}, {Attard},
  {Arzoumanian}, {Wilson}, {Ade}, {Aussel}, {Baluteau}, {Benedettini},
  {Bernard}, {Blommaert}, {Cambr{\'e}sy}, {Cox}, {di Giorgio}, {Hargrave},
  {Hennemann}, {Huang}, {Kirk}, {Krause}, {Launhardt}, {Leeks}, {Le Pennec},
  {Li}, {Martin}, {Maury}, {Olofsson}, {Omont}, {Peretto}, {Pezzuto}, {Prusti},
  {Roussel}, {Russeil}, {Sauvage}, {Sibthorpe}, {Sicilia-Aguilar}, {Spinoglio},
  {Waelkens}, {Woodcraft}, \& {Zavagno}}]{2010A&A...518L.102A}
{Andr{\'e}}, P., {Men'shchikov}, A., {Bontemps}, S., {et~al.} 2010, \aap, 518,
  L102, \dodoi{10.1051/0004-6361/201014666}

\bibitem[{{Andr{\'e}} {et~al.}(2019){Andr{\'e}}, {Hughes}, {Guillet},
  {Boulanger}, {Bracco}, {Ntormousi}, {Arzoumanian}, {Maury}, {Bernard},
  {Bontemps}, {Ristorcelli}, {Girart}, {Motte}, {Tassis}, {Pantin},
  {Montmerle}, {Johnstone}, {Gabici}, {Efstathiou}, {Basu}, {B{\'e}thermin},
  {Beuther}, {Braine}, {Francesco}, {Falgarone}, {Ferri{\`e}re}, {Fletcher},
  {Galametz}, {Giard}, {Hennebelle}, {Jones}, {Kepley}, {Kwon}, {Lagache},
  {Lesaffre}, {Levrier}, {Li}, {Li}, {Mao}, {Nakagawa}, {Onaka}, {Paladino},
  {Peretto}, {Poglitsch}, {Rev{\'e}ret}, {Rodriguez}, {Sauvage}, {Soler},
  {Spinoglio}, {Tabatabaei}, {Tritsis}, {van der Tak}, {Ward-Thompson},
  {Wiesemeyer}, {Ysard}, \& {Zhang}}]{2019PASA...36...29A}
{Andr{\'e}}, P., {Hughes}, A., {Guillet}, V., {et~al.} 2019, \pasa, 36, e029,
  \dodoi{10.1017/pasa.2019.20}

\bibitem[{{Arce} {et~al.}(2010){Arce}, {Borkin}, {Goodman}, {Pineda}, \&
  {Halle}}]{2010ApJ...715.1170A}
{Arce}, H.~G., {Borkin}, M.~A., {Goodman}, A.~A., {Pineda}, J.~E., \& {Halle},
  M.~W. 2010, \apj, 715, 1170, \dodoi{10.1088/0004-637X/715/2/1170}

\bibitem[{{Arnold} {et~al.}(2012){Arnold}, {Watson}, {Kim}, {Manoj}, {Remming},
  {Sheehan}, {Adame}, {Forrest}, {Furlan}, {Mamajek}, {McClure}, {Espaillat},
  {Ausfeld}, \& {Rapson}}]{2012ApJS..201...12A}
{Arnold}, L.~A., {Watson}, D.~M., {Kim}, K.~H., {et~al.} 2012, \apjs, 201, 12,
  \dodoi{10.1088/0067-0049/201/2/12}

\bibitem[{{Arzoumanian} {et~al.}(2011){Arzoumanian}, {Andr{\'e}}, {Didelon},
  {K{\"o}nyves}, {Schneider}, {Men'shchikov}, {Sousbie}, {Zavagno}, {Bontemps},
  {di Francesco}, {Griffin}, {Hennemann}, {Hill}, {Kirk}, {Martin}, {Minier},
  {Molinari}, {Motte}, {Peretto}, {Pezzuto}, {Spinoglio}, {Ward-Thompson},
  {White}, \& {Wilson}}]{2011A&A...529L...6A}
{Arzoumanian}, D., {Andr{\'e}}, P., {Didelon}, P., {et~al.} 2011, \aap, 529,
  L6, \dodoi{10.1051/0004-6361/201116596}

\bibitem[{{Arzoumanian} {et~al.}(2019){Arzoumanian}, {Andr{\'e}},
  {K{\"o}nyves}, {Palmeirim}, {Roy}, {Schneider}, {Benedettini}, {Didelon}, {Di
  Francesco}, {Kirk}, \& {Ladjelate}}]{2019A&A...621A..42A}
{Arzoumanian}, D., {Andr{\'e}}, P., {K{\"o}nyves}, V., {et~al.} 2019, \aap,
  621, A42, \dodoi{10.1051/0004-6361/201832725}

\bibitem[{{Astropy Collaboration} {et~al.}(2013){Astropy Collaboration},
  {Robitaille}, {Tollerud}, {Greenfield}, {Droettboom}, {Bray}, {Aldcroft},
  {Davis}, {Ginsburg}, {Price-Whelan}, {Kerzendorf}, {Conley}, {Crighton},
  {Barbary}, {Muna}, {Ferguson}, {Grollier}, {Parikh}, {Nair}, {Unther},
  {Deil}, {Woillez}, {Conseil}, {Kramer}, {Turner}, {Singer}, {Fox}, {Weaver},
  {Zabalza}, {Edwards}, {Azalee Bostroem}, {Burke}, {Casey}, {Crawford},
  {Dencheva}, {Ely}, {Jenness}, {Labrie}, {Lim}, {Pierfederici}, {Pontzen},
  {Ptak}, {Refsdal}, {Servillat}, \& {Streicher}}]{2013A&A...558A..33A}
{Astropy Collaboration}, {Robitaille}, T.~P., {Tollerud}, E.~J., {et~al.} 2013,
  \aap, 558, A33, \dodoi{10.1051/0004-6361/201322068}

\bibitem[{{Attard} {et~al.}(2009){Attard}, {Houde}, {Novak}, {Li},
  {Vaillancourt}, {Dowell}, {Davidson}, \& {Shinnaga}}]{2009ApJ...702.1584A}
{Attard}, M., {Houde}, M., {Novak}, G., {et~al.} 2009, \apj, 702, 1584,
  \dodoi{10.1088/0004-637X/702/2/1584}

\bibitem[{{Bally}(2016)}]{2016ARA&A..54..491B}
{Bally}, J. 2016, \araa, 54, 491, \dodoi{10.1146/annurev-astro-081915-023341}

\bibitem[{{Bally} {et~al.}(1996){Bally}, {Devine}, \&
  {Reipurth}}]{1996ApJ...473L..49B}
{Bally}, J., {Devine}, D., \& {Reipurth}, B. 1996, \apjl, 473, L49,
  \dodoi{10.1086/310381}

\bibitem[{{Bally} {et~al.}(2008){Bally}, {Walawender}, {Johnstone}, {Kirk}, \&
  {Goodman}}]{2008hsf1.book..308B}
{Bally}, J., {Walawender}, J., {Johnstone}, D., {Kirk}, H., \& {Goodman}, A.
  2008, {The Perseus Cloud}, ed. B.~{Reipurth}, 308

\bibitem[{{Bastien} {et~al.}(2011){Bastien}, {Bissonnette}, {Simon},
  {Coud{\'e}}, {Ade}, {Savini}, {Pisano}, {Leclerc}, {Bernier}, {Landry},
  {Houde}, {Hezareh}, {Naylor}, {Gom}, {Jenness}, {Berry}, {Johnstone}, \&
  {Matthews}}]{2011ASPC..449...68B}
{Bastien}, P., {Bissonnette}, E., {Simon}, A., {et~al.} 2011, in Astronomical
  Society of the Pacific Conference Series, Vol. 449, Astronomical Polarimetry
  2008: Science from Small to Large Telescopes, ed. P.~{Bastien}, N.~{Manset},
  D.~P. {Clemens}, \& N.~{St-Louis}, 68

\bibitem[{{Baug} {et~al.}(2020){Baug}, {Wang}, {Liu}, {Tang}, {Zhang}, {Li},
  {Eswaraiah}, {Liu}, {Tej}, {Goldsmith}, {Bronfman}, {Qin}, {T{\'o}th}, {Li},
  \& {Kim}}]{2020ApJ...890...44B}
{Baug}, T., {Wang}, K., {Liu}, T., {et~al.} 2020, \apj, 890, 44,
  \dodoi{10.3847/1538-4357/ab66b6}

\bibitem[{{Beccari} {et~al.}(2020){Beccari}, {Boffin}, \&
  {Jerabkova}}]{2020MNRAS.491.2205B}
{Beccari}, G., {Boffin}, H. M.~J., \& {Jerabkova}, T. 2020, \mnras, 491, 2205,
  \dodoi{10.1093/mnras/stz3195}

\bibitem[{{Beccari} {et~al.}(2018){Beccari}, {Boffin}, {Jerabkova}, {Wright},
  {Kalari}, {Carraro}, {De Marchi}, \& {de Wit}}]{2018MNRAS.481L..11B}
{Beccari}, G., {Boffin}, H. M.~J., {Jerabkova}, T., {et~al.} 2018, \mnras, 481,
  L11, \dodoi{10.1093/mnrasl/sly144}

\bibitem[{{Bock} {et~al.}(2006){Bock}, {Bolatto}, {Hawkins}, {Kemball}, {Lamb},
  {Plambeck}, {Pound}, {Scott}, {Woody}, \& {Wright}}]{2006SPIE.6267E..13B}
{Bock}, D.~C.-J., {Bolatto}, A.~D., {Hawkins}, D.~W., {et~al.} 2006, in
  \procspie, Vol. 6267, Society of Photo-Optical Instrumentation Engineers
  (SPIE) Conference Series, 626713, \dodoi{10.1117/12.674051}

\bibitem[{{Cernis}(1990)}]{1990Ap&SS.166..315C}
{Cernis}, K. 1990, \apss, 166, 315, \dodoi{10.1007/BF01094902}

\bibitem[{{Chapin} {et~al.}(2013){Chapin}, {Berry}, {Gibb}, {Jenness}, {Scott},
  {Tilanus}, {Economou}, \& {Holland}}]{2013MNRAS.430.2545C}
{Chapin}, E.~L., {Berry}, D.~S., {Gibb}, A.~G., {et~al.} 2013, \mnras, 430,
  2545, \dodoi{10.1093/mnras/stt052}

\bibitem[{{Chrysostomou} {et~al.}(2004){Chrysostomou}, {Curran}, \&
  {Aitken}}]{2004Ap&SS.292..509C}
{Chrysostomou}, A., {Curran}, R., \& {Aitken}, D. 2004, \apss, 292, 509,
  \dodoi{10.1023/B:ASTR.0000045056.55646.6a}

\bibitem[{{Chuss} {et~al.}(2019){Chuss}, {Andersson}, {Bally}, {Dotson},
  {Dowell}, {Guerra}, {Harper}, {Houde}, {Jones}, {Lazarian}, {Lopez
  Rodriguez}, {Michail}, {Morris}, {Novak}, {Siah}, {Staguhn}, {Vaillancourt},
  {Volpert}, {Werner}, {Wollack}, {Benford}, {Berthoud}, {Cox}, {Crutcher},
  {Dale}, {Fissel}, {Goldsmith}, {Hamilton}, {Hanany}, {Henning}, {Looney},
  {Moseley}, {Santos}, {Stephens}, {Tassis}, {Trinh}, {Van Camp},
  {Ward-Thompson}, \& {HAWC + Science Team}}]{2019ApJ...872..187C}
{Chuss}, D.~T., {Andersson}, B.~G., {Bally}, J., {et~al.} 2019, \apj, 872, 187,
  \dodoi{10.3847/1538-4357/aafd37}

\bibitem[{{Col{\'{\i}}n} {et~al.}(2013){Col{\'{\i}}n}, {V{\'a}zquez-Semadeni},
  \& {G{\'o}mez}}]{2013MNRAS.435.1701C}
{Col{\'{\i}}n}, P., {V{\'a}zquez-Semadeni}, E., \& {G{\'o}mez}, G.~C. 2013,
  \mnras, 435, 1701, \dodoi{10.1093/mnras/stt1409}

\bibitem[{{Connelley} {et~al.}(2008){Connelley}, {Reipurth}, \&
  {Tokunaga}}]{2008AJ....135.2496C}
{Connelley}, M.~S., {Reipurth}, B., \& {Tokunaga}, A.~T. 2008, \aj, 135, 2496,
  \dodoi{10.1088/0004-6256/135/6/2496}

\bibitem[{{Coud{\'e}} {et~al.}(2019){Coud{\'e}}, {Bastien}, {Houde}, {Sadavoy},
  {Friesen}, {Di Francesco}, {Johnstone}, {Mairs}, {Hasegawa}, {Kwon}, {Lai},
  {Qiu}, {Ward-Thompson}, {Berry}, {Chen}, {Fiege}, {Franzmann}, {Hatchell},
  {Lacaille}, {Matthews}, {Moriarty-Schieven}, {Pon}, {Andr{\'e}},
  {Arzoumanian}, {Aso}, {Byun}, {Eswaraiah}, {Chen}, {Chen}, {Ching}, {Cho},
  {Choi}, {Chrysostomou}, {Chung}, {Doi}, {Drabek-Maunder}, {Dowell}, {Eyres},
  {Falle}, {Friberg}, {Fuller}, {Furuya}, {Gledhill}, {Graves}, {Greaves},
  {Griffin}, {Gu}, {Hayashi}, {Hoang}, {Holland}, {Inoue}, {Inutsuka},
  {Iwasaki}, {Jeong}, {Kanamori}, {Kataoka}, {Kang}, {Kang}, {Kang},
  {Kawabata}, {Kemper}, {Kim}, {Kim}, {Kim}, {Kim}, {Kim}, {Kim}, {Kirk},
  {Kobayashi}, {Koch}, {Kwon}, {Lee}, {Lee}, {Lee}, {Li}, {Li}, {Li}, {Liu},
  {Liu}, {Liu}, {Liu}, {van Loo}, {Lyo}, {Matsumura}, {Nagata}, {Nakamura},
  {Nakanishi}, {Ohashi}, {Onaka}, {Parsons}, {Pattle}, {Peretto}, {Pyo},
  {Qian}, {Rao}, {Rawlings}, {Retter}, {Richer}, {Rigby}, {Robitaille},
  {Saito}, {Savini}, {Scaife}, {Seta}, {Shinnaga}, {Soam}, {Tamura}, {Tang},
  {Tomisaka}, {Tsukamoto}, {Wang}, {Wang}, {Whitworth}, {Yen}, {Yoo}, {Yuan},
  {Zenko}, {Zhang}, {Zhang}, {Zhou}, {Zhu}, \& {The B-fields In STar-forming
  Regions Observations (BISTRO Collaboration}}]{2019ApJ...877...88C}
{Coud{\'e}}, S., {Bastien}, P., {Houde}, M., {et~al.} 2019, \apj, 877, 88,
  \dodoi{10.3847/1538-4357/ab1b23}

\bibitem[{{Crutcher}(2012)}]{2012ARA&A..50...29C}
{Crutcher}, R.~M. 2012, \araa, 50, 29,
  \dodoi{10.1146/annurev-astro-081811-125514}

\bibitem[{{Curran} \& {Chrysostomou}(2007)}]{2007MNRAS.382..699C}
{Curran}, R.~L., \& {Chrysostomou}, A. 2007, \mnras, 382, 699,
  \dodoi{10.1111/j.1365-2966.2007.12399.x}

\bibitem[{{Currie} {et~al.}(2014){Currie}, {Berry}, {Jenness}, {Gibb}, {Bell},
  \& {Draper}}]{2014ASPC..485..391C}
{Currie}, M.~J., {Berry}, D.~S., {Jenness}, T., {et~al.} 2014, in Astronomical
  Society of the Pacific Conference Series, Vol. 485, Astronomical Data
  Analysis Software and Systems XXIII, ed. N.~{Manset} \& P.~{Forshay}, 391

\bibitem[{{Curtis} {et~al.}(2010){Curtis}, {Richer}, {Swift}, \&
  {Williams}}]{2010MNRAS.408.1516C}
{Curtis}, E.~I., {Richer}, J.~S., {Swift}, J.~J., \& {Williams}, J.~P. 2010,
  \mnras, 408, 1516, \dodoi{10.1111/j.1365-2966.2010.17214.x}

\bibitem[{{Dempsey} {et~al.}(2013){Dempsey}, {Friberg}, {Jenness}, {Tilanus},
  {Thomas}, {Holland}, {Bintley}, {Berry}, {Chapin}, {Chrysostomou}, {Davis},
  {Gibb}, {Parsons}, \& {Robson}}]{2013MNRAS.430.2534D}
{Dempsey}, J.~T., {Friberg}, P., {Jenness}, T., {et~al.} 2013, \mnras, 430,
  2534, \dodoi{10.1093/mnras/stt090}

\bibitem[{{Doi} {et~al.}(2015){Doi}, {Takita}, {Ootsubo}, {Arimatsu}, {Tanaka},
  {Kitamura}, {Kawada}, {Matsuura}, {Nakagawa}, {Morishima}, {Hattori},
  {Komugi}, {White}, {Ikeda}, {Kato}, {Chinone}, {Etxaluze}, \&
  {Cypriano}}]{2015PASJ...67...50D}
{Doi}, Y., {Takita}, S., {Ootsubo}, T., {et~al.} 2015, \pasj, 67, 50,
  \dodoi{10.1093/pasj/psv022}

\bibitem[{{Draine} \& {Weingartner}(1996)}]{1996ApJ...470..551D}
{Draine}, B.~T., \& {Weingartner}, J.~C. 1996, \apj, 470, 551,
  \dodoi{10.1086/177887}

\bibitem[{{Draine} \& {Weingartner}(1997)}]{1997ApJ...480..633D}
---. 1997, \apj, 480, 633, \dodoi{10.1086/304008}

\bibitem[{{Enoch} {et~al.}(2006){Enoch}, {Young}, {Glenn}, {Evans}, {Golwala},
  {Sargent}, {Harvey}, {Aguirre}, {Goldin}, {Haig}, {Huard}, {Lange},
  {Laurent}, {Maloney}, {Mauskopf}, {Rossinot}, \&
  {Sayers}}]{2006ApJ...638..293E}
{Enoch}, M.~L., {Young}, K.~E., {Glenn}, J., {et~al.} 2006, \apj, 638, 293,
  \dodoi{10.1086/498678}

\bibitem[{{Ester} {et~al.}(1996){Ester}, {Kriegel}, {Sander}, \& {Xu}}]{DBSCAN}
{Ester}, M., {Kriegel}, H.-P., {Sander}, J., \& {Xu}, X. 1996, in Proceedings
  of the Second International Conference on Knowledge Discovery and Data Mining
  (KDD-96), ed. E.~{Simoudis}, J.~{Han}, \& U.~M. {Fayyad}, AAAI Press,
  226--231

\bibitem[{{Federrath}(2016)}]{2016MNRAS.457..375F}
{Federrath}, C. 2016, \mnras, 457, 375, \dodoi{10.1093/mnras/stv2880}

\bibitem[{{Fissel} {et~al.}(2019){Fissel}, {Ade}, {Angil{\`e}}, {Ashton},
  {Benton}, {Chen}, {Cunningham}, {Devlin}, {Dober}, {Friesen}, {Fukui},
  {Galitzki}, {Gandilo}, {Goodman}, {Green}, {Jones}, {Klein}, {King},
  {Korotkov}, {Li}, {Lowe}, {Martin}, {Matthews}, {Moncelsi}, {Nakamura},
  {Netterfield}, {Newmark}, {Novak}, {Pascale}, {Poidevin}, {Santos}, {Savini},
  {Scott}, {Shariff}, {Soler}, {Thomas}, {Tucker}, {Tucker}, {Ward-Thompson},
  \& {Zucker}}]{2019ApJ...878..110F}
{Fissel}, L.~M., {Ade}, P. A.~R., {Angil{\`e}}, F.~E., {et~al.} 2019, \apj,
  878, 110, \dodoi{10.3847/1538-4357/ab1eb0}

\bibitem[{{Friberg} {et~al.}(2016){Friberg}, {Bastien}, {Berry}, {Savini},
  {Graves}, \& {Pattle}}]{2016SPIE.9914E..03F}
{Friberg}, P., {Bastien}, P., {Berry}, D., {et~al.} 2016, in \procspie, Vol.
  9914, Millimeter, Submillimeter, and Far-Infrared Detectors and
  Instrumentation for Astronomy VIII, 991403, \dodoi{10.1117/12.2231943}

\bibitem[{{Galametz} {et~al.}(2018){Galametz}, {Maury}, {Girart}, {Rao},
  {Zhang}, {Gaudel}, {Valdivia}, {Keto}, \& {Lai}}]{2018A&A...616A.139G}
{Galametz}, M., {Maury}, A., {Girart}, J.~M., {et~al.} 2018, \aap, 616, A139,
  \dodoi{10.1051/0004-6361/201833004}

\bibitem[{{Girart} {et~al.}(1999){Girart}, {Crutcher}, \&
  {Rao}}]{1999ApJ...525L.109G}
{Girart}, J.~M., {Crutcher}, R.~M., \& {Rao}, R. 1999, \apjl, 525, L109,
  \dodoi{10.1086/312345}

\bibitem[{{Girart} {et~al.}(2006){Girart}, {Rao}, \&
  {Marrone}}]{2006Sci...313..812G}
{Girart}, J.~M., {Rao}, R., \& {Marrone}, D.~P. 2006, Science, 313, 812,
  \dodoi{10.1126/science.1129093}

\bibitem[{{Goldsmith} {et~al.}(2008){Goldsmith}, {Heyer}, {Narayanan}, {Snell},
  {Li}, \& {Brunt}}]{2008ApJ...680..428G}
{Goldsmith}, P.~F., {Heyer}, M., {Narayanan}, G., {et~al.} 2008, \apj, 680,
  428, \dodoi{10.1086/587166}

\bibitem[{{Goodman} {et~al.}(1990){Goodman}, {Bastien}, {Myers}, \&
  {M{\'e}nard}}]{1990ApJ...359..363G}
{Goodman}, A.~A., {Bastien}, P., {Myers}, P.~C., \& {M{\'e}nard}, F. 1990,
  \apj, 359, 363, \dodoi{10.1086/169070}

\bibitem[{{Hacar} {et~al.}(2017){Hacar}, {Tafalla}, \&
  {Alves}}]{2017A&A...606A.123H}
{Hacar}, A., {Tafalla}, M., \& {Alves}, J. 2017, \aap, 606, A123,
  \dodoi{10.1051/0004-6361/201630348}

\bibitem[{Hahsler {et~al.}(2019)Hahsler, Piekenbrock, \& Doran}]{JSSv091i01}
Hahsler, M., Piekenbrock, M., \& Doran, D. 2019, Journal of Statistical
  Software, Articles, 91, 1, \dodoi{10.18637/jss.v091.i01}

\bibitem[{{Hatchell} \& {Dunham}(2009)}]{2009A&A...502..139H}
{Hatchell}, J., \& {Dunham}, M.~M. 2009, \aap, 502, 139,
  \dodoi{10.1051/0004-6361/200911818}

\bibitem[{{Hatchell} {et~al.}(2013){Hatchell}, {Wilson}, {Drabek}, {Curtis},
  {Richer}, {Nutter}, {Di Francesco}, {Ward-Thompson}, \& {JCMT GBS
  Consortium}}]{2013MNRAS.429L..10H}
{Hatchell}, J., {Wilson}, T., {Drabek}, E., {et~al.} 2013, \mnras, 429, L10,
  \dodoi{10.1093/mnrasl/sls015}

\bibitem[{{Heiles} {et~al.}(1993){Heiles}, {Goodman}, {McKee}, \&
  {Zweibel}}]{1993prpl.conf..279H}
{Heiles}, C., {Goodman}, A.~A., {McKee}, C.~F., \& {Zweibel}, E.~G. 1993, in
  Protostars and Planets III, ed. E.~H. {Levy} \& J.~I. {Lunine}, 279--326

\bibitem[{{Hennebelle}(2013)}]{2013A&A...556A.153H}
{Hennebelle}, P. 2013, \aap, 556, A153, \dodoi{10.1051/0004-6361/201321292}

\bibitem[{{Hennebelle} \& {Falgarone}(2012)}]{2012A&ARv..20...55H}
{Hennebelle}, P., \& {Falgarone}, E. 2012, \aapr, 20, 55,
  \dodoi{10.1007/s00159-012-0055-y}

\bibitem[{{Hennebelle} \& {Inutsuka}(2019)}]{2019FrASS...6....5H}
{Hennebelle}, P., \& {Inutsuka}, S.-i. 2019, Frontiers in Astronomy and Space
  Sciences, 6, 5, \dodoi{10.3389/fspas.2019.00005}

\bibitem[{{Hildebrand}(1988)}]{1988QJRAS..29..327H}
{Hildebrand}, R.~H. 1988, \qjras, 29, 327

\bibitem[{{Hoang} \& {Lazarian}(2008)}]{2008MNRAS.388..117H}
{Hoang}, T., \& {Lazarian}, A. 2008, \mnras, 388, 117,
  \dodoi{10.1111/j.1365-2966.2008.13249.x}

\bibitem[{{Hoang} \& {Lazarian}(2014)}]{2014MNRAS.438..680H}
---. 2014, \mnras, 438, 680, \dodoi{10.1093/mnras/stt2240}

\bibitem[{{Hoang} \& {Lazarian}(2016)}]{2016ApJ...831..159H}
---. 2016, \apj, 831, 159, \dodoi{10.3847/0004-637X/831/2/159}

\bibitem[{{Holland} {et~al.}(2013){Holland}, {Bintley}, {Chapin},
  {Chrysostomou}, {Davis}, {Dempsey}, {Duncan}, {Fich}, {Friberg}, {Halpern},
  {Irwin}, {Jenness}, {Kelly}, {MacIntosh}, {Robson}, {Scott}, {Ade},
  {Atad-Ettedgui}, {Berry}, {Craig}, {Gao}, {Gibb}, {Hilton}, {Hollister},
  {Kycia}, {Lunney}, {McGregor}, {Montgomery}, {Parkes}, {Tilanus}, {Ullom},
  {Walther}, {Walton}, {Woodcraft}, {Amiri}, {Atkinson}, {Burger}, {Chuter},
  {Coulson}, {Doriese}, {Dunare}, {Economou}, {Niemack}, {Parsons},
  {Reintsema}, {Sibthorpe}, {Smail}, {Sudiwala}, \&
  {Thomas}}]{2013MNRAS.430.2513H}
{Holland}, W.~S., {Bintley}, D., {Chapin}, E.~L., {et~al.} 2013, \mnras, 430,
  2513, \dodoi{10.1093/mnras/sts612}

\bibitem[{{Hull} \& {Zhang}(2019)}]{2019FrASS...6....3H}
{Hull}, C. L.~H., \& {Zhang}, Q. 2019, Frontiers in Astronomy and Space
  Sciences, 6, 3, \dodoi{10.3389/fspas.2019.00003}

\bibitem[{{Hull} {et~al.}(2013){Hull}, {Plambeck}, {Bolatto}, {Bower},
  {Carpenter}, {Crutcher}, {Fiege}, {Franzmann}, {Hakobian}, {Heiles}, {Houde},
  {Hughes}, {Jameson}, {Kwon}, {Lamb}, {Looney}, {Matthews}, {Mundy}, {Pillai},
  {Pound}, {Stephens}, {Tobin}, {Vaillancourt}, {Volgenau}, \&
  {Wright}}]{2013ApJ...768..159H}
{Hull}, C.~L.~H., {Plambeck}, R.~L., {Bolatto}, A.~D., {et~al.} 2013, \apj,
  768, 159, \dodoi{10.1088/0004-637X/768/2/159}

\bibitem[{{Hull} {et~al.}(2014){Hull}, {Plambeck}, {Kwon}, {Bower},
  {Carpenter}, {Crutcher}, {Fiege}, {Franzmann}, {Hakobian}, {Heiles}, {Houde},
  {Hughes}, {Lamb}, {Looney}, {Marrone}, {Matthews}, {Pillai}, {Pound},
  {Rahman}, {Sandell}, {Stephens}, {Tobin}, {Vaillancourt}, {Volgenau}, \&
  {Wright}}]{2014ApJS..213...13H}
{Hull}, C.~L.~H., {Plambeck}, R.~L., {Kwon}, W., {et~al.} 2014, \apjs, 213, 13,
  \dodoi{10.1088/0067-0049/213/1/13}

\bibitem[{{Inoue} {et~al.}(2018){Inoue}, {Hennebelle}, {Fukui}, {Matsumoto},
  {Iwasaki}, \& {Inutsuka}}]{2018PASJ...70S..53I}
{Inoue}, T., {Hennebelle}, P., {Fukui}, Y., {et~al.} 2018, \pasj, 70, S53,
  \dodoi{10.1093/pasj/psx089}

\bibitem[{{Inutsuka} {et~al.}(2015){Inutsuka}, {Inoue}, {Iwasaki}, \&
  {Hosokawa}}]{2015A&A...580A..49I}
{Inutsuka}, S.-i., {Inoue}, T., {Iwasaki}, K., \& {Hosokawa}, T. 2015, \aap,
  580, A49, \dodoi{10.1051/0004-6361/201425584}

\bibitem[{{Inutsuka} \& {Miyama}(1997)}]{1997ApJ...480..681I}
{Inutsuka}, S.-i., \& {Miyama}, S.~M. 1997, \apj, 480, 681,
  \dodoi{10.1086/303982}

\bibitem[{{Jenness} {et~al.}(2002){Jenness}, {Stevens}, {Archibald},
  {Economou}, {Jessop}, \& {Robson}}]{2002MNRAS.336...14J}
{Jenness}, T., {Stevens}, J.~A., {Archibald}, E.~N., {et~al.} 2002, \mnras,
  336, 14, \dodoi{10.1046/j.1365-8711.2002.05604.x}

\bibitem[{{Jerabkova} {et~al.}(2019){Jerabkova}, {Boffin}, {Beccari}, \&
  {Anderson}}]{2019MNRAS.489.4418J}
{Jerabkova}, T., {Boffin}, H. M.~J., {Beccari}, G., \& {Anderson}, R.~I. 2019,
  \mnras, 489, 4418, \dodoi{10.1093/mnras/stz2315}

\bibitem[{{J{\o}rgensen} {et~al.}(2008){J{\o}rgensen}, {Johnstone}, {Kirk},
  {Myers}, {Allen}, \& {Shirley}}]{2008ApJ...683..822J}
{J{\o}rgensen}, J.~K., {Johnstone}, D., {Kirk}, H., {et~al.} 2008, \apj, 683,
  822, \dodoi{10.1086/589956}

\bibitem[{{Juvela} {et~al.}(2012){Juvela}, {Pelkonen}, {White}, {K{\"o}nyves},
  {Kirk}, \& {Andr{\'e}}}]{2012A&A...544A..14J}
{Juvela}, M., {Pelkonen}, V.-M., {White}, G.~J., {et~al.} 2012, \aap, 544, A14,
  \dodoi{10.1051/0004-6361/201219084}

\bibitem[{{Klassen} {et~al.}(2017){Klassen}, {Pudritz}, \&
  {Kirk}}]{2017MNRAS.465.2254K}
{Klassen}, M., {Pudritz}, R.~E., \& {Kirk}, H. 2017, \mnras, 465, 2254,
  \dodoi{10.1093/mnras/stw2889}

\bibitem[{{Knee} \& {Sandell}(2000)}]{2000A&A...361..671K}
{Knee}, L.~B.~G., \& {Sandell}, G. 2000, \aap, 361, 671

\bibitem[{{Koch} \& {Rosolowsky}(2015)}]{2015MNRAS.452.3435K}
{Koch}, E.~W., \& {Rosolowsky}, E.~W. 2015, \mnras, 452, 3435,
  \dodoi{10.1093/mnras/stv1521}

\bibitem[{{K{\"o}nyves} {et~al.}(2010){K{\"o}nyves}, {Andr{\'e}},
  {Men'shchikov}, {Schneider}, {Arzoumanian}, {Bontemps}, {Attard}, {Motte},
  {Didelon}, {Maury}, {Abergel}, {Ali}, {Baluteau}, {Bernard}, {Cambr{\'e}sy},
  {Cox}, {di Francesco}, {di Giorgio}, {Griffin}, {Hargrave}, {Huang}, {Kirk},
  {Li}, {Martin}, {Minier}, {Molinari}, {Olofsson}, {Pezzuto}, {Russeil},
  {Roussel}, {Saraceno}, {Sauvage}, {Sibthorpe}, {Spinoglio}, {Testi},
  {Ward-Thompson}, {White}, {Wilson}, {Woodcraft}, \&
  {Zavagno}}]{2010A&A...518L.106K}
{K{\"o}nyves}, V., {Andr{\'e}}, P., {Men'shchikov}, A., {et~al.} 2010, \aap,
  518, L106, \dodoi{10.1051/0004-6361/201014689}

\bibitem[{{K{\"o}nyves} {et~al.}(2015){K{\"o}nyves}, {Andr{\'e}},
  {Men'shchikov}, {Palmeirim}, {Arzoumanian}, {Schneider}, {Roy}, {Didelon},
  {Maury}, {Shimajiri}, {Di Francesco}, {Bontemps}, {Peretto}, {Benedettini},
  {Bernard}, {Elia}, {Griffin}, {Hill}, {Kirk}, {Ladjelate}, {Marsh}, {Martin},
  {Motte}, {Nguy{\^e}n Luong}, {Pezzuto}, {Roussel}, {Rygl}, {Sadavoy},
  {Schisano}, {Spinoglio}, {Ward-Thompson}, \& {White}}]{2015A&A...584A..91K}
---. 2015, \aap, 584, A91, \dodoi{10.1051/0004-6361/201525861}

\bibitem[{{Kriegel} {et~al.}(2011){Kriegel}, {Kröger}, {Sander}, \&
  {Zimek}}]{dbcluster}
{Kriegel}, H.-P., {Kröger}, P., {Sander}, J., \& {Zimek}, A. 2011, in WIREs
  Data Mining and Knowledge Discovery, Vol. 1 (3), 231--240,
  \dodoi{10.1002/widm.30}

\bibitem[{{Kudoh} \& {Basu}(2008)}]{2008ApJ...679L..97K}
{Kudoh}, T., \& {Basu}, S. 2008, \apjl, 679, L97, \dodoi{10.1086/589618}

\bibitem[{{Kudoh} \& {Basu}(2011)}]{2011ApJ...728..123K}
---. 2011, \apj, 728, 123, \dodoi{10.1088/0004-637X/728/2/123}

\bibitem[{{Kwon} {et~al.}(2018){Kwon}, {Doi}, {Tamura}, {Matsumura}, {Pattle},
  {Berry}, {Sadavoy}, {Matthews}, {Ward-Thompson}, {Hasegawa}, {Furuya}, {Pon},
  {Di Francesco}, {Arzoumanian}, {Hayashi}, {Kawabata}, {Onaka}, {Choi},
  {Kang}, {Hoang}, {Lee}, {Lee}, {Liu}, {Liu}, {Inutsuka}, {Eswaraiah},
  {Bastien}, {Kwon}, {Lai}, {Qiu}, {Coud{\'e}}, {Franzmann}, {Friberg},
  {Graves}, {Greaves}, {Houde}, {Johnstone}, {Kirk}, {Koch}, {Li}, {Parsons},
  {Rao}, {Rawlings}, {Shinnaga}, {van Loo}, {Aso}, {Byun}, {Chen}, {Chen},
  {Chen}, {Ching}, {Cho}, {Chrysostomou}, {Chung}, {Drabek-Maunder}, {Eyres},
  {Fiege}, {Friesen}, {Fuller}, {Gledhill}, {Griffin}, {Gu}, {Hatchell},
  {Holland}, {Inoue}, {Iwasaki}, {Jeong}, {Kang}, {Kang}, {Kemper}, {Kim},
  {Kim}, {Kim}, {Kim}, {Kim}, {Kim}, {Lacaille}, {Lee}, {Li}, {Li}, {Liu},
  {Liu}, {Lyo}, {Mairs}, {Moriarty-Schieven}, {Nakamura}, {Nakanishi},
  {Ohashi}, {Peretto}, {Pyo}, {Qian}, {Retter}, {Richer}, {Rigby},
  {Robitaille}, {Savini}, {Scaife}, {Soam}, {Tang}, {Tomisaka}, {Wang}, {Wang},
  {Whitworth}, {Yen}, {Yoo}, {Yuan}, {Zhang}, {Zhang}, {Zhou}, {Zhu},
  {Andr{\'e}}, {Dowell}, {Falle}, {Tsukamoto}, {Nakagawa}, {Kanamori},
  {Kataoka}, {Kobayashi}, {Nagata}, {Saito}, {Seta}, \&
  {Zenko}}]{2018ApJ...859....4K}
{Kwon}, J., {Doi}, Y., {Tamura}, M., {et~al.} 2018, \apj, 859, 4,
  \dodoi{10.3847/1538-4357/aabd82}

\bibitem[{{Lazarian}(2007)}]{2007JQSRT.106..225L}
{Lazarian}, A. 2007, \jqsrt, 106, 225, \dodoi{10.1016/j.jqsrt.2007.01.038}

\bibitem[{{Lazarian} \& {Hoang}(2007)}]{2007MNRAS.378..910L}
{Lazarian}, A., \& {Hoang}, T. 2007, \mnras, 378, 910,
  \dodoi{10.1111/j.1365-2966.2007.11817.x}

\bibitem[{{Lazarian} \& {Hoang}(2008)}]{2008ApJ...676L..25L}
---. 2008, \apjl, 676, L25, \dodoi{10.1086/586706}

\bibitem[{{Lazarian} \& {Hoang}(2019)}]{2019ApJ...883..122L}
---. 2019, \apj, 883, 122, \dodoi{10.3847/1538-4357/ab3d39}

\bibitem[{{Lee} {et~al.}(2017){Lee}, {Hull}, \& {Offner}}]{2017ApJ...834..201L}
{Lee}, J. W.~Y., {Hull}, C. L.~H., \& {Offner}, S. S.~R. 2017, \apj, 834, 201,
  \dodoi{10.3847/1538-4357/834/2/201}

\bibitem[{{Leisawitz} {et~al.}(2019){Leisawitz}, {Amatucci}, {Allen},
  {Arenberg}, {Armus}, {Battersby}, {Beaman}, {Bauer}, {Bell}, {Beltran},
  {Benford}, {Bergin}, {Bolognese}, {Bradford}, {Bradley}, {Burgarella},
  {Carey}, {Carter}, {Chi}, {Cooray}, {Corsetti}, {D'Asto}, {De Beck}, {Denis},
  {Derkacz}, {Dewell}, {DiPirro}, {Earle}, {East}, {Edgington}, {Ennico},
  {Fantano}, {Feller}, {Flores}, {Folta}, {Fortney}, {Gavares}, {Generie},
  {Gerin}, {Granger}, {Greene}, {Griffiths}, {Harpole}, {Harvey}, {Helmich},
  {Helou}, {Hilliard}, {Howard}, {Jacoby}, {Jamil}, {Jamison}, {Kaltenegger},
  {Kataria}, {Knight}, {Knollenberg}, {Lawrence}, {Lightsey}, {Lipscy},
  {Lynch}, {Mamajek}, {Martins}, {Mather}, {Meixner}, {Melnick}, {Milam},
  {Mooney}, {Moseley}, {Narayanan}, {Neff}, {Nguyen}, {Nordt}, {Olson},
  {Padgett}, {Petach}, {Petro}, {Pohner}, {Pontoppidan}, {Pope}, {Ramspacher},
  {Rao}, {Rieke}, {Rieke}, {Roellig}, {Sakon}, {Sand in}, {Sandstrom}, {Scott},
  {Seals}, {Sheth}, {Staguhn}, {Steeves}, {Stevenson}, {Stokowski},
  {Stoneking}, {Su}, {Tajdaran}, {Tompkins}, {Turner}, {Vieira}, {Webster},
  {Wiedner}, {Wright}, {Wu}, \& {Zmuidzinas}}]{2019SPIE11115E..0QL}
{Leisawitz}, D., {Amatucci}, E., {Allen}, L., {et~al.} 2019, in Society of
  Photo-Optical Instrumentation Engineers (SPIE) Conference Series, Vol. 11115,
  \procspie, 111150Q, \dodoi{10.1117/12.2530514}

\bibitem[{{Li} \& {Klein}(2019)}]{2019MNRAS.485.4509L}
{Li}, P.~S., \& {Klein}, R.~I. 2019, \mnras, 485, 4509,
  \dodoi{10.1093/mnras/stz653}

\bibitem[{{Liu} {et~al.}(2019){Liu}, {Qiu}, {Berry}, {Di Francesco}, {Bastien},
  {Koch}, {Furuya}, {Kim}, {Coud{\'e}}, {Lee}, {Soam}, {Eswaraiah}, {Li},
  {Hwang}, {Lyo}, {Pattle}, {Hasegawa}, {Kwon}, {Lai}, {Ward-Thompson},
  {Ching}, {Chen}, {Gu}, {Li}, {Li}, {Liu}, {Qian}, {Wang}, {Yuan}, {Zhang},
  {Zhang}, {Zhang}, {Zhou}, {Zhu}, {Andr{\'e}}, {Arzoumanian}, {Aso}, {Byun},
  {Chen}, {Chen}, {Chen}, {Cho}, {Choi}, {Chrysostomou}, {Chung}, {Doi},
  {Drabek-Maunder}, {Dowell}, {Eyres}, {Falle}, {Fanciullo}, {Fiege},
  {Franzmann}, {Friberg}, {Friesen}, {Fuller}, {Gledhill}, {Graves}, {Greaves},
  {Griffin}, {Han}, {Hatchell}, {Hayashi}, {Hoang}, {Holland}, {Houde},
  {Inoue}, {Inutsuka}, {Iwasaki}, {Jeong}, {Johnstone}, {Kanamori}, {Kang},
  {Kang}, {Kang}, {Kataoka}, {Kawabata}, {Kemper}, {Kim}, {Kim}, {Kim}, {Kim},
  {Kim}, {Kirk}, {Kobayashi}, {Kusune}, {Kwon}, {Lacaille}, {Lee}, {Lee},
  {Lee}, {Lee}, {Liu}, {Liu}, {van Loo}, {Mairs}, {Matsumura}, {Matthews},
  {Moriarty-Schieven}, {Nagata}, {Nakamura}, {Nakanishi}, {Ohashi}, {Onaka},
  {Parker}, {Parsons}, {Pascale}, {Peretto}, {Pon}, {Pyo}, {Rao}, {Rawlings},
  {Retter}, {Richer}, {Rigby}, {Robitaille}, {Sadavoy}, {Saito}, {Savini},
  {Scaife}, {Seta}, {Shinnaga}, {Tamura}, {Tang}, {Tomisaka}, {Tsukamoto},
  {Wang}, {Whitworth}, {Yen}, {Yoo}, \& {Zenko}}]{2019ApJ...877...43L}
{Liu}, J., {Qiu}, K., {Berry}, D., {et~al.} 2019, \apj, 877, 43,
  \dodoi{10.3847/1538-4357/ab0958}

\bibitem[{{Liu} {et~al.}(2018){Liu}, {Li}, {Juvela}, {Kim}, {Evans}, {Di
  Francesco}, {Liu}, {Yuan}, {Tatematsu}, {Zhang}, {Ward-Thompson}, {Fuller},
  {Goldsmith}, {Koch}, {Sanhueza}, {Ristorcelli}, {Kang}, {Chen}, {Hirano},
  {Wu}, {Sokolov}, {Lee}, {White}, {Wang}, {Eden}, {Li}, {Thompson}, {Pattle},
  {Soam}, {Nasedkin}, {Kim}, {Kim}, {Lai}, {Park}, {Qiu}, {Zhang}, {Alina},
  {Eswaraiah}, {Falgarone}, {Fich}, {Greaves}, {Gu}, {Kwon}, {Li}, {Malinen},
  {Montier}, {Parsons}, {Qin}, {Rawlings}, {Ren}, {Tang}, {Tang}, {Toth},
  {Wang}, {Wouterloot}, {Yi}, \& {Zhang}}]{2018ApJ...859..151L}
{Liu}, T., {Li}, P.~S., {Juvela}, M., {et~al.} 2018, \apj, 859, 151,
  \dodoi{10.3847/1538-4357/aac025}

\bibitem[{{Mardia} \& {Jupp}(1999)}]{1999Directional}
{Mardia}, K., \& {Jupp}, P.~E. 1999, Directional Statistics (Wiley)

\bibitem[{{Marsh} {et~al.}(2016){Marsh}, {Kirk}, {Andr{\'e}}, {Griffin},
  {K{\"o}nyves}, {Palmeirim}, {Men'shchikov}, {Ward-Thompson}, {Benedettini},
  {Bresnahan}, {di Francesco}, {Elia}, {Motte}, {Peretto}, {Pezzuto}, {Roy},
  {Sadavoy}, {Schneider}, {Spinoglio}, \& {White}}]{2016MNRAS.459..342M}
{Marsh}, K.~A., {Kirk}, J.~M., {Andr{\'e}}, P., {et~al.} 2016, \mnras, 459,
  342, \dodoi{10.1093/mnras/stw301}

\bibitem[{{Matthews} {et~al.}(2009){Matthews}, {McPhee}, {Fissel}, \&
  {Curran}}]{2009ApJS..182..143M}
{Matthews}, B.~C., {McPhee}, C.~A., {Fissel}, L.~M., \& {Curran}, R.~L. 2009,
  \apjs, 182, 143, \dodoi{10.1088/0067-0049/182/1/143}

\bibitem[{{M{\'e}nard} \& {Duch{\^e}ne}(2004)}]{2004A&A...425..973M}
{M{\'e}nard}, F., \& {Duch{\^e}ne}, G. 2004, \aap, 425, 973,
  \dodoi{10.1051/0004-6361:20041338}

\bibitem[{{Men'shchikov} {et~al.}(2010){Men'shchikov}, {Andr{\'e}}, {Didelon},
  {K{\"o}nyves}, {Schneider}, {Motte}, {Bontemps}, {Arzoumanian}, {Attard},
  {Abergel}, {Baluteau}, {Bernard}, {Cambr{\'e}sy}, {Cox}, {di Francesco}, {di
  Giorgio}, {Griffin}, {Hargrave}, {Huang}, {Kirk}, {Li}, {Martin}, {Minier},
  {Miville-Desch{\^e}nes}, {Molinari}, {Olofsson}, {Pezzuto}, {Roussel},
  {Russeil}, {Saraceno}, {Sauvage}, {Sibthorpe}, {Spinoglio}, {Testi},
  {Ward-Thompson}, {White}, {Wilson}, {Woodcraft}, \&
  {Zavagno}}]{2010A&A...518L.103M}
{Men'shchikov}, A., {Andr{\'e}}, P., {Didelon}, P., {et~al.} 2010, \aap, 518,
  L103, \dodoi{10.1051/0004-6361/201014668}

\bibitem[{{Minchin} {et~al.}(1995){Minchin}, {Sandell}, \&
  {Murray}}]{1995A&A...293L..61M}
{Minchin}, N.~R., {Sandell}, G., \& {Murray}, A.~G. 1995, \aap, 293, L61

\bibitem[{{Miville-Desch{\^e}nes} {et~al.}(2010){Miville-Desch{\^e}nes},
  {Martin}, {Abergel}, {Bernard}, {Boulanger}, {Lagache}, {Anderson},
  {Andr{\'e}}, {Arab}, {Baluteau}, {Blagrave}, {Bontemps}, {Cohen},
  {Compiegne}, {Cox}, {Dartois}, {Davis}, {Emery}, {Fulton}, {Gry}, {Habart},
  {Huang}, {Joblin}, {Jones}, {Kirk}, {Lim}, {Madden}, {Makiwa}, {Menshchikov},
  {Molinari}, {Moseley}, {Motte}, {Naylor}, {Okumura}, {Pinheiro Gon{\c
  c}alves}, {Polehampton}, {Rod{\'o}n}, {Russeil}, {Saraceno}, {Schneider},
  {Sidher}, {Spencer}, {Swinyard}, {Ward-Thompson}, {White}, \&
  {Zavagno}}]{2010A&A...518L.104M}
{Miville-Desch{\^e}nes}, M.-A., {Martin}, P.~G., {Abergel}, A., {et~al.} 2010,
  \aap, 518, L104, \dodoi{10.1051/0004-6361/201014678}

\bibitem[{{Molinari} {et~al.}(2010){Molinari}, {Swinyard}, {Bally}, {Barlow},
  {Bernard}, {Martin}, {Moore}, {Noriega-Crespo}, {Plume}, {Testi}, {Zavagno},
  {Abergel}, {Ali}, {Anderson}, {Andr{\'e}}, {Baluteau}, {Battersby},
  {Beltr{\'a}n}, {Benedettini}, {Billot}, {Blommaert}, {Bontemps}, {Boulanger},
  {Brand}, {Brunt}, {Burton}, {Calzoletti}, {Carey}, {Caselli}, {Cesaroni},
  {Cernicharo}, {Chakrabarti}, {Chrysostomou}, {Cohen}, {Compiegne}, {de
  Bernardis}, {de Gasperis}, {di Giorgio}, {Elia}, {Faustini}, {Flagey},
  {Fukui}, {Fuller}, {Ganga}, {Garcia-Lario}, {Glenn}, {Goldsmith}, {Griffin},
  {Hoare}, {Huang}, {Ikhenaode}, {Joblin}, {Joncas}, {Juvela}, {Kirk},
  {Lagache}, {Li}, {Lim}, {Lord}, {Marengo}, {Marshall}, {Masi}, {Massi},
  {Matsuura}, {Minier}, {Miville-Desch{\^e}nes}, {Montier}, {Morgan}, {Motte},
  {Mottram}, {M{\"u}ller}, {Natoli}, {Neves}, {Olmi}, {Paladini}, {Paradis},
  {Parsons}, {Peretto}, {Pestalozzi}, {Pezzuto}, {Piacentini}, {Piazzo},
  {Polychroni}, {Pomar{\`e}s}, {Popescu}, {Reach}, {Ristorcelli}, {Robitaille},
  {Robitaille}, {Rod{\'o}n}, {Roy}, {Royer}, {Russeil}, {Saraceno}, {Sauvage},
  {Schilke}, {Schisano}, {Schneider}, {Schuller}, {Schulz}, {Sibthorpe},
  {Smith}, {Smith}, {Spinoglio}, {Stamatellos}, {Strafella}, {Stringfellow},
  {Sturm}, {Taylor}, {Thompson}, {Traficante}, {Tuffs}, {Umana}, {Valenziano},
  {Vavrek}, {Veneziani}, {Viti}, {Waelkens}, {Ward-Thompson}, {White},
  {Wilcock}, {Wyrowski}, {Yorke}, \& {Zhang}}]{2010A&A...518L.100M}
{Molinari}, S., {Swinyard}, B., {Bally}, J., {et~al.} 2010, \aap, 518, L100,
  \dodoi{10.1051/0004-6361/201014659}

\bibitem[{{Myers}(2009)}]{2009ApJ...700.1609M}
{Myers}, P.~C. 2009, \apj, 700, 1609, \dodoi{10.1088/0004-637X/700/2/1609}

\bibitem[{{Nagai} {et~al.}(1998){Nagai}, {Inutsuka}, \&
  {Miyama}}]{1998ApJ...506..306N}
{Nagai}, T., {Inutsuka}, S.-i., \& {Miyama}, S.~M. 1998, \apj, 506, 306,
  \dodoi{10.1086/306249}

\bibitem[{{Naghizadeh-Khouei} \& {Clarke}(1993)}]{1993A&A...274..968N}
{Naghizadeh-Khouei}, J., \& {Clarke}, D. 1993, \aap, 274, 968

\bibitem[{{Nakamura} \& {Li}(2008)}]{2008ApJ...687..354N}
{Nakamura}, F., \& {Li}, Z.-Y. 2008, \apj, 687, 354, \dodoi{10.1086/591641}

\bibitem[{{Nguyen} {et~al.}(2018){Nguyen}, {Dawson}, {Miville-Desch{\^e}nes},
  {Tang}, {Li}, {Heiles}, {Murray}, {Stanimirovi{\'c}}, {Gibson},
  {McClure-Griffiths}, {Troland}, {Bronfman}, \&
  {Finger}}]{2018ApJ...862...49N}
{Nguyen}, H., {Dawson}, J.~R., {Miville-Desch{\^e}nes}, M.~A., {et~al.} 2018,
  \apj, 862, 49, \dodoi{10.3847/1538-4357/aac82b}

\bibitem[{{Ortiz-Le{\'o}n} {et~al.}(2018){Ortiz-Le{\'o}n}, {Loinard}, {Dzib},
  {Galli}, {Kounkel}, {Mioduszewski}, {Rodr{\'{\i}}guez}, {Torres}, {Hartmann},
  {Boden}, {Evans}, {Brice{\~n}o}, \& {Tobin}}]{2018ApJ...865...73O}
{Ortiz-Le{\'o}n}, G.~N., {Loinard}, L., {Dzib}, S.~A., {et~al.} 2018, \apj,
  865, 73, \dodoi{10.3847/1538-4357/aada49}

\bibitem[{{Ostriker}(1964)}]{1964ApJ...140.1056O}
{Ostriker}, J. 1964, \apj, 140, 1056, \dodoi{10.1086/148005}

\bibitem[{{Palmeirim} {et~al.}(2013){Palmeirim}, {Andr{\'e}}, {Kirk},
  {Ward-Thompson}, {Arzoumanian}, {K{\"o}nyves}, {Didelon}, {Schneider},
  {Benedettini}, {Bontemps}, {Di Francesco}, {Elia}, {Griffin}, {Hennemann},
  {Hill}, {Martin}, {Men'shchikov}, {Molinari}, {Motte}, {Nguyen Luong},
  {Nutter}, {Peretto}, {Pezzuto}, {Roy}, {Rygl}, {Spinoglio}, \&
  {White}}]{2013A&A...550A..38P}
{Palmeirim}, P., {Andr{\'e}}, P., {Kirk}, J., {et~al.} 2013, \aap, 550, A38,
  \dodoi{10.1051/0004-6361/201220500}

\bibitem[{{Parsons} {et~al.}(2018){Parsons}, {Berry}, {Rawlings}, \&
  {Graves}}]{Parsons2018}
{Parsons}, H. A.~L., {Berry}, D.~S., {Rawlings}, M.~G., \& {Graves}, S.~F.
  2018, {The POL-2 Data Reduction Cookbook}, 1st edn., Starlink Project, East
  Asian Observatory

\bibitem[{{Pattle} {et~al.}(2017){Pattle}, {Ward-Thompson}, {Berry},
  {Hatchell}, {Chen}, {Pon}, {Koch}, {Kwon}, {Kim}, {Bastien}, {Cho},
  {Coud{\'e}}, {Di Francesco}, {Fuller}, {Furuya}, {Graves}, {Johnstone},
  {Kirk}, {Kwon}, {Lee}, {Matthews}, {Mottram}, {Parsons}, {Sadavoy},
  {Shinnaga}, {Soam}, {Hasegawa}, {Lai}, {Qiu}, \&
  {Friberg}}]{2017ApJ...846..122P}
{Pattle}, K., {Ward-Thompson}, D., {Berry}, D., {et~al.} 2017, \apj, 846, 122,
  \dodoi{10.3847/1538-4357/aa80e5}

\bibitem[{{Pattle} {et~al.}(2018){Pattle}, {Ward-Thompson}, {Hasegawa},
  {Bastien}, {Kwon}, {Lai}, {Qiu}, {Furuya}, {Berry}, \& {JCMT BISTRO Survey
  Team}}]{2018ApJ...860L...6P}
{Pattle}, K., {Ward-Thompson}, D., {Hasegawa}, T., {et~al.} 2018, \apjl, 860,
  L6, \dodoi{10.3847/2041-8213/aac771}

\bibitem[{{Peretto} {et~al.}(2012){Peretto}, {Andr{\'e}}, {K{\"o}nyves},
  {Schneider}, {Arzoumanian}, {Palmeirim}, {Didelon}, {Attard}, {Bernard}, {Di
  Francesco}, {Elia}, {Hennemann}, {Hill}, {Kirk}, {Men'shchikov}, {Motte},
  {Nguyen Luong}, {Roussel}, {Sousbie}, {Testi}, {Ward-Thompson}, {White}, \&
  {Zavagno}}]{2012A&A...541A..63P}
{Peretto}, N., {Andr{\'e}}, P., {K{\"o}nyves}, V., {et~al.} 2012, \aap, 541,
  A63, \dodoi{10.1051/0004-6361/201118663}

\bibitem[{{Planck Collaboration} {et~al.}(2014){Planck Collaboration},
  {Abergel}, {Ade}, {Aghanim}, {Alves}, {Aniano}, {Armitage-Caplan}, {Arnaud},
  {Ashdown}, {Atrio-Barandela}, \& et~al.}]{2014A&A...571A..11P}
{Planck Collaboration}, {Abergel}, A., {Ade}, P.~A.~R., {et~al.} 2014, \aap,
  571, A11, \dodoi{10.1051/0004-6361/201323195}

\bibitem[{{Planck Collaboration} {et~al.}(2015){Planck Collaboration}, {Ade},
  {Aghanim}, {Alina}, {Alves}, {Armitage-Caplan}, {Arnaud}, {Arzoumanian},
  {Ashdown}, {Atrio-Barandela}, \& et~al.}]{2015A&A...576A.104P}
{Planck Collaboration}, {Ade}, P.~A.~R., {Aghanim}, N., {et~al.} 2015, \aap,
  576, A104, \dodoi{10.1051/0004-6361/201424082}

\bibitem[{{Planck Collaboration} {et~al.}(2016{\natexlab{a}}){Planck
  Collaboration}, {Adam}, {Ade}, {Aghanim}, {Alves}, {Arnaud}, {Arzoumanian},
  {Ashdown}, {Aumont}, {Baccigalupi}, \& et~al.}]{2016A&A...586A.135P}
{Planck Collaboration}, {Adam}, R., {Ade}, P.~A.~R., {et~al.}
  2016{\natexlab{a}}, \aap, 586, A135, \dodoi{10.1051/0004-6361/201425044}

\bibitem[{{Planck Collaboration} {et~al.}(2016{\natexlab{b}}){Planck
  Collaboration}, {Ade}, {Aghanim}, {Alves}, {Arnaud}, {Arzoumanian}, {Aumont},
  {Baccigalupi}, {Banday}, {Barreiro}, {Bartolo}, {Battaner}, {Benabed},
  {Benoit-L{\'e}vy}, {Bernard}, {Bern{\'e}}, {Bersanelli}, {Bielewicz},
  {Bonaldi}, {Bonavera}, {Bond}, {Borrill}, {Bouchet}, {Boulanger}, {Bracco},
  {Burigana}, {Calabrese}, {Cardoso}, {Catalano}, {Chamballu}, {Chiang},
  {Christensen}, {Clements}, {Colombi}, {Colombo}, {Combet}, {Couchot},
  {Crill}, {Curto}, {Cuttaia}, {Danese}, {Davies}, {Davis}, {de Bernardis}, {de
  Rosa}, {de Zotti}, {Delabrouille}, {Dickinson}, {Diego}, {Donzelli},
  {Dor{\'e}}, {Douspis}, {Ducout}, {Dupac}, {Elsner}, {En{\ss}lin}, {Eriksen},
  {Falgarone}, {Ferri{\`e}re}, {Finelli}, {Forni}, {Frailis}, {Fraisse},
  {Franceschi}, {Frejsel}, {Galeotta}, {Galli}, {Ganga}, {Ghosh}, {Giard},
  {Giraud-H{\'e}raud}, {Gjerl{\o}w}, {Gonz{\'a}lez-Nuevo}, {G{\'o}rski},
  {Gregorio}, {Gruppuso}, {Guillet}, {Hansen}, {Hanson}, {Harrison},
  {Hern{\'a}ndez-Monteagudo}, {Herranz}, {Hildebrandt}, {Hivon}, {Hobson},
  {Holmes}, {Huffenberger}, {Hurier}, {Jaffe}, {Jaffe}, {Jones}, {Juvela},
  {Keskitalo}, {Kisner}, {Knoche}, {Kunz}, {Kurki-Suonio}, {Lagache},
  {Lamarre}, {Lasenby}, {Lawrence}, {Leonardi}, {Levrier}, {Liguori}, {Lilje},
  {Linden-V{\o}rnle}, {L{\'o}pez-Caniego}, {Lubin}, {Mac{\'{\i}}as-P{\'e}rez},
  {Maffei}, {Mandolesi}, {Mangilli}, {Maris}, {Martin},
  {Mart{\'{\i}}nez-Gonz{\'a}lez}, {Masi}, {Matarrese}, {Mazzotta},
  {Melchiorri}, {Mendes}, {Mennella}, {Migliaccio}, {Mitra},
  {Miville-Desch{\^e}nes}, {Moneti}, {Montier}, {Morgante}, {Mortlock},
  {Munshi}, {Murphy}, {Naselsky}, {Nati}, {Natoli}, {N{\o}rgaard-Nielsen},
  {Noviello}, {Novikov}, {Novikov}, {Oppermann}, {Pagano}, {Pajot}, {Paladini},
  {Paoletti}, {Pasian}, {Perrotta}, {Pettorino}, {Piacentini}, {Piat},
  {Pierpaoli}, {Pietrobon}, {Plaszczynski}, {Pointecouteau}, {Polenta},
  {Pratt}, {Puget}, {Rachen}, {Rebolo}, {Reinecke}, {Remazeilles}, {Renault},
  {Renzi}, {Ricciardi}, {Ristorcelli}, {Rocha}, {Rosset}, {Rossetti},
  {Roudier}, {Rubi{\~n}o-Mart{\'{\i}}n}, {Rusholme}, {Sandri}, {Savelainen},
  {Savini}, {Scott}, {Soler}, {Stolyarov}, {Sutton}, {Suur-Uski}, {Sygnet},
  {Tauber}, {Terenzi}, {Toffolatti}, {Tomasi}, {Tristram}, {Tucci},
  {Valenziano}, {Valiviita}, {Van Tent}, {Vielva}, {Villa}, {Wade}, {Wandelt},
  {Yvon}, {Zacchei}, \& {Zonca}}]{2016A&A...586A.136P}
{Planck Collaboration}, {Ade}, P.~A.~R., {Aghanim}, N., {et~al.}
  2016{\natexlab{b}}, \aap, 586, A136, \dodoi{10.1051/0004-6361/201425305}

\bibitem[{{Planck Collaboration} {et~al.}(2016{\natexlab{c}}){Planck
  Collaboration}, {Ade}, {Aghanim}, {Alves}, {Arnaud}, {Arzoumanian},
  {Ashdown}, {Aumont}, {Baccigalupi}, {Banday}, {Barreiro}, {Bartolo},
  {Battaner}, {Benabed}, {Beno{\^i}t}, {Benoit-L{\'e}vy}, {Bernard},
  {Bersanelli}, {Bielewicz}, {Bock}, {Bonavera}, {Bond}, {Borrill}, {Bouchet},
  {Boulanger}, {Bracco}, {Burigana}, {Calabrese}, {Cardoso}, {Catalano},
  {Chiang}, {Christensen}, {Colombo}, {Combet}, {Couchot}, {Crill}, {Curto},
  {Cuttaia}, {Danese}, {Davies}, {Davis}, {de Bernardis}, {de Rosa}, {de
  Zotti}, {Delabrouille}, {Dickinson}, {Diego}, {Dole}, {Donzelli}, {Dor{\'e}},
  {Douspis}, {Ducout}, {Dupac}, {Efstathiou}, {Elsner}, {En{\ss}lin},
  {Eriksen}, {Falceta-Gon{\c c}alves}, {Falgarone}, {Ferri{\`e}re}, {Finelli},
  {Forni}, {Frailis}, {Fraisse}, {Franceschi}, {Frejsel}, {Galeotta}, {Galli},
  {Ganga}, {Ghosh}, {Giard}, {Gjerl{\o}w}, {Gonz{\'a}lez-Nuevo}, {G{\'o}rski},
  {Gregorio}, {Gruppuso}, {Gudmundsson}, {Guillet}, {Harrison}, {Helou},
  {Hennebelle}, {Henrot-Versill{\'e}}, {Hern{\'a}ndez-Monteagudo}, {Herranz},
  {Hildebrandt}, {Hivon}, {Holmes}, {Hornstrup}, {Huffenberger}, {Hurier},
  {Jaffe}, {Jaffe}, {Jones}, {Juvela}, {Keih{\"a}nen}, {Keskitalo}, {Kisner},
  {Knoche}, {Kunz}, {Kurki-Suonio}, {Lagache}, {Lamarre}, {Lasenby},
  {Lattanzi}, {Lawrence}, {Leonardi}, {Levrier}, {Liguori}, {Lilje},
  {Linden-V{\o}rnle}, {L{\'o}pez-Caniego}, {Lubin}, {Mac{\'{\i}}as-P{\'e}rez},
  {Maino}, {Mandolesi}, {Mangilli}, {Maris}, {Martin},
  {Mart{\'{\i}}nez-Gonz{\'a}lez}, {Masi}, {Matarrese}, {Melchiorri}, {Mendes},
  {Mennella}, {Migliaccio}, {Miville-Desch{\^e}nes}, {Moneti}, {Montier},
  {Morgante}, {Mortlock}, {Munshi}, {Murphy}, {Naselsky}, {Nati},
  {Netterfield}, {Noviello}, {Novikov}, {Novikov}, {Oppermann}, {Oxborrow},
  {Pagano}, {Pajot}, {Paladini}, {Paoletti}, {Pasian}, {Perotto}, {Pettorino},
  {Piacentini}, {Piat}, {Pierpaoli}, {Pietrobon}, {Plaszczynski},
  {Pointecouteau}, {Polenta}, {Ponthieu}, {Pratt}, {Prunet}, {Puget}, {Rachen},
  {Reinecke}, {Remazeilles}, {Renault}, {Renzi}, {Ristorcelli}, {Rocha},
  {Rossetti}, {Roudier}, {Rubi{\~n}o-Mart{\'{\i}}n}, {Rusholme}, {Sandri},
  {Santos}, {Savelainen}, {Savini}, {Scott}, {Soler}, {Stolyarov}, {Sudiwala},
  {Sutton}, {Suur-Uski}, {Sygnet}, {Tauber}, {Terenzi}, {Toffolatti}, {Tomasi},
  {Tristram}, {Tucci}, {Umana}, {Valenziano}, {Valiviita}, {Van Tent},
  {Vielva}, {Villa}, {Wade}, {Wandelt}, {Wehus}, {Ysard}, {Yvon}, \&
  {Zonca}}]{2016A&A...586A.138P}
---. 2016{\natexlab{c}}, \aap, 586, A138, \dodoi{10.1051/0004-6361/201525896}

\bibitem[{{Planck Collaboration} {et~al.}(2018{\natexlab{a}}){Planck
  Collaboration}, {Akrami}, {Ashdown}, {Aumont}, {Baccigalupi}, {Ballardini},
  {Banday}, {Barreiro}, {Bartolo}, {Basak}, {Benabed}, {Bernard}, {Bersanelli},
  {Bielewicz}, {Bond}, {Borrill}, {Bouchet}, {Boulanger}, {Bracco}, {Bucher},
  {Burigana}, {Calabrese}, {Cardoso}, {Carron}, {Chiang}, {Combet}, {Crill},
  {de Bernardis}, {de Zotti}, {Delabrouille}, {Delouis}, {Di Valentino},
  {Dickinson}, {Diego}, {Ducout}, {Dupac}, {Efstathiou}, {Elsner},
  {En{\ss}lin}, {Falgarone}, {Fantaye}, {Ferri{\`e}re}, {Finelli},
  {Forastieri}, {Frailis}, {Fraisse}, {Franceschi}, {Frolov}, {Galeotta},
  {Galli}, {Ganga}, {G{\'e}nova-Santos}, {Ghosh}, {Gonz{\'a}lez-Nuevo},
  {G{\'o}rski}, {Gruppuso}, {Gudmundsson}, {Guillet}, {Handley}, {Hansen},
  {Herranz}, {Huang}, {Jaffe}, {Jones}, {Keih{\"a}nen}, {Keskitalo}, {Kiiveri},
  {Kim}, {Krachmalnicoff}, {Kunz}, {Kurki-Suonio}, {Lamarre}, {Lasenby}, {Le
  Jeune}, {Levrier}, {Liguori}, {Lilje}, {Lindholm}, {L{\'o}pez-Caniego},
  {Lubin}, {Ma}, {Mac{\'{\i}}as-P{\'e}rez}, {Maggio}, {Maino}, {Mandolesi},
  {Mangilli}, {Martin}, {Mart{\'{\i}}nez-Gonz{\'a}lez}, {Matarrese}, {McEwen},
  {Meinhold}, {Melchiorri}, {Migliaccio}, {Miville-Desch{\^e}nes}, {Molinari},
  {Moneti}, {Montier}, {Morgante}, {Natoli}, {Pagano}, {Paoletti}, {Pettorino},
  {Piacentini}, {Polenta}, {Puget}, {Rachen}, {Reinecke}, {Remazeilles},
  {Renzi}, {Rocha}, {Rosset}, {Roudier}, {Rubi{\~n}o-Mart{\'{\i}}n},
  {Ruiz-Granados}, {Salvati}, {Sandri}, {Savelainen}, {Scott}, {Soler},
  {Spencer}, {Tauber}, {Tavagnacco}, {Toffolatti}, {Tomasi}, {Trombetti},
  {Valiviita}, {Vansyngel}, {Van Tent}, {Vielva}, {Villa}, {Vittorio}, {Wehus},
  {Zacchei}, \& {Zonca}}]{2018arXiv180104945P}
{Planck Collaboration}, {Akrami}, Y., {Ashdown}, M., {et~al.}
  2018{\natexlab{a}}, ArXiv e-prints.
\newblock \doarXiv{1801.04945}

\bibitem[{{Planck Collaboration} {et~al.}(2018{\natexlab{b}}){Planck
  Collaboration}, {Aghanim}, {Akrami}, {Alves}, {Ashdown}, {Aumont},
  {Baccigalupi}, {Ballardini}, {Banday}, {Barreiro}, {Bartolo}, {Basak},
  {Benabed}, {Bernard}, {Bersanelli}, {Bielewicz}, {Bock}, {Bond}, {Borrill},
  {Bouchet}, {Boulanger}, {Bracco}, {Bucher}, {Burigana}, {Calabrese},
  {Cardoso}, {Carron}, {Chary}, {Chiang}, {Colombo}, {Combet}, {Crill},
  {Cuttaia}, {de Bernardis}, {de Zotti}, {Delabrouille}, {Delouis}, {Di
  Valentino}, {Dickinson}, {Diego}, {Dor{\'e}}, {Douspis}, {Ducout}, {Dupac},
  {Efstathiou}, {Elsner}, {En{\ss}lin}, {Eriksen}, {Fantaye},
  {Fernandez-Cobos}, {Ferri{\`e}re}, {Forastieri}, {Frailis}, {Fraisse},
  {Franceschi}, {Frolov}, {Galeotta}, {Galli}, {Ganga}, {G{\'e}nova-Santos},
  {Gerbino}, {Ghosh}, {Gonz{\'a}lez-Nuevo}, {G{\'o}rski}, {Gratton}, {Green},
  {Gruppuso}, {Gudmundsson}, {Guillet}, {Handley}, {Hansen}, {Helou},
  {Herranz}, {Hivon}, {Huang}, {Jaffe}, {Jones}, {Keih{\"a}nen}, {Keskitalo},
  {Kiiveri}, {Kim}, {Krachmalnicoff}, {Kunz}, {Kurki-Suonio}, {Lagache},
  {Lamarre}, {Lasenby}, {Lattanzi}, {Lawrence}, {Le Jeune}, {Levrier},
  {Liguori}, {Lilje}, {Lindholm}, {L{\'o}pez-Caniego}, {Lubin}, {Ma},
  {Mac{\'{\i}}as-P{\'e}rez}, {Maggio}, {Maino}, {Mandolesi}, {Mangilli},
  {Marcos-Caballero}, {Maris}, {Martin}, {Mart{\'{\i}}nez-Gonz{\'a}lez},
  {Matarrese}, {Mauri}, {McEwen}, {Melchiorri}, {Mennella}, {Migliaccio},
  {Miville-Desch{\^e}nes}, {Molinari}, {Moneti}, {Montier}, {Morgante}, {Moss},
  {Natoli}, {Pagano}, {Paoletti}, {Patanchon}, {Perrotta}, {Pettorino},
  {Piacentini}, {Polastri}, {Polenta}, {Puget}, {Rachen}, {Reinecke},
  {Remazeilles}, {Renzi}, {Ristorcelli}, {Rocha}, {Rosset}, {Roudier},
  {Rubi{\~n}o-Mart{\'{\i}}n}, {Ruiz-Granados}, {Salvati}, {Sandri},
  {Savelainen}, {Scott}, {Sirignano}, {Sunyaev}, {Suur-Uski}, {Tauber},
  {Tavagnacco}, {Tenti}, {Toffolatti}, {Tomasi}, {Trombetti}, {Valiviita}, {Van
  Tent}, {Vielva}, {Villa}, {Vittorio}, {Wandelt}, {Wehus}, {Zacchei}, \&
  {Zonca}}]{2018arXiv180706212P}
{Planck Collaboration}, {Aghanim}, N., {Akrami}, Y., {et~al.}
  2018{\natexlab{b}}, ArXiv e-prints.
\newblock \doarXiv{1807.06212}

\bibitem[{{Plunkett} {et~al.}(2013){Plunkett}, {Arce}, {Corder}, {Mardones},
  {Sargent}, \& {Schnee}}]{2013ApJ...774...22P}
{Plunkett}, A.~L., {Arce}, H.~G., {Corder}, S.~A., {et~al.} 2013, \apj, 774,
  22, \dodoi{10.1088/0004-637X/774/1/22}

\bibitem[{{Poidevin} {et~al.}(2010){Poidevin}, {Bastien}, \&
  {Matthews}}]{2010ApJ...716..893P}
{Poidevin}, F., {Bastien}, P., \& {Matthews}, B.~C. 2010, \apj, 716, 893,
  \dodoi{10.1088/0004-637X/716/2/893}

\bibitem[{{Pudritz} \& {Ray}(2019)}]{2019FrASS...6...54P}
{Pudritz}, R.~E., \& {Ray}, T.~P. 2019, Frontiers in Astronomy and Space
  Sciences, 6, 54, \dodoi{10.3389/fspas.2019.00054}

\bibitem[{{Sancisi}(1974)}]{1974IAUS...60..115S}
{Sancisi}, R. 1974, in IAU Symposium, Vol.~60, Galactic Radio Astronomy, ed.
  F.~J. {Kerr} \& S.~C. {Simonson}, 115

\bibitem[{{Sandell} \& {Knee}(2001)}]{2001ApJ...546L..49S}
{Sandell}, G., \& {Knee}, L.~B.~G. 2001, \apjl, 546, L49,
  \dodoi{10.1086/318060}

\bibitem[{{Santos} {et~al.}(2019){Santos}, {Chuss}, {Dowell}, {Houde},
  {Looney}, {Lopez Rodriguez}, {Novak}, {Ward-Thompson}, {Berthoud}, {Dale},
  {Guerra}, {Hamilton}, {Hanany}, {Harper}, {Henning}, {Jones}, {Lazarian},
  {Michail}, {Morris}, {Staguhn}, {Stephens}, {Tassis}, {Trinh}, {Van Camp},
  {Volpert}, \& {Wollack}}]{2019ApJ...882..113S}
{Santos}, F.~P., {Chuss}, D.~T., {Dowell}, C.~D., {et~al.} 2019, \apj, 882,
  113, \dodoi{10.3847/1538-4357/ab3407}

\bibitem[{{Schneider} \& {Elmegreen}(1979)}]{1979ApJS...41...87S}
{Schneider}, S., \& {Elmegreen}, B.~G. 1979, \apjs, 41, 87,
  \dodoi{10.1086/190609}

\bibitem[{{Schuller} {et~al.}(2009){Schuller}, {Menten}, {Contreras},
  {Wyrowski}, {Schilke}, {Bronfman}, {Henning}, {Walmsley}, {Beuther},
  {Bontemps}, {Cesaroni}, {Deharveng}, {Garay}, {Herpin}, {Lefloch}, {Linz},
  {Mardones}, {Minier}, {Molinari}, {Motte}, {Nyman}, {Reveret}, {Risacher},
  {Russeil}, {Schneider}, {Testi}, {Troost}, {Vasyunina}, {Wienen}, {Zavagno},
  {Kovacs}, {Kreysa}, {Siringo}, \& {Wei{\ss}}}]{2009A&A...504..415S}
{Schuller}, F., {Menten}, K.~M., {Contreras}, Y., {et~al.} 2009, \aap, 504,
  415, \dodoi{10.1051/0004-6361/200811568}

\bibitem[{Serkowski(1962)}]{SERKOWSKI1962289}
Serkowski, K. 1962, Advances in Astronomy and Astrophysics, 1, 289,
  \dodoi{https://doi.org/10.1016/B978-1-4831-9919-1.50009-1}

\bibitem[{{Shimajiri} {et~al.}(2019){Shimajiri}, {Andr{\'e}}, {Palmeirim},
  {Arzoumanian}, {Bracco}, {K{\"o}nyves}, {Ntormousi}, \&
  {Ladjelate}}]{2019A&A...623A..16S}
{Shimajiri}, Y., {Andr{\'e}}, P., {Palmeirim}, P., {et~al.} 2019, \aap, 623,
  A16, \dodoi{10.1051/0004-6361/201834399}

\bibitem[{{Soam} {et~al.}(2018){Soam}, {Pattle}, {Ward-Thompson}, {Lee},
  {Sadavoy}, {Koch}, {Kim}, {Kwon}, {Kwon}, {Arzoumanian}, {Berry}, {Hoang},
  {Tamura}, {Lee}, {Liu}, {Kim}, {Johnstone}, {Nakamura}, {Lyo}, {Onaka},
  {Kim}, {Furuya}, {Hasegawa}, {Lai}, {Bastien}, {Chung}, {Kim}, {Parsons},
  {Rawlings}, {Mairs}, {Graves}, {Robitaille}, {Liu}, {Whitworth}, {Eswaraiah},
  {Rao}, {Yoo}, {Houde}, {Kang}, {Doi}, {Choi}, {Kang}, {Coud{\'e}}, {Li},
  {Matsumura}, {Matthews}, {Pon}, {Di Francesco}, {Hayashi}, {Kawabata},
  {Inutsuka}, {Qiu}, {Franzmann}, {Friberg}, {Greaves}, {Kirk}, {Li},
  {Shinnaga}, {van Loo}, {Aso}, {Byun}, {Chen}, {Chen}, {Chen}, {Ching}, {Cho},
  {Chrysostomou}, {Drabek-Maunder}, {Eyres}, {Fiege}, {Friesen}, {Fuller},
  {Gledhill}, {Griffin}, {Gu}, {Hatchell}, {Holland}, {Inoue}, {Iwasaki},
  {Jeong}, {Kang}, {Kemper}, {Kim}, {Kim}, {Lacaille}, {Lee}, {Li}, {Liu},
  {Liu}, {Moriarty-Schieven}, {Nakanishi}, {Ohashi}, {Peretto}, {Pyo}, {Qian},
  {Retter}, {Richer}, {Rigby}, {Savini}, {Scaife}, {Tang}, {Tomisaka}, {Wang},
  {Wang}, {Yen}, {Yuan}, {Zhang}, {Zhang}, {Zhou}, {Zhu}, {Andr{\'e}},
  {Dowell}, {Falle}, {Tsukamoto}, {Kanamori}, {Kataoka}, {Kobayashi}, {Nagata},
  {Saito}, {Seta}, {Hwang}, {Han}, {Lee}, \& {Zenko}}]{2018ApJ...861...65S}
{Soam}, A., {Pattle}, K., {Ward-Thompson}, D., {et~al.} 2018, \apj, 861, 65,
  \dodoi{10.3847/1538-4357/aac4a6}

\bibitem[{{Soam} {et~al.}(2019){Soam}, {Liu}, {Andersson}, {Lee}, {Liu},
  {Juvela}, {Li}, {Goldsmith}, {Zhang}, {Koch}, {Kim}, {Qiu}, {Evans},
  {Johnstone}, {Thompson}, {Ward-Thompson}, {Di Francesco}, {Tang},
  {Montillaud}, {Kim}, {Mairs}, {Sanhueza}, {Kim}, {Berry}, {Gordon},
  {Tatematsu}, {Liu}, {Pattle}, {Eden}, {McGehee}, {Wang}, {Ristorcelli},
  {Graves}, {Alina}, {Lacaille}, {Montier}, {Park}, {Kwon}, {Chung},
  {Pelkonen}, {Micelotta}, {Saajasto}, \& {Fuller}}]{2019ApJ...883...95S}
{Soam}, A., {Liu}, T., {Andersson}, B.~G., {et~al.} 2019, \apj, 883, 95,
  \dodoi{10.3847/1538-4357/ab39dd}

\bibitem[{{Soler} {et~al.}(2013){Soler}, {Hennebelle}, {Martin},
  {Miville-Desch{\^e}nes}, {Netterfield}, \& {Fissel}}]{2013ApJ...774..128S}
{Soler}, J.~D., {Hennebelle}, P., {Martin}, P.~G., {et~al.} 2013, \apj, 774,
  128, \dodoi{10.1088/0004-637X/774/2/128}

\bibitem[{{Soler} {et~al.}(2017){Soler}, {Ade}, {Angil{\`e}}, {Ashton},
  {Benton}, {Devlin}, {Dober}, {Fissel}, {Fukui}, {Galitzki}, {Gandilo},
  {Hennebelle}, {Klein}, {Li}, {Korotkov}, {Martin}, {Matthews}, {Moncelsi},
  {Netterfield}, {Novak}, {Pascale}, {Poidevin}, {Santos}, {Savini}, {Scott},
  {Shariff}, {Thomas}, {Tucker}, {Tucker}, \&
  {Ward-Thompson}}]{2017A&A...603A..64S}
{Soler}, J.~D., {Ade}, P.~A.~R., {Angil{\`e}}, F.~E., {et~al.} 2017, \aap, 603,
  A64, \dodoi{10.1051/0004-6361/201730608}

\bibitem[{{Stein}(1966)}]{1966ApJ...144..318S}
{Stein}, W. 1966, \apj, 144, 318, \dodoi{10.1086/148606}

\bibitem[{{Stephens} {et~al.}(2017){Stephens}, {Dunham}, {Myers}, {Pokhrel},
  {Sadavoy}, {Vorobyov}, {Tobin}, {Pineda}, {Offner}, {Lee}, {Kristensen},
  {J{\o}rgensen}, {Goodman}, {Bourke}, {Arce}, \&
  {Plunkett}}]{2017ApJ...846...16S}
{Stephens}, I.~W., {Dunham}, M.~M., {Myers}, P.~C., {et~al.} 2017, \apj, 846,
  16, \dodoi{10.3847/1538-4357/aa8262}

\bibitem[{{Stod{\'o}lkiewicz}(1963)}]{1963AcA....13...30S}
{Stod{\'o}lkiewicz}, J.~S. 1963, \actaa, 13, 30

\bibitem[{{Stutz} \& {Gould}(2016)}]{2016A&A...590A...2S}
{Stutz}, A.~M., \& {Gould}, A. 2016, \aap, 590, A2,
  \dodoi{10.1051/0004-6361/201527979}

\bibitem[{{Sun} {et~al.}(2006){Sun}, {Kramer}, {Ossenkopf}, {Bensch},
  {Stutzki}, \& {Miller}}]{2006A&A...451..539S}
{Sun}, K., {Kramer}, C., {Ossenkopf}, V., {et~al.} 2006, \aap, 451, 539,
  \dodoi{10.1051/0004-6361:20054256}

\bibitem[{{Tahani} {et~al.}(2018){Tahani}, {Plume}, {Brown}, \&
  {Kainulainen}}]{2018A&A...614A.100T}
{Tahani}, M., {Plume}, R., {Brown}, J.~C., \& {Kainulainen}, J. 2018, \aap,
  614, A100, \dodoi{10.1051/0004-6361/201732219}

\bibitem[{{Tahani} {et~al.}(2019){Tahani}, {Plume}, {Brown}, {Soler}, \&
  {Kainulainen}}]{2019A&A...632A..68T}
{Tahani}, M., {Plume}, R., {Brown}, J.~C., {Soler}, J.~D., \& {Kainulainen}, J.
  2019, \aap, 632, A68, \dodoi{10.1051/0004-6361/201936280}

\bibitem[{{Tamura} {et~al.}(1995){Tamura}, {Hough}, \&
  {Hayashi}}]{1995ApJ...448..346T}
{Tamura}, M., {Hough}, J.~H., \& {Hayashi}, S.~S. 1995, \apj, 448, 346,
  \dodoi{10.1086/175965}

\bibitem[{{Tamura} {et~al.}(1988){Tamura}, {Yamashita}, {Sato}, {Nagata}, \&
  {Gatley}}]{1988MNRAS.231..445T}
{Tamura}, M., {Yamashita}, T., {Sato}, S., {Nagata}, T., \& {Gatley}, I. 1988,
  \mnras, 231, 445, \dodoi{10.1093/mnras/231.2.445}

\bibitem[{{Tang} {et~al.}(2019){Tang}, {Koch}, {Peretto}, {Novak},
  {Duarte-Cabral}, {Chapman}, {Hsieh}, \& {Yen}}]{2019ApJ...878...10T}
{Tang}, Y.-W., {Koch}, P.~M., {Peretto}, N., {et~al.} 2019, \apj, 878, 10,
  \dodoi{10.3847/1538-4357/ab1484}

\bibitem[{{Targon} {et~al.}(2011){Targon}, {Rodrigues}, {Cerqueira}, \&
  {Hickel}}]{2011ApJ...743...54T}
{Targon}, C.~G., {Rodrigues}, C.~V., {Cerqueira}, A.~H., \& {Hickel}, G.~R.
  2011, \apj, 743, 54, \dodoi{10.1088/0004-637X/743/1/54}

\bibitem[{{Tobin} {et~al.}(2016){Tobin}, {Looney}, {Li}, {Chandler}, {Dunham},
  {Segura-Cox}, {Sadavoy}, {Melis}, {Harris}, {Kratter}, \&
  {Perez}}]{2016ApJ...818...73T}
{Tobin}, J.~J., {Looney}, L.~W., {Li}, Z.-Y., {et~al.} 2016, \apj, 818, 73,
  \dodoi{10.3847/0004-637X/818/1/73}

\bibitem[{{Tomisaka}(2015)}]{2015ApJ...807...47T}
{Tomisaka}, K. 2015, \apj, 807, 47, \dodoi{10.1088/0004-637X/807/1/47}

\bibitem[{{Ungerechts} \& {Thaddeus}(1987)}]{1987ApJS...63..645U}
{Ungerechts}, H., \& {Thaddeus}, P. 1987, \apjs, 63, 645,
  \dodoi{10.1086/191176}

\bibitem[{{Vaillancourt}(2006)}]{2006PASP..118.1340V}
{Vaillancourt}, J.~E. 2006, \pasp, 118, 1340, \dodoi{10.1086/507472}

\bibitem[{{V{\'a}zquez-Semadeni} {et~al.}(2011){V{\'a}zquez-Semadeni},
  {Banerjee}, {G{\'o}mez}, {Hennebelle}, {Duffin}, \&
  {Klessen}}]{2011MNRAS.414.2511V}
{V{\'a}zquez-Semadeni}, E., {Banerjee}, R., {G{\'o}mez}, G.~C., {et~al.} 2011,
  \mnras, 414, 2511, \dodoi{10.1111/j.1365-2966.2011.18569.x}

\bibitem[{{Walawender} {et~al.}(2008){Walawender}, {Bally}, {Francesco},
  {J{\o}rgensen}, \& {Getman}}]{2008hsf1.book..346W}
{Walawender}, J., {Bally}, J., {Francesco}, J.~D., {J{\o}rgensen}, J., \&
  {Getman}, K.~. 2008, {NGC 1333: A Nearby Burst of Star Formation}, ed.
  B.~{Reipurth}, 346

\bibitem[{{Wang} {et~al.}(2019){Wang}, {Lai}, {Eswaraiah}, {Pattle}, {Di
  Francesco}, {Johnstone}, {Koch}, {Liu}, {Tamura}, {Furuya}, {Onaka},
  {Ward-Thompson}, {Soam}, {Kim}, {Lee}, {Lee}, {Mairs}, {Arzoumanian}, {Kim},
  {Hoang}, {Hwang}, {Liu}, {Berry}, {Bastien}, {Hasegawa}, {Kwon}, {Qiu},
  {Andr{\'e}}, {Aso}, {Byun}, {Chen}, {Chen}, {Chen}, {Ching}, {Cho}, {Choi},
  {Chrysostomou}, {Chung}, {Coud{\'e}}, {Doi}, {Dowell}, {Drabek-Maunder},
  {Duan}, {Eyres}, {Falle}, {Fanciullo}, {Fiege}, {Franzmann}, {Friberg},
  {Friesen}, {Fuller}, {Gledhill}, {Graves}, {Greaves}, {Griffin}, {Gu}, {Han},
  {Hatchell}, {Hayashi}, {Holland}, {Houde}, {Inoue}, {Inutsuka}, {Iwasaki},
  {Jeong}, {Kanamori}, {Kang}, {Kang}, {Kang}, {Kataoka}, {Kawabata}, {Kemper},
  {Kim}, {Kim}, {Kim}, {Kim}, {Kirk}, {Kobayashi}, {Konyves}, {Kwon},
  {Lacaille}, {Lee}, {Lee}, {Lee}, {Lee}, {Li}, {Li}, {Li}, {Liu}, {Liu},
  {Lyo}, {Matsumura}, {Matthews}, {Moriarty-Schieven}, {Nagata}, {Nakamura},
  {Nakanishi}, {Ohashi}, {Park}, {Parsons}, {Pascale}, {Peretto}, {Pon}, {Pyo},
  {Qian}, {Rao}, {Rawlings}, {Retter}, {Richer}, {Rigby}, {Robitaille},
  {Sadavoy}, {Saito}, {Savini}, {Scaife}, {Seta}, {Shinnaga}, {Tang},
  {Tomisaka}, {Tsukamoto}, {van Loo}, {Wang}, {Whitworth}, {Yen}, {Yoo},
  {Yuan}, {Yun}, {Zenko}, {Zhang}, {Zhang}, {Zhang}, {Zhou}, \&
  {Zhu}}]{2019ApJ...876...42W}
{Wang}, J.-W., {Lai}, S.-P., {Eswaraiah}, C., {et~al.} 2019, \apj, 876, 42,
  \dodoi{10.3847/1538-4357/ab13a2}

\bibitem[{{Ward-Thompson} {et~al.}(2017){Ward-Thompson}, {Pattle}, {Bastien},
  {Furuya}, {Kwon}, {Lai}, {Qiu}, {Berry}, {Choi}, {Coud{\'e}}, {Di Francesco},
  {Hoang}, {Franzmann}, {Friberg}, {Graves}, {Greaves}, {Houde}, {Johnstone},
  {Kirk}, {Koch}, {Kwon}, {Lee}, {Li}, {Matthews}, {Mottram}, {Parsons}, {Pon},
  {Rao}, {Rawlings}, {Shinnaga}, {Sadavoy}, {van Loo}, {Aso}, {Byun},
  {Eswaraiah}, {Chen}, {Chen}, {Chen}, {Ching}, {Cho}, {Chrysostomou}, {Chung},
  {Doi}, {Drabek-Maunder}, {Eyres}, {Fiege}, {Friesen}, {Fuller}, {Gledhill},
  {Griffin}, {Gu}, {Hasegawa}, {Hatchell}, {Hayashi}, {Holland}, {Inoue},
  {Inutsuka}, {Iwasaki}, {Jeong}, {Kang}, {Kang}, {Kang}, {Kawabata}, {Kemper},
  {Kim}, {Kim}, {Kim}, {Kim}, {Kim}, {Kim}, {Lacaille}, {Lee}, {Lee}, {Li},
  {Li}, {Liu}, {Liu}, {Liu}, {Liu}, {Lyo}, {Mairs}, {Matsumura},
  {Moriarty-Schieven}, {Nakamura}, {Nakanishi}, {Ohashi}, {Onaka}, {Peretto},
  {Pyo}, {Qian}, {Retter}, {Richer}, {Rigby}, {Robitaille}, {Savini}, {Scaife},
  {Soam}, {Tamura}, {Tang}, {Tomisaka}, {Wang}, {Wang}, {Whitworth}, {Yen},
  {Yoo}, {Yuan}, {Zhang}, {Zhang}, {Zhou}, {Zhu}, {Andr{\'e}}, {Dowell},
  {Falle}, \& {Tsukamoto}}]{2017ApJ...842...66W}
{Ward-Thompson}, D., {Pattle}, K., {Bastien}, P., {et~al.} 2017, \apj, 842, 66,
  \dodoi{10.3847/1538-4357/aa70a0}

\bibitem[{{Wardle} \& {Kronberg}(1974)}]{1974ApJ...194..249W}
{Wardle}, J.~F.~C., \& {Kronberg}, P.~P. 1974, \apj, 194, 249,
  \dodoi{10.1086/153240}

\bibitem[{{Wenger} {et~al.}(2000){Wenger}, {Ochsenbein}, {Egret}, {Dubois},
  {Bonnarel}, {Borde}, {Genova}, {Jasniewicz}, {Lalo{\"e}}, {Lesteven}, \&
  {Monier}}]{2000A&AS..143....9W}
{Wenger}, M., {Ochsenbein}, F., {Egret}, D., {et~al.} 2000, \aaps, 143, 9,
  \dodoi{10.1051/aas:2000332}

\bibitem[{{Zari} {et~al.}(2016){Zari}, {Lombardi}, {Alves}, {Lada}, \&
  {Bouy}}]{2016A&A...587A.106Z}
{Zari}, E., {Lombardi}, M., {Alves}, J., {Lada}, C.~J., \& {Bouy}, H. 2016,
  \aap, 587, A106, \dodoi{10.1051/0004-6361/201526597}

\bibitem[{{Zucker} {et~al.}(2018){Zucker}, {Schlafly}, {Speagle}, {Green},
  {Portillo}, {Finkbeiner}, \& {Goodman}}]{2018ApJ...869...83Z}
{Zucker}, C., {Schlafly}, E.~F., {Speagle}, J.~S., {et~al.} 2018, \apj, 869,
  83, \dodoi{10.3847/1538-4357/aae97c}

\bibitem[{{Zucker} {et~al.}(2019){Zucker}, {Speagle}, {Schlafly}, {Green},
  {Finkbeiner}, {Goodman}, \& {Alves}}]{2019ApJ...879..125Z}
{Zucker}, C., {Speagle}, J.~S., {Schlafly}, E.~F., {et~al.} 2019, \apj, 879,
  125, \dodoi{10.3847/1538-4357/ab2388}

\end{thebibliography}
\bibliographystyle{aasjournal}


\listofchanges

\end{document}